\documentclass[journal]{IEEEtran}

\IEEEoverridecommandlockouts

\usepackage{amsmath}
\usepackage{amsfonts}
\usepackage{amssymb}
\usepackage{graphics}
\usepackage{pstricks}
\usepackage{array}
\usepackage{float}
\usepackage{epsf}
\usepackage{graphicx}
\usepackage{multirow}
\usepackage{cite}
\usepackage{acronym}
\usepackage{comment}
\usepackage{upgreek}
\usepackage{nicefrac}
\usepackage[font=footnotesize]{caption}
\usepackage{subcaption}

\acrodef{rocof}[RoCoF]{Rate of Change of Frequency}

\begin{document}

\newcommand{\bfg}[1]{\boldsymbol{#1}}
\newcommand{\bfp}[1]{\boldsymbol{#1}'}
\newcommand{\bfpp}[1]{\boldsymbol{#1}''}
\newcommand{\bfppp}[1]{\boldsymbol{#1}'''}
\newcommand{\bfd}[1]{\dot{\boldsymbol{#1}}}
\newcommand{\bfdd}[1]{\ddot{\boldsymbol{#1}}}
\newcommand{\bfddd}[1]{\dddot{\boldsymbol{#1}}}
\newcommand{\bfb}[1]{\boldsymbol{\rm #1}}
\newcommand{\flux}{\bfg \varphi}
\newcommand{\ii}{\imath}
\newcommand{\jj}{\jmath}
\newcommand{\geom}[1]{\bfg #1 \bfp #1}
\newcommand{\gw}[1]{\bfb \Omega_{#1}}
\newcommand{\gr}[1]{\frac{\geom{#1}}{#1^2}}
\newcommand{\gf}[1]{\rho_{#1} + \gw{#1}}
\newcommand{\inner}[1]{\frac{\bfg #1}{#1} \cdot \frac{\bfp #1}{#1}}
\newcommand{\out}[1]{\frac{\bfg #1}{#1} \wedge \frac{\bfp #1}{#1}}
\newcommand{\e}[1]{\boldsymbol{\rm e}_{\rm #1}}
\newcommand{\ep}[1]{\boldsymbol{\rm e}'_{\rm #1}}
\newcommand{\eh}[1]{\hat{\boldsymbol{\rm e}}_{\rm #1}}
\newcommand{\sw}{\bfg \omega_u}
\newcommand{\dt}{\partial_t}
\newcommand{\darboux}{\bfg \omega_{\rm d}}
\newcommand{\T}{{\scriptstyle \bfb T}}
\newcommand{\B}{{\scriptstyle \bfb B}}
\newcommand{\N}{{\scriptstyle \bfb N}}
\newcommand{\Td}{\dot{\bfb T}}
\newcommand{\Bd}{\dot{\bfb B}}
\newcommand{\Nd}{\dot{\bfb N}}
\newcommand{\Tp}{\bfb T'}
\newcommand{\Bp}{\bfb B'}
\newcommand{\Np}{\bfb N'}
\newcommand{\vd}{v_{\rm d}}
\newcommand{\vq}{v_{\rm q}}
\newcommand{\vo}{v_{\rm o}}
\newcommand{\vdp}{v'_{\rm d}}
\newcommand{\vqp}{v'_{\rm q}}
\newcommand{\vop}{v'_{\rm o}}
\newcommand{\vrp}{\hat{\bfg v}'}
\renewcommand{\th}[1]{\theta_{#1}}
\newcommand{\thp}[1]{\theta'_{#1}}
\newcommand{\V}[1]{V_{#1}}
\newcommand{\Vp}[1]{V'_{#1}}
\renewcommand{\ss}[2]{{\rm ss}(\th{#1}, \th{#2})}
\renewcommand{\sc}[2]{{\rm sc}(\th{#1}, \th{#2})}
\newcommand{\wo}{\omega_o}
\newcommand{\ur}{\varrho_{r}}
\newcommand{\ar}{\alpha_{u}}
\renewcommand{\wr}{\varrho_{u}}
\newcommand{\wk}{\omega_{\kappa}}
\newcommand{\wt}{\omega_{\tau}}
\newcommand{\lancret}{\omega_{\rm d}}
\newcommand{\w}[1]{w_{#1}}
\newcommand{\F}{\mathbb{F}}
\newcommand{\tnb}{(\T, \N, \B)}
\newcommand{\vtnb}{\bfg v_{\scriptsize \rm TNB}}
\newcommand{\vtnbt}{\hat{\bfg v}'_{\scriptsize \rm TNB}}
\newcommand{\poisson}[2]{\bfg #2 + \darboux \times \bfg #1}
\newcommand{\coriolis}[2]{%
  \bfg #2' + 2 \, \darboux \times \bfg #2 +
  \darboux' \times \bfg #1 +
  \darboux \times (\darboux \times \bfg #1)%
}
\newcommand{\triple}[3]{\bfg #1 \times (\bfg #2 \times \bfg #3)}
\newcommand{\Wgeom}[1]{\mathit{\hat{\Omega}}_{\bfg #1}}
\newcommand{\radial}{\partial_t}
\newcommand{\orbit}{\bfg \omega_r}
\newcommand{\mom}{\bfb p}
\newcommand{\inertia}[1]{\bfg I_{#1}}
\newcommand{\direc}[1]{\nabla_{\bfg #1}}
\newcommand{\unitvec}[1]{\bfb e_{\bfg #1}}
\newcommand{\Tr}{^{\intercal}}

\newtheorem{proposition}{Proposition}
\newtheorem{definition}{Definition}

\title{Instantaneous Power Theory Revisited \\ with Classical Mechanics}


\author{Federico~Milano,~\IEEEmembership{Fellow,~IEEE}, Georgios Tzounas,~\IEEEmembership{Member,~IEEE}, and Ioannis Dassios%
  \thanks{F.~Milano, G.~Tzounas and I.~Dassios are
    with the School of Electrical and Electronic Engineering,
    University College Dublin, Dublin, D04V1W8, Ireland.  e-mails:
    \{federico.milano, georgios.tzounas, ioannis.dassios\}@ucd.ie}%
  \thanks{This work is supported by the Sustainable Energy Authority
    of Ireland (SEAI) by funding F.~Milano and I.~Dassios under project FRESLIPS,
    Grant No.~RDD/00681.}%
}

\maketitle

\begin{abstract}
  The paper revisits the concepts of instantaneous active and reactive powers and provides a novel definition for basic circuit elements based on quantities utilized in classical mechanics, such as absolute and relative velocity, momentum density, angular momentum and apparent forces.  The discussion leverages from recent publications by the authors that interpret the voltage and current as \textit{velocities} in generalized Lagrangian coordinates.  The main result of the paper is a general and compact expression for the instantaneous active and reactive power of inductances, capacitances and resistances as a multivector proportional to the generalized kinetic energy and the geometric frequency multivector.  Several numerical examples considering stationary and transient sinusoidal and non-sinusoidal conditions are discussed in the case study.
\end{abstract}


\IEEEpeerreviewmaketitle

\section*{Notation}

Unless otherwise indicated, scalars are represented in Italic, e.g.,
$x$, $X$; vectors in lower case bold, e.g.,
$\bfg x = (x_1, x_2, x_3)$; pseudovectors in upper case bold, e.g.,
$\bfg X$; and multivectors in upper case Italic with a hat, e.g.,
$\hat{X}$.  Vectors and pseudovectors have order 3.

\vspace{3mm}

\subsubsection*{Scalars}

\begin{itemize}[\IEEEiedlabeljustifyl \IEEEsetlabelwidth{Z} \labelsep 0.9cm]
\item[$C$] capacitance
\item[$\ell$] momentum density
\item[$I$] moment of inertia
\item[$\mathcal{L}$] Lagrangian
\item[$L$] inductance
\item[$m$] mass
\item[$p$] instantaneous active power
\item[$s$] arc length of a curve
\item[$R$] resistance
\item[$t$] time
\item[$T$] kinetic energy
\item[$U$] potential energy
\item[$\theta$] voltage phase angle
\item[$\alpha_r, \ar$] radial components of $2^{\rm nd}$ time derivatives
\item[$\ur, \wr$] radial components of $1^{\rm st}$ time derivatives
\item[$\omega_{\rm d}^2$] Lancret curvature ($=\wk^2 + \wt^2$)
\item[$\wk$] azimuthal frequency (related to curvature)
\item[$\wt$] torsional frequency (related to torsion)
\end{itemize}

\vspace{3mm}

\subsubsection*{Vectors}

\begin{itemize}[\IEEEiedlabeljustifyl \IEEEsetlabelwidth{Z} \labelsep 0.9cm]
\item[$\bfg 0$] null vector %
\item[$\bfg f$] force
\item[$\bfg \ii$] current
\item[$\bfg n$] normal vector before normalization
\item[$\mom$] momentum 
\item[$\bfg q$] electric charge 
\item[$\bfg r$] position  
\item[$\bfg u$] velocity
\item[$\bfg v$] voltage 
\item[$\bfg \alpha$] relative acceleration
\item[$\bfg \nu$] relative velocity
\item[$\bfg \nu_{\rm d}$] projection of $\bfg \xi$ onto the Darboux vector
\item[$\bfg \xi$] relative position
\item[$\bfg \pi$] relative momentum
\item[$\flux$] magnetic flux
\end{itemize}

\vspace{3mm}

\subsubsection*{Pseudovectors}

\begin{itemize}[\IEEEiedlabeljustifyl \IEEEsetlabelwidth{Z} \labelsep 0.9cm]
\item[$\bfg L$] angular momentum %
\item[$\bfg N$] torque %
\item[$\bfg R$] rotatum residual %
\item[$\bfg Q$] instantaneous reactive power %
\item[$\boldsymbol{\mathit{\Lambda}}$] relative angular momentum
\item[$\darboux$] Darboux vector
\item[$\orbit$] orbital angular velocity vector
\item[$\sw$] binormal vector before normalization
\item[$\bfg \omega_{\xi}$] relative rotation vector
\end{itemize}

\vspace{3mm}

\subsubsection*{Multivectors}

\begin{itemize}[\IEEEiedlabeljustifyl \IEEEsetlabelwidth{Z} \labelsep 0.9cm]
\item[$\hat{E}$] energy multivector %
\item[$\hat{L}$] momentum multivector %
\item[$\hat{R}$] power residual multivector %
\item[$\hat{S}$] instantaneous power multivector %
\item[$\hat{W}$] power multivector %
\end{itemize}

\vspace{3mm}

\subsubsection*{Operators}

\begin{itemize}[\IEEEiedlabeljustifyl \IEEEsetlabelwidth{Z} \labelsep 0.9cm]
\item[$\direc{a}$] directional derivative
\item[$\bfb F$] Frenet-frame transformation matrix
\item[$\inertia{\bfg a}$] moment of inertia tensor %
\item[$\bfg \Omega_{\rm d}$] rotation matrix ($= \darboux \times$)
\item[$\Wgeom{a}$] geometric frequency multivector %
\end{itemize}

\vspace{3mm}

\subsubsection*{Coordinates}

\begin{itemize}[\IEEEiedlabeljustifyl \IEEEsetlabelwidth{Z} \labelsep 0.9cm]
\item[$\boldsymbol{\rm e}_i$] $i$-th vector of an orthonormal basis
\item[$\B$]  binormal vector of the Frenet frame %
\item[$\N$]  normal vector of the Frenet frame %
\item[$\T$]  tangent vector of the Frenet frame %
\end{itemize}

\section{Introduction}
\label{sec:intro}

\subsection{Motivation}

The definition of instantaneous power, in particular, of the reactive one, has been object of study since the early years of electrification.  It still periodically raises intense 
discussions on public fora of the power system community.  The aim of most definitions and theories proposed so far has been pragmatic, that is, 
to properly calculate losses and the power factor.  This is, for example, the main goal of international standards, e.g., \cite{IEEEStd1459}.  Despite all these discussions, there remain unresolved issues \cite{Kirkham:2022}.  For example, the definitions 
in \cite{IEEEStd1459} require stationary conditions and the knowledge of the period $T$ of voltages and currents.  But the period is not defined during transients and, even in stationary conditions, it is rarely exactly the nominal one.  Another underlying issue is that, in most works, instantaneous power is formulated through equations without an attempt to define its physical meaning.  This is particularly true for reactive power which is often defined as what is not, e.g., \textit{non-active} or \textit{powerless} \cite{Depenbrock:1996}, rather than what it is.  This work addresses these issues and proposes a physical meaning of the instantaneous power based on classical mechanics, variational principles and differential geometry.

\subsection{Literature Review}

The literature and discussions on active and reactive power are as old as AC power systems themselves.  The topic has always been controversial.  Already in \cite{Steinmetz:1900}, Steinmetz expresses his concern on the definition of $P$ and $Q$ in terms of the product of the voltage and current phasor magnitudes.  The issue is that, as the instantaneous power has the double of the frequency of the voltage and current, the product of the phasors, which 
are defined at the fundamental frequency, cannot be a consistent quantity.  This impasse was later solved by defining the active and reactive powers as averages over a period, but as mentioned above, this requires assuming stationary conditions and to know \textit{a priori} the period.

To solve these inconsistencies, developments on the theory of active and reactive power recognized that the key of the problem was the identification of two components of the current, one parallel and one in quadrature with the voltage.  This led to the concepts of \textit{active} and \textit{non-active} current \cite{Depenbrock:1996}, \textit{power} and \textit{powerless} current \cite{Volker:2008}; $\alpha$ and $\beta$ current components \cite{Akagi:1984}; and \textit{instantaneous active} and \textit{instantaneous reactive} currents \cite{Peng:1996}.  All these approaches, starting from Fryze in 1932 \cite{Depenbrock:1996}, are substantially attempts to define a set of coordinates.  However, it is 
more
recent the proposal of considering voltage and currents as generalized vectors \cite{Willems:1992}.  This has put the instantaneous power theory developed in \cite{IPT} in the context of vector algebra and paved the way to the application of geometric algebra \cite{4450060504, Stankovic:2000, 1308315, 4126799, 4512338, 5316097, 6161614, 2015talebi, 2016barry, 2018ishihara, math9111295}.  In this work, we use as starting point the geometric approach  of these works and combine it with two related but, so far, not considered together, concepts.  

The first concept is the recent interpretation by the first author of ``frequency'' as a geometric quantity \cite{freqgeom}.  This definition requires the utilization of differential geometry, which, differently from geometric algebra, studies the kinematics of a curve (or a surface) in three or higher dimensions.  The ultimate goal of differential geometry is to determine the \textit{invariants} of a curve.  The authors have shown that certain invariants such as arc length, curvature, and torsion are intrinsically related to the frequency and transient behavior of voltages and currents in three- and multi-phase electric circuits \cite{freqfrenet, paradoxes, parkfrenet}.  Another feature of differential geometry is that it considers instantaneous quantities, and, thus it is fully consistent with the instantaneous power theory.  More importantly for the developments of this work, differential geometry allows defining a set of intrinsic coordinates, called Frenet frame, that identifies the parallel and quadrature components of voltage and current as they evolve in time.  In this work, we exploit this property to define active and reactive components of the instantaneous power.

The second concept is the Lagrangian approach to describe the dynamics of a physical system in terms of generalized coordinates, namely, positions and momenta.  That is, not only we assume that voltages and currents are vectors with respect to some set of orthogonal coordinates, but that the components of the voltage and current vectors are the generalized coordinates in the sense of Lagrange.   Using the Lagrangian equation to describe electric circuits can be found in several textbooks, e.g., \cite{Goldstein:1980}.  However, for our aims, we need that voltages and currents have to be both velocities and forces, depending on the ``domain'', electrical or magnetic.  This approach, which is not usual, was proposed for the first time in \cite{Chua:1974} and assumes that the energy stored in capacitances and inductances are both kinetic, and then defines potential energy using fluxes and electric charges as generalized positions.  This formulation has never really become popular, mostly because --- we believe --- the Lagrangian function provides the same information that can be obtained from Kirchhoff laws, only in more involved way.  However, the formulation in \cite{Chua:1974} appears as the missing link between the kinematic approach given by differential geometry and the dynamic (i.e., generalized laws of motion) of an electric circuit.

\subsection{Contributions}

This work proposes a ``physical'' interpretation and definition of instantaneous active and reactive powers for basic circuit elements, i.e., capacitances, inductances and resistances, based on differential geometry and classical mechanics.  This definition is compatible with that provided by the well-known instantaneous power theory \cite{Peng:1996} and, thus also with the FBD-method \cite{Depenbrock:1996}.  Considering voltages and currents as generalized velocities and forces, the proposed approach allows decomposing the instantaneous active power in a variety of terms linked to the time-derivative of generalized kinetic energy and the Lagrangian function.  We also show that the instantaneous active and reactive powers can be viewed as components of the second time derivative of generalized momentum density and angular momentum, respectively.  The main contribution of the paper is a formula that links the instantaneous power, both active and reactive, with the generalized kinetic energy and the geometric frequency defined by the second author in \cite{freqgeom}.  The use of differential geometry and, in concrete, of the Frenet frame, allows describing the various terms in which the instantaneous power is decomposed in terms of geometrical invariants, that is, quantities that are independent from the coordinate system utilized to measure them.  With this aim, we show the link between the orbital angular velocity that appears in the expression of the angular momentum and of the kinetic energy and invariants obtained from the Frenet frame apparatus.   Referring voltages and currents onto the Frenet frame also allows expressing the instantaneous power components in terms of generalized apparent accelerations.  


\subsection{Organization}

The remainder of the paper is organized as follows.  Section~\ref{sec:generalized} states the correspondence between mechanical, electrical and magnetic domains.  Electrical and magnetic quantities are treated as generalized positions, velocities, momenta and forces. 
Based on this framework,  Section~\ref{sec:power} focuses on the mechanical domain, recalls the definitions of momentum density and angular momentum, and defines the instantaneous active and reactive powers in terms of the kinetic energy and the second time derivative of the angular momentum and momentum density.  
Section~\ref{sec:examples} illustrates the formulas  derived in Section~\ref{sec:power} through a series of examples.  The examples are aimed at showing relevant special cases, including stationary balanced and unbalanced, as well as sinusoidal and non-sinusoidal systems.  Section~\ref{sec:conc} draws conclusions and outlines future work.

\section{Generalized Quantities}
\label{sec:generalized}

The definition and interpretation of instantaneous active and reactive power given in this work are based on an analogy with quantities that appear in classical mechanics, such as angular momentum and torque. Before presenting the proposed interpretation, we first provide the concept of generalized quantities, which is typical of the Lagrangian formulation, but we extend it to electrical and magnetic quantities.

In the Lagrangian approach, generalized quantities are positions and
momenta.  With this aim, we use as starting point the correspondences
proposed in \cite{Chua:1974} and, more recently, in \cite{PHbook},
where the generalized quantities are defined as shown in Table~\ref{tab:gen}.  In the table, conventional mechanical quantities are
as position $\bfg r$, velocity $\bfg u$, momentum $\mom$, and force
$\bfg f$.  The electrical and magnetic quantities are the electric
charge $\bfg q$ and current $\bfg \ii$, the flux $\bfg \varphi$, and the
voltage $\bfg v$.  All these quantities are considered as
\textit{vectors} in some opportunely defined coordinates.  The
definition of the coordinates is a critical aspect of the generalized
approach and of this work in particular and is discussed in detail in
following sections.  Finally, the physical properties of 
mechanical, magnetic and electrical media are represented by mass $m$,
inductance $L$ and capacitance $C$, which, for simplicity, are assumed constant.

\begin{table}[b!]
  \centering
  \caption{\small Generalized positions, velocities, momenta and forces}
  \label{tab:gen}
  \small
  \begin{tabular}{ccccc}
    \hline
    Domain & Position & Velocity & Momentum & Force \\
    \hline
    Mechanical & $\bfg r$ & $\bfg u = \bfg r'$ & $\mom = m \bfg u$ & $\bfg f = \mom'$ \\
    Electrical & $\bfg \varphi$ & $\bfg v=\bfg \varphi'$ & $\bfg q=C\bfg v$ & $\bfg \imath = \bfg q'$ \\
    Magnetic & $\bfg q$ & $\bfg \imath=\bfg q'$ & $\bfg \varphi=L\bfg \ii$ & $\bfg v= \bfg \varphi'$ \\
    \hline
  \end{tabular}
\end{table}

The duality of electric and magnetic equations is apparent in this formulation.
The constitutive equations of condensers and inductors can be all viewed as generalized Newton's second law of motion, where $C$ and $L$
take the meaning of generalized masses.  Moreover, considering that the conventional mechanical kinetic and potential energies are defined
as:
\begin{equation}
  \label{eq:TU}
  T = \frac{1}{2} m |\bfg u|^2 \, , \qquad
  U = -\bfg f \cdot \bfg r \, , 
\end{equation}
the electrical kinetic and potential energies are:
\begin{equation}
  \label{eq:TUe}
  T_e = \frac{1}{2} C |\bfg v|^2 \, , \qquad
  U_e = -\bfg \ii \cdot \bfg \varphi \, , 
\end{equation}
and the magnetic kinetic and potential energies are:
\begin{equation}
  \label{eq:TUm}
  T_m = \frac{1}{2} L |\bfg \ii|^2 \, , \qquad
  U_m = -\bfg v \cdot \bfg q \, .
\end{equation}

It is important to note that the quantities that appear in the magnetic domain cannot be mixed up with the quantities that appear in the electrical domain.  For example, it is conceptually incorrect to take the expression of the magnetic flux from the magnetic domain and substitute it into the expression of the potential energy in the electrical domain.  This is because the expressions above apply, for each domain, to a specific component.  The interested reader can find more details on this point in \cite{PHbook}, which describes port-Hamiltonian systems and extensively discusses the analogies shown in Table~\ref{tab:gen}. 

It is also worth noting that this is not the usual notation utilized in most works and textbooks.  More often, the correspondence between mechanical and electrical systems is done by assuming only the charge as position and the current as velocity, and defining $T_e$ as a potential energy (see, for example, Chapter~13 of \cite{Goldstein:1980}).  The advantages of the notation adopted in \cite{Chua:1974} will be evident in the remainder of this paper.

\section{Instantaneous Power}
\label{sec:power}

This section defines the instantaneous active and reactive powers based on classical mechanics and differential geometry.  We have
chosen to present the mathematical developments 
in Sections~\ref{sub:mommulti} to \ref{sub:power}
using 
mechanical 
quantities, since these 
are more intuitive and
can be visualized 
better than electrical and magnetic ones.  The extension to electrical and magnetic quantities is then carried out in Section~\ref{sub:EMdomain}
based on the correspondences given in Table~\ref{tab:gen}.  Moreover,
before introducing the proposed definition of instantaneous active and
reactive powers, we need to introduce another set of mechanical
quantities, namely momentum, angular momentum and momentum density of
a point particle with mass $m$.  We use the latter two quantities to
define a momentum multivector.  We then define the first and second
time derivatives of the angular momentum and momentum density, as well as the
corresponding
energy and power multivectors.  Finally, we show that the power multivector includes 
the instantaneous active and reactive powers and that these are functions of the kinetic energy and the curvature of the particle's trajectory.

\subsection{Momentum Multivector}
\label{sub:mommulti}

\begin{figure}[ht!]
  \centering
  \resizebox{0.6\linewidth}{!}{\includegraphics{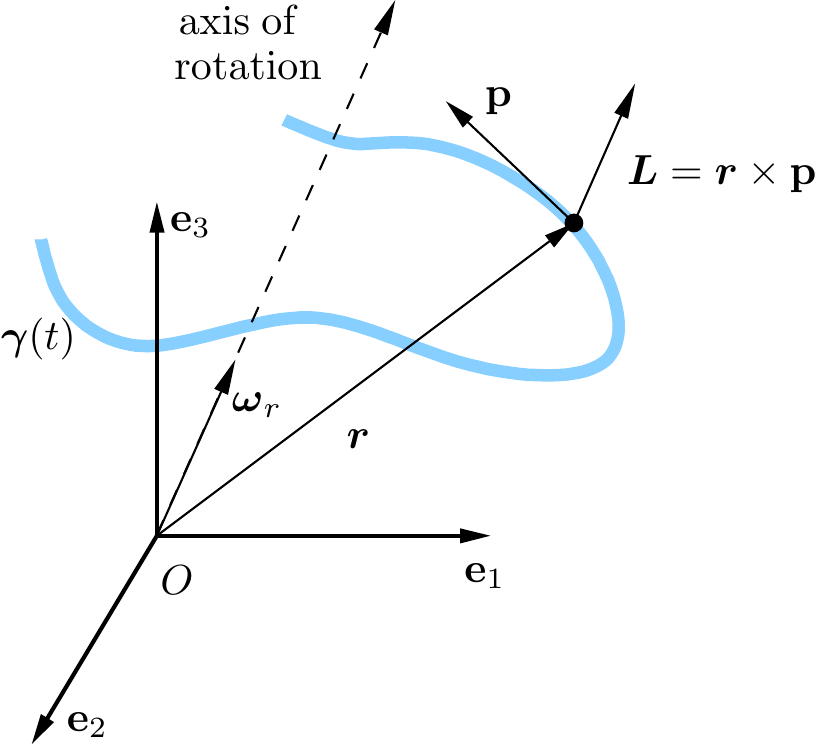}}
  \caption{Representation of momentum $\mom$ and angular momentum
    $\bfg L$ for a point particle of mass $m$ and position vector
    $\bfg r$. The particle rotates with angular velocity vector
    $\orbit$ with respect to the origin $O$.  By construction,
    $\bfg L \, \| \, \bfg \orbit$.}
  \label{fig:general}
\end{figure}

Figure \ref{fig:general} illustrates the relationships between
position vector $\bfg r$, momentum $\mom = m \bfg u = m \bfg r'$ and
angular momentum $\bfg L$ for a point particle of mass $m$ moving
along a space curve $\bfg \gamma(t)$.  The orbital angular velocity
vector $\orbit$ with respect to the origin $O$ is given by:
\begin{equation}
  \label{eq:darboux2}
  \orbit = \frac{\bfg r \times \bfg u}{|\bfg r|^2} \, .
\end{equation}

We can write the angular momentum as follows:
\begin{equation}
  \label{eq:L}
  \begin{aligned}
  \bfg L =
  \bfg r \times \mom &=
  m \, \bfg r \times \bfg u = m 
  |\bfg r|^2 \, \orbit \\ &= I \, \orbit  \, ,
  \end{aligned}
\end{equation}
where $I = m \, |\bfg r|^2$ is the moment of inertia with respect to
the origin $O$ of the coordinates where the vector $\bfg r$ is
defined.

The \textit{momentum density} $\ell$ is defined as the scalar quantity:
\begin{equation}
  \label{eq:ell}
  \ell = 
  \bfg r \cdot \mom =
  m \, \bfg r \cdot \bfg u =
  \frac{1}{2} I' \, ,
\end{equation}
and, introducing the \textit{radial speed} as:
\begin{equation}
  \label{eq:radial}
  \ur= \frac{\bfg r \cdot \bfg u}{|\bfg r|^2} = \frac{|\bfg r|'}{|\bfg r|} \, ,
\end{equation}
the momentum density can be written as:
\begin{equation}
  \label{eq:ell2}
  \ell = I \, \ur \, .
\end{equation}
%
%
%
%

Putting together the momentum density and the angular momentum, one
can define the following multivector\footnote{Note that in this work
  we use three dimensions, and thus the multivectors considered are
  equivalent to Hamiltonian quaternions.  Yet, the applicability of
  the proposed theory is not limited to three dimensions and can be,
  in principle, applied to multivectors of arbitrary dimensions.  }
(see also Appendix~\ref{app:clifford}):
\begin{equation}
  \hat{L} = \ell + \bfg L = I (\ur + \orbit) \, ,
  \nonumber
\end{equation}
or, equivalently:
%
\begin{equation}
  \label{eq:Lhat2}
  \hat{L} = I \, \Wgeom{r} \, ,
\end{equation}
where $\Wgeom{r}$ is the \textit{geometric frequency} operator proposed in \cite{freqgeom}.  In the remainder of this work, we utilize the geometric frequency as an operator that, when applied to a smooth vector $\bfg a(t)$, returns the following expression:
\begin{equation}
  \label{eq:Wgeom}
  \Wgeom{a} = \varrho_a + \bfg \omega_a =
  \frac{\bfg a \cdot \bfg a'}{|\bfg a|^2} +
  \frac{\bfg a \times \bfg a'}{|\bfg a|^2} \, .
\end{equation}
%

\subsection{Energy Multivector}

The time derivative of the angular momentum gives:
\begin{equation}
  \label{eq:dL}
  \bfg L' =
  \bfg u \times \mom + \bfg r \times \mom' =
  \bfg r \times \bfg f = \bfg N \, ,
\end{equation}
where $\bfg u \times \mom = \bfg 0$ because the momentum is parallel to the velocity and $\bfg N$ is the resultant torque applied to
the particle.

The time derivative of the momentum density satisfies the identity:
\begin{equation}
  \label{eq:dell}
  \begin{aligned}
    \ell' &= \frac{1}{2}I'' =
            \bfg u \cdot \mom + \bfg r \cdot \mom' \\
          &= m |\bfg u|^2 + \bfg r \cdot \bfg f \\
          &=  2T - U \, ,
  \end{aligned}
\end{equation}
where 
\begin{equation}
   \label{eq:T0}
   \begin{aligned}
    T &= \frac{1}{2} m |\bfg u|^2  = \frac{1}{2} I (\ur^2 + |\orbit|^2) \\
    &= \frac{1}{2} I |\Wgeom{r}|^2 = \frac{1}{2} \frac{|\hat{L}|^2}{I} \, ,
    \end{aligned}
\end{equation}
is the kinetic energy of the rotating mass; and $U$ is the potential energy from \eqref{eq:TU}.  This form appears also from the expressions of the potential energy given in \eqref{eq:TUe} and \eqref{eq:TUm} for the electrical and magnetic domains, respectively.

We define the \textit{energy multivector} $\hat{E}$ as the time derivative $\hat{L}'$ of
the momentum multivector $\hat{L}$:
\begin{equation}
  \label{eq:dLhat}
  \begin{aligned}
    \hat{E} = \hat{L}' &= \ell' + \bfg L' \\
    &= 2T - U + \bfg N \, .
  \end{aligned}
\end{equation}
All terms that appear in \eqref{eq:dLhat} have the units of an
energy, although the torque $\bfg N$ is conventionally expressed in
Nm.\footnote{Alternatively, one can interpret the torque as a
\textit{rotating energy}, not to be confused with the kinetic energy,
which is due to the rotation of the particle and is a scalar.}


\subsection{Power Multivector}
\label{sub:power}

Differentiating $\ell'$ and $\bfg L'$ with respect to time, one obtains:
\begin{equation}
  \label{eq:ddell}
  \begin{aligned}
    \ell'' =
    2T' - U'
    &= \bfg u' \cdot \mom + \bfg u \cdot \mom' +
      \bfg u \cdot \mom' + \bfg r \cdot \mom'' \\
    &= m \, (3 \, \bfg u \cdot \bfg u' + \bfg r \cdot \bfg u'' ) \, ,
  \end{aligned}
\end{equation}
where we have used the identity $\mom = m \bfg u$,
and
\begin{equation}
  \label{eq:ddL}
  \begin{aligned}
    \bfg L'' = \bfg N'
    &= \bfg u \times \mom'  + \bfg r \times \mom'' \\
    &= m ( \bfg u \times \bfg u'  + \bfg r \times \bfg u'' ) \, .
  \end{aligned}
\end{equation}
Both $\ell''$ and $\bfg L''$ have the dimensions of power.\footnote{We note that as early as in \cite{Steinmetz:1900}, Steinmetz noted that the reactive power is linked to the torque of a rotating machine.  However, Steinmetz links the torque itself, not its time derivative to the reactive power and does not consider generalized quantities nor the angular momentum and momentum density.}  The latter term, i.e.~the time derivative of the torque, is sometimes called \textit{rotatum}.  We define as \textit{instantaneous active power} the
term:
\begin{equation}
   \label{eq:pdef}
  \boxed{ p \equiv \bfg u \cdot \bfg f }
\end{equation}
where we have used Newton's 2nd law $\bfg f = \mom'$.  Note that:
\begin{equation}
  \label{eq:p2}
  p =
  \bfg u \cdot m \, \bfg u' =
  \frac{d}{dt} \left ( \frac{1}{2} m \, \bfg u \cdot \bfg u \right ) =
  T' \, ,
\end{equation}
and hence:
\begin{equation}
\label{eq:elld}
  \ell'' = p + T' - U' = p + \mathcal{L}' \, , 
\end{equation}
where $\mathcal{L}$ is the Lagrangian of the system.

Substituting in \eqref{eq:p2} the expression of $\bfg u'$ given in
\eqref{eq:du} (see Appendices \ref{app:clifford} and \ref{app:frenet}), namely:
\begin{equation}
  \bfg u' =
  \wr \, \bfg u + \sw \times \bfg u = \Wgeom{u} \otimes \bfg u  \, ,
  \nonumber
\end{equation}
the instantaneous active power and, hence, the time derivative of the kinetic energy, can be written as:
\begin{equation}
  \label{eq:dT}
  p = T' = m \, \bfg u \cdot \bfg u' = m \, |\bfg u|^2 \wr =
  2 \, T \, \wr \, .
\end{equation}
%
%

Also, since $\bfg u = \Wgeom{r} \otimes \bfg r$, the acceleration can be written as:
\begin{equation}
  \bfg u' = \Wgeom{u} \otimes (\Wgeom{r} \otimes \bfg r) \, ,
  \nonumber
\end{equation}
or, equivalently:
\begin{align}
   \label{eq:coriolis}
   \bfg u' &=  \frac{d}{dt} (\ur \, \bfg r + \orbit \times \bfg r) \\
   \nonumber
   &= \ur' \, \bfg r + \ur \, \bfg u + \orbit' \times \bfg r + \orbit \times \bfg u \\
   \nonumber
   & = \alpha_r \bfg r + 2 \orbit  \times \bfg u_{\|} +
  \orbit' \times \bfg r +
  \orbit \times (\orbit \times \bfg r) \, , \\ \nonumber
   & = \beta_r \bfg u_{\|} + 2 \orbit \times \bfg u_{\|} +
  \orbit' \times \bfg r +
  \orbit \times (\orbit \times \bfg r) \, , 
\end{align}
where we have defined:
\begin{align}
\nonumber
\bfg u_{\|} &= \ur \bfg r \, , \\
\nonumber
\alpha_r &= \frac{|\bfg r|''}{|\bfg r|} = \ur' + \ur^2 \, , \\
\beta_r &= \frac{\alpha_r}{\ur} = \frac{|\bfg r|''}{|\bfg r|'} = 2\ur + \frac{\bfg r \cdot \bfg u'}{\bfg r \cdot \bfg u} \, , 
\nonumber
\end{align}
and we have utilized the following identities:
\begin{equation*}
  \begin{aligned}
  \ur' \bfg r &= ( \alpha_r - \ur^2 ) \bfg r = (\beta_r \ur - \ur^2) \bfg r \, , \\
  \ur \, \bfg u &= \ur^2 \, \bfg r + \ur \, \orbit \times \bfg r  
  = \ur^2 \, \bfg r + \orbit \times (\ur \,  \bfg r) \, , \\
  \orbit \times \bfg u &= \orbit \times (\ur \, \bfg r + \orbit \times \bfg r) \, .
  \end{aligned}
\end{equation*}
%

Equation \eqref{eq:coriolis} represents the well-known Coriolis'
theorem for the accelerations in relative non-inertial coordinates,
which, we recall, are the vectors $(\T, \N, \B)$ of the Frenet frame
(please also refer to Appendix~\ref{app:frenet}).  In the right-hand
side of the third line of \eqref{eq:coriolis}, $\alpha_r \bfg r$ is
the relative time derivative of the velocity in Frenet-frame
coordinates; $2 \orbit \times \bfg u_{\|} $ is the Coriolis
acceleration; $\orbit' \times \bfg r$ is the Euler acceleration; and
$\orbit \times (\orbit \times \bfg r)$ is the centrifugal
acceleration.  Separating and identifying these terms is relevant to
characterize the various components of the active and reactive power.
This point is comprehensively discussed in the examples given in
Section~\ref{sec:examples}.

The time derivative of the Lagrangian $\mathcal{L}'$ is:
\begin{equation}
  \begin{aligned}
    \mathcal{L}' = T' - U'
    &= 
    \dfrac{1}{2}\bfg u' \cdot \mom + 
    \dfrac{1}{2}\bfg u \cdot \mom' +
    \bfg u \cdot \mom' + \bfg r \cdot \mom''  \, , 
  \end{aligned}
\end{equation}
which includes the term:
\begin{equation}
  \label{eq:xddp}
  \bfg r \cdot \mom'' =
  \bfg r \cdot \bfg f' =
  m \, \bfg r \cdot \bfg u'' \, . 
\end{equation}
This term depends on the second time derivative of the velocity.  
\color{black}
Using
\eqref{eq:frenet2}, \eqref{eq:du},
the latter can be expressed in $(\T, \N, \B)$ as:
\begin{equation}
  \label{eq:ddu}
  \begin{aligned}
    \bfg u''
    =& \; (\ar - \wk^2) | \bfg u| \, \T \, + \\
      & \; (\wk' + 2 \wk \wr) | \bfg u| \, \N \, + \\
      & \; (\wt \wk) |\bfg u|\, \B \, ,
  \end{aligned}
\end{equation}
where:
%
\begin{equation}
  \ar = \frac{|\bfg u|''}{|\bfg u|} = \wr' + \wr^2 \, .
\end{equation}
Note also that the terms $2 \wk \wr |\bfg u| \, \N$, $-\wk^2 |\bfg u| \, \T + \wt \wk |\bfg u|
\, \B$ and $\wk' |\bfg u| \, \N$ result from the Coriolis, centrifugal and Euler,
respectively, accelerations cross-multiplied by the velocity vector $\bfg u = |\bfg u| \T$.
Then, \eqref{eq:xddp} can be rewritten as:
\begin{equation}
  \label{eq:xddp2}
  \begin{aligned}
    \bfg r \cdot \bfg f'
    =& \; m \, (\ar - \wk^2) \, \bfg r \cdot \bfg u \, + \\
    & \; m \, (\wk' + 2\wk \wr) \, |\bfg u| \, \bfg r \cdot \N \, + \\
    & \; m \, (\wt \wk) |\bfg u| \, \bfg r \cdot \B \, . 
    \\
  \end{aligned}
\end{equation}

Given that $\N = -\T \times \B$ and $\B = \T \times \N$ and applying identity 
\eqref{eq:triple} for the scalar triple product --- see in Appendix~\ref{app:identities} ---, one has:
\begin{equation*}
  \begin{aligned}
    m \, |\bfg u| \, \bfg r \cdot \N
    &= - m \, \B \cdot (\bfg r \times \bfg u) 
    = - \bfg L \cdot \B = -I \, \orbit \cdot \B \, , \\
    m \, |\bfg u| \, \bfg r \cdot \B
    &= m \, \N \cdot (\bfg r \times \bfg u) 
    = \bfg L \cdot \N = I \, \orbit \cdot \N \, . 
  \end{aligned}
\end{equation*}
Finally, substituting the expression of $m \, \bfg r \cdot \bfg u$ from
\eqref{eq:ell2}, equation \eqref{eq:xddp2} can be written as:
\begin{equation}
  \label{eq:xddp3}
  \begin{aligned}
  \bfg r \cdot \bfg f' =& \; 
  I \, (\ar - \wk^2) \, \ur \, + \\
  & \; I \, (\wt \wk) \, \orbit \cdot \N \, - \\
  & \; I \, (\wk' + 2\wk \wr) \, \orbit \cdot \B  \, .
  \end{aligned}
\end{equation}


We define as \textit{instantaneous reactive power} pseudo-vector the cross product of the velocity by the force, namely:
\begin{equation}
  \label{eq:Q}
  \boxed{\bfg Q \equiv
  \bfg u \times \bfg f } 
\end{equation}
Expression \eqref{eq:Q} appears in a variety of works based on geometric algebra, e.g., \cite{6161614, 4450060504, 4450060503, 1308315, 5316097}, although all these works use voltage and current instead of velocity and force.  Most importantly, in these references, the reactive power is defined without providing the physical rationale of its expression.  Equation \eqref{eq:ddL} shows that $\bfg Q$ is in effect a component of the time derivative of the torque.  More specifically, recalling that $\bfg f = m\bfg u'$ and the expression of the vector $\sw$ given in \eqref{eq:frenet2} in Appendix~\ref{app:frenet}, \eqref{eq:Q} can be equivalently written as:
\begin{equation}
  \label{eq:Q2}
  \begin{aligned}
    \bfg Q
    &= m \, \bfg u \times \bfg u'
      = m \, |\bfg u|^2 \sw \\
    &= 2T \, \sw = 2T \, \wk \, \B \\
    &= I \, |\orbit|^2 \, \wk \, \B \, ,
  \end{aligned}
\end{equation}
that is, \textit{the instantaneous reactive power vector lays along
  the direction of the binormal and its magnitude is given by twice the
  kinetic energy by the azimuthal frequency}.

Finally, we define the cross product of the position and the
\textit{yank}, i.e., the time derivative of the force, as the
\textit{rotatum residual} pseudo-vector, namely:
\begin{equation}
  \label{eq:R}
  \bfg R \equiv \bfg r \times \bfg f' \, .
\end{equation}
Observing that $\bfg f' = m \bfg u''$ and substituting the expression of $\bfg u''$ from \eqref{eq:ddu} into \eqref{eq:R} and recalling that $\bfg u = u \, \T$, we obtain:
\begin{equation}
  \label{eq:R2}
  \begin{aligned}
    \bfg R =& \; m \, (\ar - \wk^2) \, |\bfg u| \, \bfg r \times \T \, + \\
    & \; m \, (\wk' + 2\wk \wr) \, |\bfg u| \, \bfg r \times \N \, + \\
    & \; m \, (\wt \wk) \, |\bfg u| \, \bfg r \times \B \, .
  \end{aligned}
\end{equation}
Then, substituting $\N = - \T \times \B$ and $\B = \T \times \N$, and
using the Lagrange and Jacobi identities \eqref{eq:lagrange} and \eqref{eq:jacobi}, respectively, given in Appendix \ref{app:identities}, one has:
\begin{equation*}
  \begin{aligned}
    |\bfg u| \, \bfg r \times \T &= \bfg r \times \bfg u  \\
    &= |\bfg r|^2 \orbit \, ,  \\   
    |\bfg u| \, \bfg r \times \N &= - \bfg r \times (\bfg u \times \B) \\
    &= \bfg u \times (\B \times \bfg r) + \B \times (\bfg r \times \bfg u) \\
    &= (\bfg r \cdot \bfg u) \, \B - |\bfg r|^2 \orbit \times \B \\
    &= |\bfg r|^2 (\ur \, \B - \orbit \times \B) \, , \\
    |\bfg u| \, \bfg r \times \B &= \bfg r \times ( \bfg u \times \N) \\
    &= - \bfg u \times (\N \times \bfg r) - \N \times (\bfg r \times \bfg u) \\
    &= -(\bfg r \cdot \bfg u) \, \N + |\bfg r|^2 \orbit \times \N \\
    &= - |\bfg r|^2 (\ur \, \N - \orbit \times \N) \, ,
  \end{aligned}
\end{equation*}
where we have used the fact that $\bfg u$ is perpendicular to $\N$ and $\B$ by construction of the Frenet frame.  Substituting the expressions above into \eqref{eq:R2} yields:
\begin{equation}
  \label{eq:R3}
  \begin{aligned}
    \bfg R =& \; I \, (\ar - \wk^2) \, \orbit \, - \\
    & \; I \, (\wk \wt) \, (\ur \, \N - \orbit \times \N) \, + \\
    & \; I \, (\wk' + 2\wk \wr ) \,  (\ur \, \B - \orbit \times \B) \, .
  \end{aligned}
\end{equation}
%

We define the \textit{instantaneous power multivector} as follows:
\begin{equation}
  \label{eq:Shat}
  \begin{aligned}
    \hat{W} \equiv \hat{E}'
    &= \ell'' + \bfg L'' \\
    &= 2T' - U' + \bfg N' \\
    &= p + \bfg Q + \mathcal{L}' + \bfg R \\
    &= \hat{S} + \hat{R} \, ,
  \end{aligned}
\end{equation}
which is composed of the sum of two multivectors,
$\hat{S} = p + \bfg Q$ and $\hat{R} = \mathcal{L}' + \bfg R$.
Specifically, from \eqref{eq:dT} and \eqref{eq:Q2}, we obtain:
\begin{equation}
    \hat{S} = 2T \, (\wr + \sw) \, , 
    \nonumber
\end{equation}
or, equivalently:
\begin{equation}
  \label{eq:hatS2}
  \boxed{\boxed{ \hat{S} = 2T \, \Wgeom{u} }}
\end{equation}
or, equivalently, using \eqref{eq:T0}:
\begin{equation}
    \hat{S} = I \, |\Wgeom{r}|^2 \, \Wgeom{u} = \frac{|\hat{L}|^2}{I} \, \Wgeom{u} \, ,
\end{equation}
and
\begin{equation}\label{eq:Rhat}
\begin{aligned}
\hat{R} =& \; I \, (\ar - \wk^2) \, \Wgeom{r} \, - \\
& \; I \, (\wk \wt) \, \N \otimes \Wgeom{r} \, + \\
& \; I \, (\wk' + 2\wk \wr ) \, \B \otimes \Wgeom{r} \, .
\end{aligned}
\end{equation}

Equation \eqref{eq:hatS2} is the main result of this paper: it
represents the sought link among the internal stored energy, the
geometric frequency and the instantaneous power.  In
Appendix~\ref{sec:relative} we demonstrate how $T$, $\orbit$, and thus
also $\hat{S}$, can be expressed in terms of differential geometric
invariants.

The multivector $\hat{R}$ is also defined for the first time in this
work and is composed of the rate of change of the Lagrangian and the
rotatum residual.  The expression of $\hat{R}$ can be written in an
alternative form.  Observe that:
\begin{equation}
   \label{eq:Rhat2}
  \hat{R} = 2p + \bfg r \cdot \bfg f' + \bfg r \times \bfg f'  = 2p + \bfg r \otimes \bfg f' \, ,
\end{equation}
where, as discussed above, $2p = 4 T \varrho_u$.  The first time derivative of the force can be written as:
\begin{equation}
  \label{eq:fdot}
  \bfg f' = (\varrho_f + \bfg \omega_f \times) \bfg f = \Wgeom{f} \otimes \bfg f \, . 
\end{equation}
Then, the multivector product in \eqref{eq:Rhat2} can be rewritten as:
\begin{align}
  \label{eq:rfdot}
  \bfg r \otimes \bfg f' =& \; \bfg r \cdot (\varrho_f \bfg f) + \bfg r \cdot (\bfg \omega_f \times \bfg f) \\ \nonumber
  & \; + \bfg r \times (\varrho_f \bfg f) + \bfg r \times (\bfg \omega_f \times \bfg f) \\ \nonumber
  =& \; -\varrho_f U - \bfg \omega_f \cdot (\bfg f \times \bfg r) + \varrho_f \bfg N \\ \nonumber
  & \; -\bfg \omega_f \times (\bfg f \times \bfg r) - \bfg f \times (\bfg r \times \bfg \omega_f) \\  \nonumber
  =& \; -\varrho_f U - \bfg \omega_f \cdot \bfg N + \varrho_f \bfg N \\ \nonumber
  & \; + \bfg \omega_f \times \bfg N - (\bfg f \cdot \bfg \omega_f) \, \bfg r + (\bfg f \cdot \bfg r) \, \bfg \omega_f \\ \nonumber
  =& \; -\varrho_f U - \bfg \omega_f \cdot \bfg N + \varrho_f \bfg N + \bfg \omega_f \times \bfg N - \bfg \omega_f U \\ \nonumber
  =& \; - U \Wgeom{f} + \Wgeom{f} \otimes \bfg N \\
  =& \; - \Wgeom{f} \otimes (U - \bfg N) = \Wgeom{f} \otimes (\bfg r \otimes \bfg f) \, , \nonumber
\end{align}
where we have utilised the definitions of $U = -\bfg r \cdot \bfg f$ and $\bfg N = \bfg r \times \bfg f$, the multivector product \eqref{eq:geomprod}, as well as the vector product identities \eqref{eq:triple}-\eqref{eq:jacobi}.

\color{black}

\subsection{Electrical and Magnetic Domains}
\label{sub:EMdomain}

We can now express the
generalized momentum, energy and power multivectors in terms of electrical and magnetic quantities.  With
this aim, we use the correspondences among position, velocity, momenta and
forces given in Table \ref{tab:gen}.

For the electrical domain, we obtain:
\begin{equation}
  \label{eq:elec1}
  \begin{aligned}
    \hat{L}_e &= \bfg \varphi \cdot \bfg q +
                 \bfg \varphi \times \bfg q
               = I_e \Wgeom{\varphi} \, , \\
    \hat{E}_e &= 2 T_e - U_e + \bfg N_e \, , \\
    \hat{W}_e &= 2 T_e \, \Wgeom{v} + \hat{R}_e =
                 p_e + \mathcal{L}'_e + \bfg Q_e + \bfg R_e \, , \\
  \end{aligned}
\end{equation}
where $I_e = C \, |\bfg \varphi|^2$ and $T_e$, $U_e$ are defined in \eqref{eq:TUe}, and:
\begin{equation}
  \label{eq:elec2}
  \begin{aligned}
    \bfg N_e &= \bfg \varphi \times \bfg \ii \, , &\qquad
    p_e &= \bfg v \cdot \bfg \ii \, , \\
    \bfg Q_e &= \bfg v \times \bfg \ii \, , &\qquad
    \bfg R_e &= \bfg \varphi \times \bfg \ii' \, .
  \end{aligned}
\end{equation}
The force, in the electrical domain corresponding to the current, can be expressed in terms of the Coriolis theorem --- see \eqref{eq:coriolis}:
\begin{equation}
\begin{aligned}
   \label{eq:coriolis:electric}
   \bfg \ii =  C \bfg v' =
   & \, C \, \beta_\varphi \bfg v_{\|} + 2C \, \bfg\omega_\varphi \times \bfg v_{\|} \, + \\
   & \, C \, \bfg\omega_\varphi' \times \flux +
   C \, \bfg\omega_\varphi \times (\bfg\omega_\varphi \times \flux) \, , 
\end{aligned}
\end{equation}
where the right-hand side terms, express, in order, the relative, Coriolis, Euler, and centrifugal components of the current.

For the magnetic domain, we obtain:
\begin{equation}
  \label{eq:magn1}
  \begin{aligned}
    \hat{L}_m &= \bfg q \cdot \bfg \varphi +
                 \bfg q \times \bfg \varphi
               = I_m \Wgeom{q} \, , \\
    \hat{E}_m &= 2 T_m - U_m + \bfg N_m \, , \\
    \hat{W}_m &= 2 T_m \, \Wgeom{\ii} + \hat{R}_m =
                 p_m + \mathcal{L}'_m + \bfg Q_m + \bfg R_m \, , \\
  \end{aligned}
\end{equation}
where $I_m = L \, |\bfg q|^2$ and $T_m$, $U_m$ are defined in \eqref{eq:TUm} and:
\begin{equation}
  \label{eq:magn2}
  \begin{aligned}
    \bfg N_m &= \bfg q \times \bfg v \, , &\qquad
    p_m &= \bfg \ii \cdot \bfg v \, , \\
    \bfg Q_m &= \bfg \ii \times \bfg v \, , &\qquad
    \bfg R_m &= \bfg q \times \bfg v' \, .
  \end{aligned}
\end{equation}
The force in the magnetic domain corresponds to the voltage and can be also expressed in terms of the Coriolis theorem:
\begin{equation}
\begin{aligned}
   \label{eq:coriolis:magnetic}
   \bfg v =  L \, \bfg\ii' =
   & \, L \, \beta_q \bfg \ii_{\|} + 2L \, \bfg\omega_q \times \bfg \ii_{\|} \, + \\
   & \, L \, \bfg\omega_q' \times \bfg q +
   L \, \bfg\omega_q \times (\bfg\omega_q \times \bfg q) \, , 
\end{aligned}
\end{equation}
where the right-hand side terms, express, in order, the relative, Coriolis, Euler, and centrifugal components of the voltage.

It is relevant to note that from the expressions of the reactive power in the electrical domain, one obtains:
\begin{equation}
  \bfg Q_e = \bfg v \times \bfg \ii = - \bfg \ii \times \bfg v \, ,
\end{equation}
which has thus opposite sign with respect to the expression of $\bfg Q_m$ for the magnetic domain, as it is the usual convention.  Interestingly, with the proposed approach, the opposite sign is obtained as a natural consequence of the different meaning of $\bfg v$ and $\bfg \ii$ in the electrical and magnetic domains, that is, velocity/force and force/velocity, respectively.  

\subsection{Inclusion of Losses}

\begin{figure}[ht!]
  \centering
  \resizebox{0.95\linewidth}{!}{\includegraphics{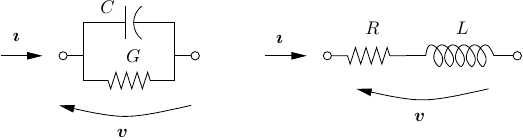}}
  \caption{Left: $\bfg v$-controlled conductance; Right: $\bfg \ii$-controlled resistance.}
  \label{fig:rlc}
\end{figure}

In the formulation provided in \cite{Chua:1974}, losses are either
voltage-controlled conductances or current-controlled resistances,
depending on whether these elements are in parallel with a capacitance
or in series with an inductance.  Similarly, we can account for the
cases illustrated in Fig.~\ref{fig:rlc}.

The definitions of $p$ and $\bfg Q$ given in \eqref{eq:pdef} and \eqref{eq:Q}, respectively, still apply as long as the generalized force is given by $\bfg f_e = \bfg \ii - G \, \bfg v$ for the capacitance and $\bfg f_m = \bfg v - R \, \bfg \ii$ for the inductance.  The terms $G \, \bfg v$ and $R \, \bfg \ii$, which are generalized non-conservative forces, only affect the active power, as expected, as $\bfg v \times G \, \bfg v= \bfg 0$ and $\bfg \ii \times R \, \bfg \ii= \bfg 0$.

We note also that nonlinear circuit elements, in the same vein as discussed in \cite{Chua:1974}, can be captured as long as $C$, $L$ and $R$ are represented with adequate nonlinear functions.  The analysis of nonlinear components is, however, beyond the scope of this paper and will be considered in future work.

\section{Examples}
\label{sec:examples}

This section illustrates theoretical results of this work through examples 
on three-phase AC systems.  These consider a variety of cases,
including balanced, unbalanced, non-sinusoisal (with harmonics) and
non-stationary (in transient conditions).  In all examples, we assume
that fluxes/charges are the coordinates of three-dimensional curves on
a Cartesian coordinate system:
\begin{equation}
  \begin{aligned}
    \e{1} &= (1, 0, 0) \, , \\ 
    \e{2} &= (0, 1, 0) \, , \\
    \e{3} &= (0, 0, 1) \, . 
  \end{aligned}
  \nonumber
\end{equation}

\subsection{Stationary Balanced Sinusoidal Case}
\label{sec:balsin}

We start by considering a stationary balanced sinusoidal three-phase
voltage with constant angular frequency $\wo$ and constant magnitude.
The voltage vector is:
\begin{equation}
    \bfg v = v_a \, \e{1} + v_b \, \e{2} + v_c \e{3} \, ,
    \label{eq:3ph:vector}
\end{equation}
where
%
\begin{equation}
  \begin{aligned}
v_a &= V \cos (\wo t) \, , \\
v_b &= V \cos (\wo t - {2\pi}/{3}) \, , \\
v_c &= V \cos (\wo t + {2\pi}/{3}) \, .
  \end{aligned}
  \nonumber
\end{equation} 

\begin{figure}[ht!]
  \centering
  \resizebox{0.99\linewidth}{!}{\includegraphics{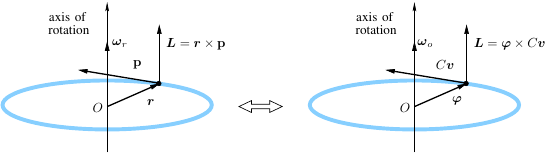}}
  \caption{A point spherical particle that rotates with constant
    radius and angular speed is analogous to a stationary balanced
    sinusoidal 3-phase system.}
  \label{fig:momentum}
\end{figure}

This case is analogous to a mechanical system where a point spherical particle follows 
a circular trajectory 
(see Fig.~\ref{fig:momentum}).

%
The magnetic flux, i.e.,~the primitive of $\bfg v$, is given by:
\begin{equation}
  \flux = \frac{V}{\wo} [
  \sin(\wo t ) \, \e{1} +
  \sin(\wo t - \dfrac{2\pi}{3}) \, \e{2} +
  \sin(\wo t + \dfrac{2\pi}{3}) \, \e{3} ] \, ,
  \nonumber
\end{equation}
and hence:
\begin{equation}
  \flux = -\frac{v}{\wo} \, \N \, , \quad
  \bfg v = v \, \T \, , \quad
  \bfg \omega_v = \wo \, \B \, .
  \nonumber
\end{equation}
Then, applying the formulas given in Appendix \ref{app:frenet}, we
obtain:
\begin{equation*}
  v = \sqrt{\tfrac{3}{2}} V \, , \qquad 
  \wk = \wo \, , \qquad \wt = 0 \, ,
\end{equation*}
and $\varrho_\varphi = \varrho_v = 0$, as the magnitudes of the flux
and the voltage are constant.

Moreover, from \eqref{eq:orbit}, and observing that
$\bfg \omega_{\xi} = \bfg \nu_{\rm d} = \bfg 0$, the following
identity holds (please refer to Appendix~\ref{sec:relative}):
\begin{equation}
   \label{eq:wvwf}
  \bfg \omega_v= \bfg\omega_{\varphi} \, , 
\end{equation}
where, in this case,
$\bfb F \, \bfg\omega_{\varphi} = \bfg\omega_{\varphi}$, as
$\bfg\omega_{\varphi}$ is parallel to $\B$.

\begin{figure}[ht!]
  \centering
  \resizebox{0.7\linewidth}{!}{\includegraphics{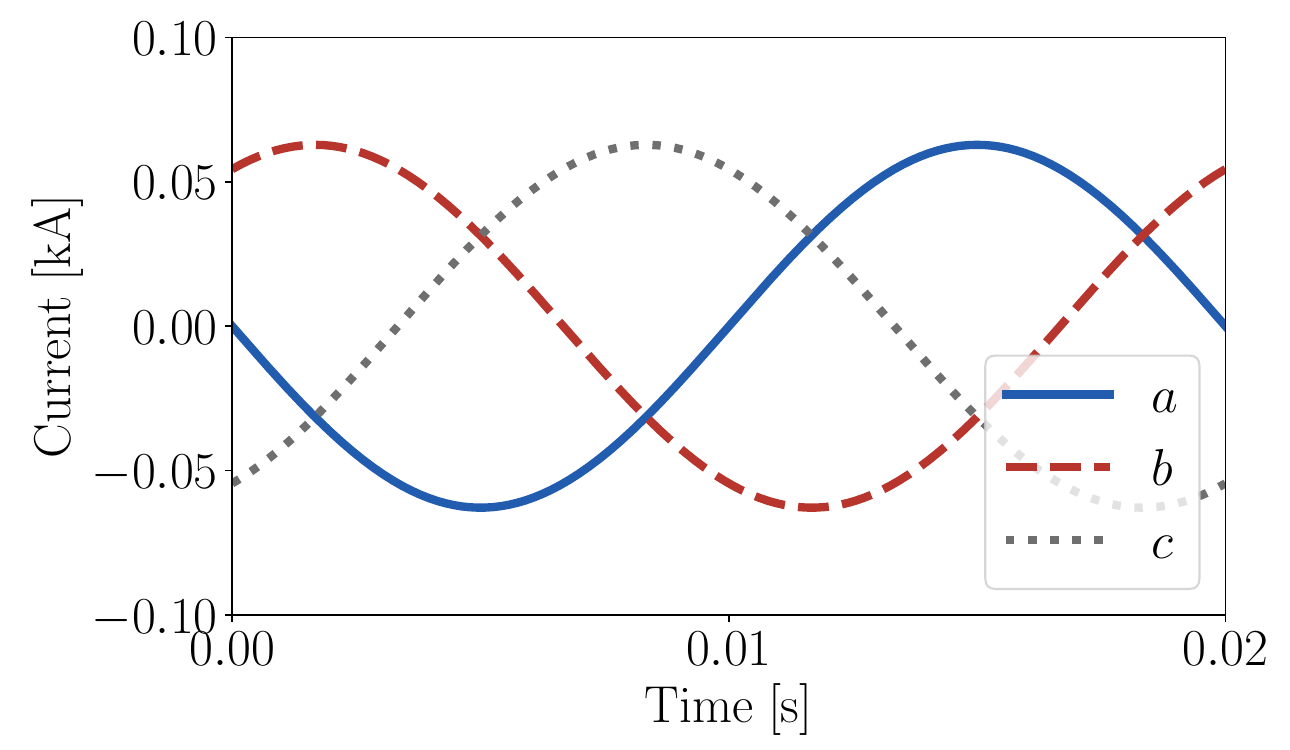}}
  \caption{A 3-phase capacitor with stationary balanced sinusoidal AC
    voltage is always subject to a current that is centrifugal:
    $\bfg \ii = C\bfg\omega_{\varphi} \times (\bfg\omega_{\varphi}
    \times \flux)$. The relative, Coriolis, and Euler components are
    zero.}
  \label{fig:E0:force}
  \vspace{-3mm}
\end{figure}

Assume a balanced three-phase capacitor (with capacitance $C$ per
phase) with voltage vector $\bfg v$. Since $\rho_\varphi = 0$, the
capacitor is subject to a current vector:
\begin{equation}
\nonumber
  \bfg\ii = C \, \bfg v' = C \, \bfg \omega_v \times \bfg v = \wo C v \, \N \, ,
\end{equation}
or, equivalently, using \eqref{eq:wvwf} and the identity
$\bfg v = \bfg\omega_{\varphi} \times \flux$:
\begin{equation}
  \nonumber
  \begin{aligned}
    \bfg\ii &= C \, \bfg\omega_{\varphi} \times (\bfg\omega_{\varphi} \times \flux) \, ,
  \end{aligned}
\end{equation}
which, using the correspondences of Table~\ref{tab:gen}, is a
generalized centrifugal force. In other words, the only non-null
component of $\bfg v'$ in \eqref{eq:coriolis} is the centrifugal
acceleration.  The relative, Coriolis and Euler components in
\eqref{eq:coriolis:electric} are zero in this case.  Considering a
numerical example with $\wo=100\pi$~rad/s, $V=20$~kV,
$C=10$~$\mu \rm F$, the current $\bfg \ii$, or, more precisely, the
phase currents $\ii_a$, $\ii_b$, $\ii_c$, where:
\begin{equation}
  \bfg \ii = \ii_a \, \e{1} + \ii_b \, \e{2} + \ii_c \e{3} \, ,
  \nonumber
\end{equation}
are shown in Fig.~\ref{fig:E0:force}.
%

%
%

The generalized momentum is defined as:
\begin{equation}
  \bfg L =
  \flux \times C \bfg v =
  C \frac{v^2}{\wo} \, \B =
  C \frac{v^2}{\wo^2} \, \sw =
  I_e \, \sw \, , 
  \nonumber
\end{equation}
where $I_e$ is the generalized moment of inertia of the condenser and
$\darboux = \sw$ because $\wt = 0$.  Then, observing that
$\bfg L' = \bfg N = \bfg 0$ and hence $\bfg L'' = \bfg N' = \bfg 0$,
the instantaneous reactive power of the capacitor becomes:
\begin{equation}
  \begin{aligned}
    \bfg Q = - \flux \times \bfg \ii' = 
    \bfg v \times \bfg \ii = 
    C v^2 \, \bfg \omega =
    I_e \, \wo^2 \, \bfg \omega =
    2 T_e \, \bfg \omega \, ,
  \end{aligned}
  \nonumber
\end{equation}
where $T_e$ is the capacitor's stored kinetic energy, as defined in
\eqref{eq:TUe}.  In turn, in balanced stationary conditions, the
capacitor's reactive power is due to the centrifugal force
$\bfg \ii = C \bfg v'$.  Note that the magnitude of $\bfg Q$ is
equivalent to the common definition, in fact:
\begin{equation}
  \begin{aligned}
    |\bfg Q| &= | \bfg v \times \bfg \ii | = | v\, \T \times \wo C v \, \N | \\
    &= \wo C v^2 \, |\T \times \N | = \wo C v^2 \, ,
  \end{aligned}
\end{equation}
as $\T \times \N = \B$ and $|\B| = 1$ by construction of the Frenet coordinate vectors.

In the numerical example considered, the instantaneous reactive power pseudovector is:
\begin{equation}
  \begin{aligned}
    \bfg Q = 1.088 \, \e{1} +
    1.088 \, \e{2} +
    1.088 \, \e{3} \ \ {\rm MVAr} \, .
  \end{aligned}
   \nonumber
 \end{equation}

Finally, we note that, as expected, the instantaneous active power of the condenser is null, in fact:
\begin{equation}
  p = \bfg v \cdot \bfg \ii = \bfg v \cdot C \bfg v' = 0 \, ,
  \nonumber
\end{equation}
as $\bfg v \perp \bfg v'$.  This is also consistent with the common knowledge that the inner product is $\bfg u \cdot \bfg f = 0$ for a mass rotating at constant angular velocity along a circle of constant radius, as the centrifugal force is parallel to the position vector $\bfg r$ and, hence, perpendicular to the velocity (see Fig.~\ref{fig:momentum}).

Note that, in this case, $\hat{E} = \rm const.$ and $\hat{E}' = 0$ as the momentum density and the angular momentum of a system are conserved. Hence, one has:
\begin{equation}
  \label{eq:stationary}
  \begin{aligned}
    p &= -\mathcal{L}' \, , \\
    \bfg Q &= - \bfg R \, .
  \end{aligned}
\end{equation}

\subsection{Stationary Unbalanced Sinusoidal Case}
\label{sec:unbalsin}

Let us now consider that the same three-phase capacitor has an unbalanced voltage.  In this example, the voltage vector is given by \eqref{eq:3ph:vector}, where:
\begin{equation}
  \begin{aligned}
v_a &= V_a \cos (\wo t) \, , \\
v_b &= V_b \cos (\wo t - {2\pi}/{3}) \, , \\
v_c &= V_c \cos (\wo t + {1.6\pi}/{3}) \, ,
  \end{aligned}
  \nonumber
\end{equation} 
with $V_a=20$~kV, $V_b=19$~kV, $V_c=23$~kV.

\begin{figure}[ht!]
\centering
  \begin{subfigure}{.49\linewidth}
  \centering
  \resizebox{\linewidth}{!}{\includegraphics{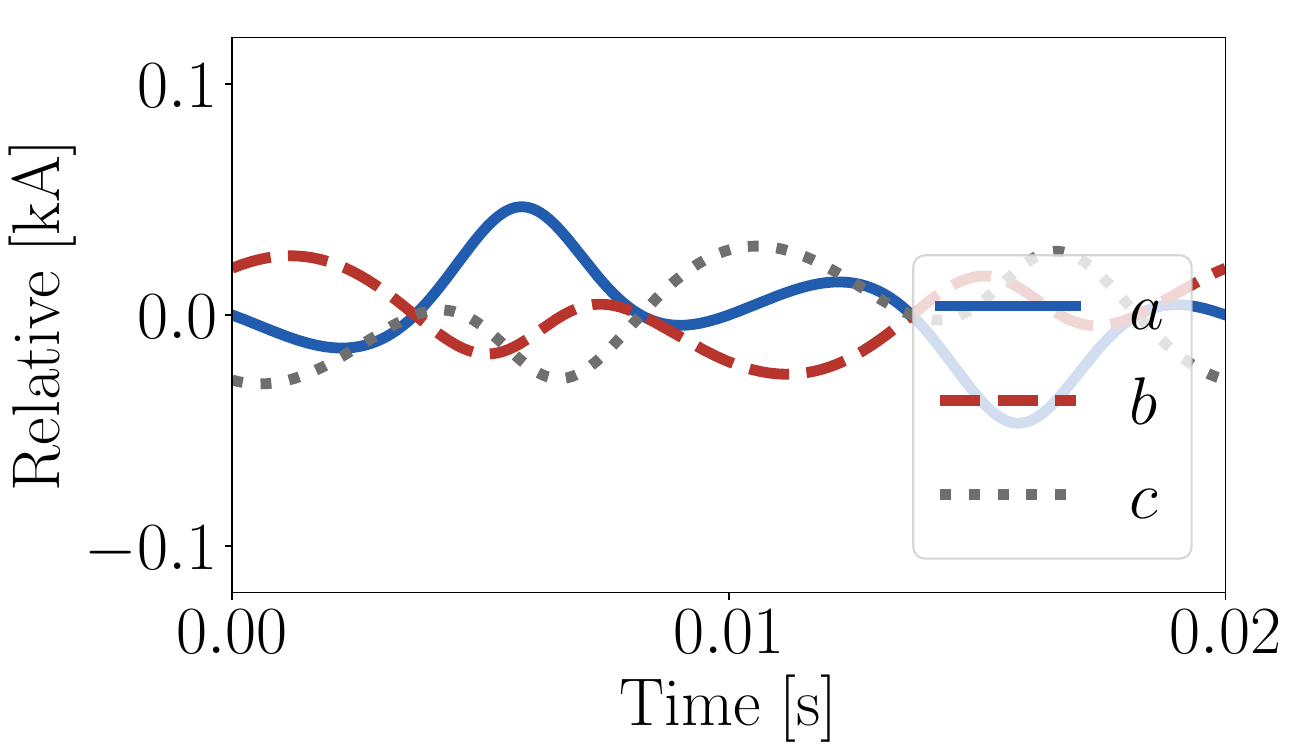}}
  \caption{Relative $C \beta_\varphi \bfg v_{\|}$}
  \label{fig:E1:frel}
  \end{subfigure}
  \begin{subfigure}{.49\linewidth}
  \centering
  \resizebox{\linewidth}{!}{\includegraphics{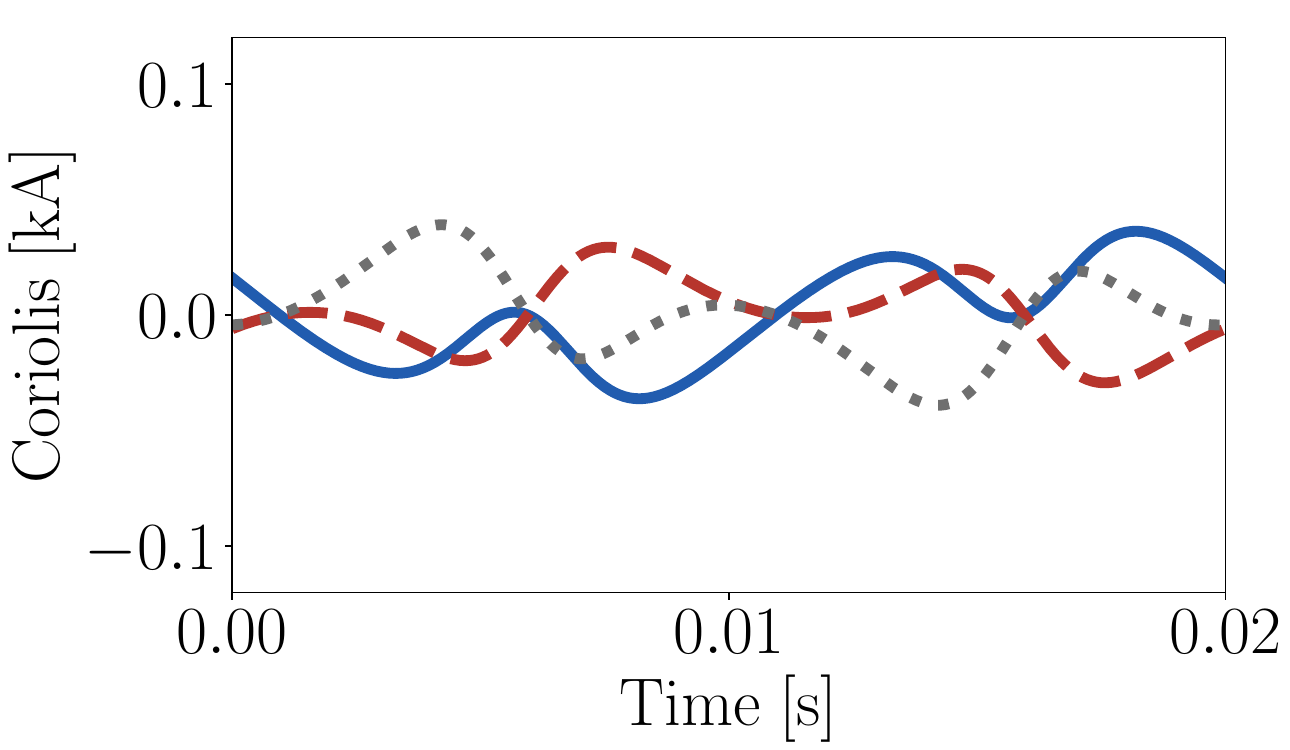}}
  \caption{Coriolis $
  2C \bfg\omega_\varphi \times \bfg v_{\|} $}
  \label{fig:E1:fcor}
  \end{subfigure}
   \begin{subfigure}{.49\linewidth}
  \centering
  \resizebox{\linewidth}{!}{\includegraphics{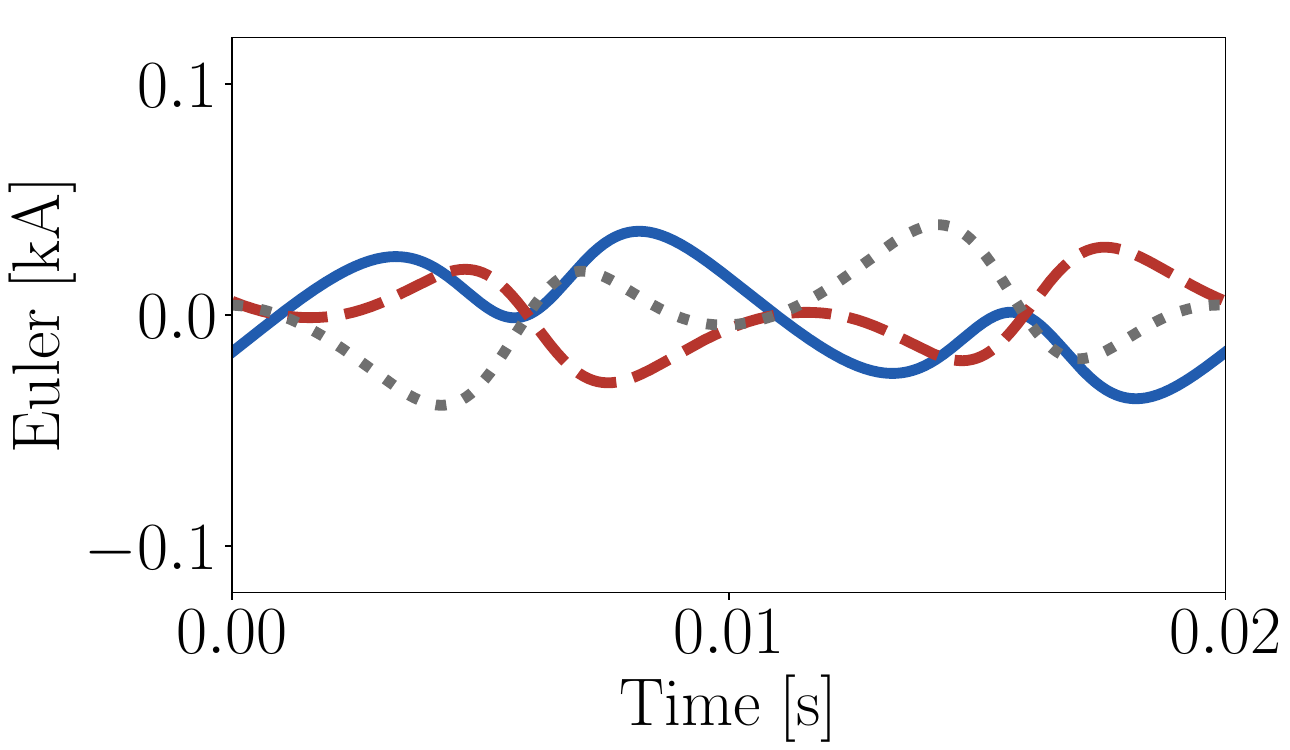}}
  \caption{Euler 
  $C\bfg\omega_\varphi' \times \flux $}
  \label{fig:E1:feul}
  \end{subfigure}
  \begin{subfigure}{.49\linewidth}
  \centering
  \resizebox{\linewidth}{!}{\includegraphics{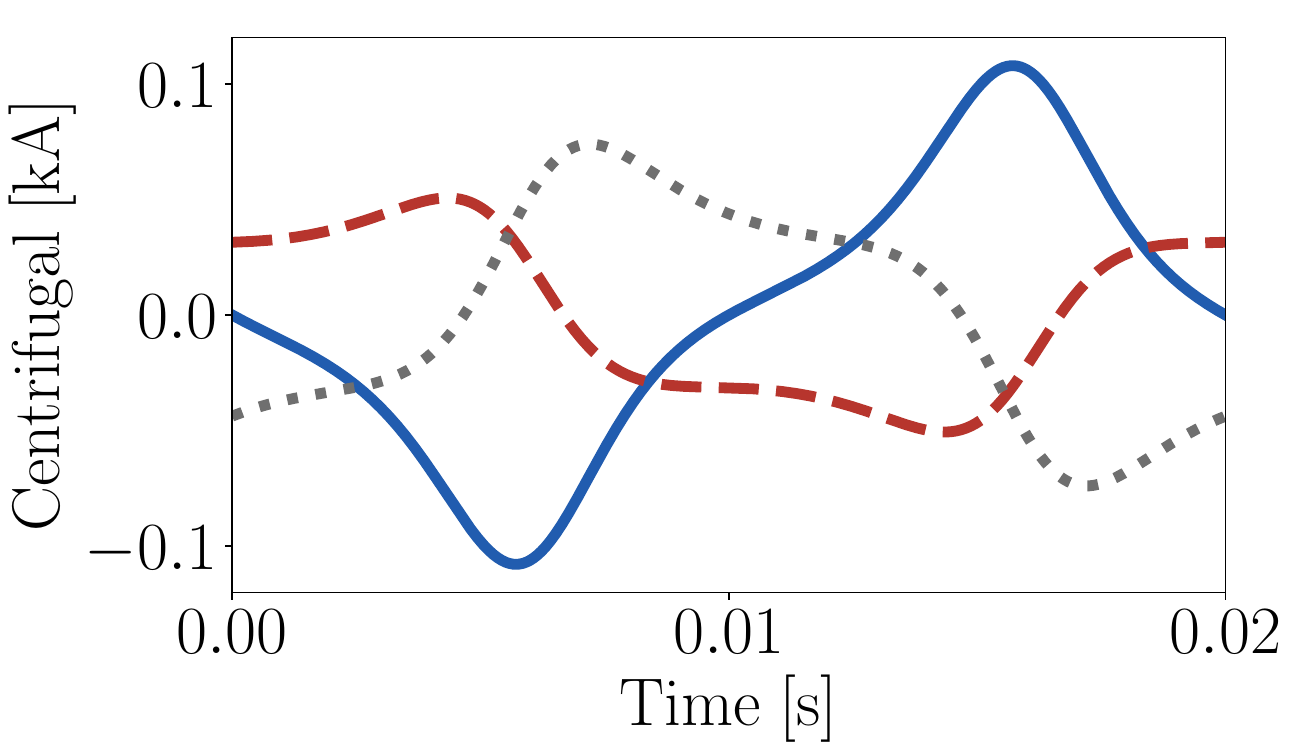}}
  \caption{Centrifugal $
  C\bfg\omega_\varphi \times (\bfg\omega_\varphi \times \flux)$}
  \label{fig:E1:fcen}
  \end{subfigure}
  \caption{Balanced 3-phase capacitor with stationary unbalanced
    sinusoidal AC voltage: Relative, Coriolis, Euler and centrifugal
    components of current. The Coriolis and Euler components are
    opposite to each other.}
  \label{fig:E1:coriolis_theorem}
  \vspace{-3mm}
\end{figure}

\begin{figure}[ht!]
  \centering
  \resizebox{0.7\linewidth}{!}{\includegraphics{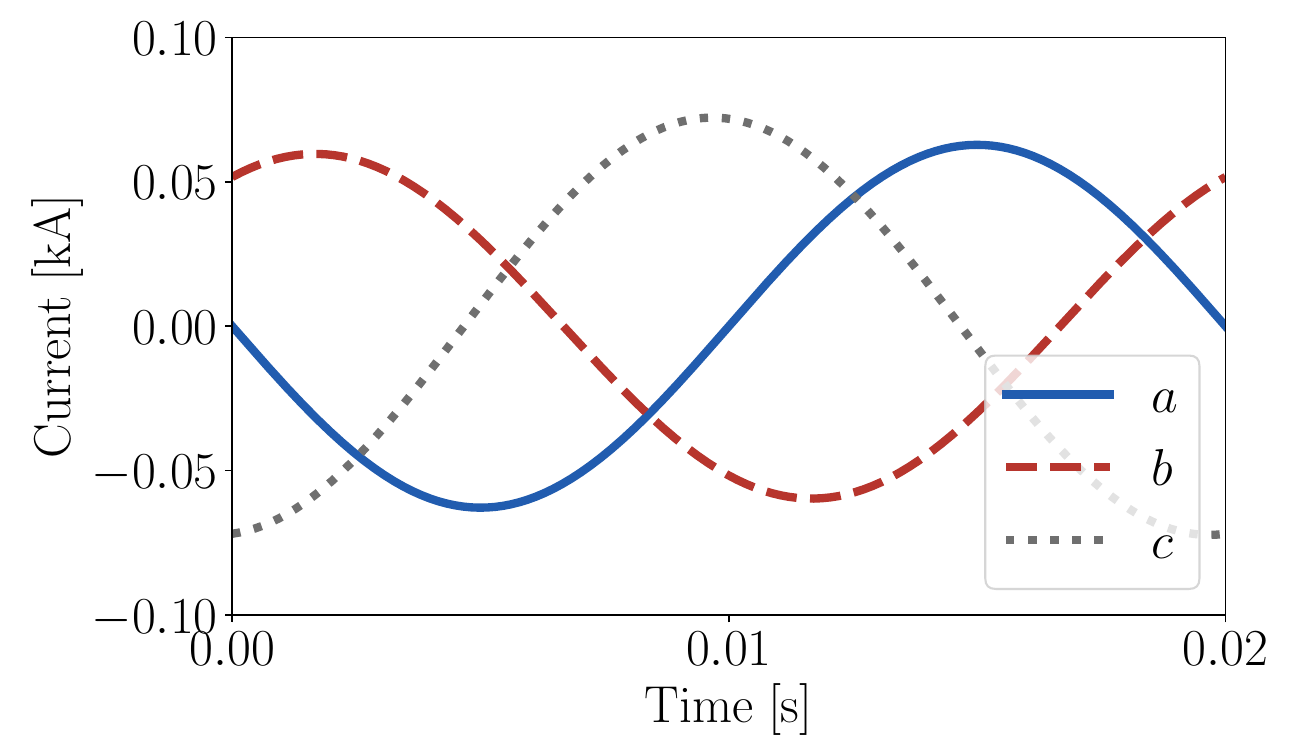}}
  \caption{A 3-phase capacitor with stationary unbalanced sinusoidal
    AC voltage is subject to the sum of a relative and a centrifugal
    current:
    $\bfg \ii = C \beta_\varphi \bfg v_{\|} + C\bfg\omega_{\varphi}
    \times (\bfg\omega_{\varphi} \times \flux)$. The Coriolis and
    Euler components cancel out each other.}
  \label{fig:E1:force}
\end{figure}

The relative, Coriolis, Euler, and centrifugal components of the current applied to the capacitor in this case are illustrated in Fig.~\ref{fig:E1:coriolis_theorem}. 
We observe that:
\begin{equation}
\label{eq:coreulopp}
2C \bfg\omega_\varphi \times \bfg v_{\|} = - C\bfg\omega_\varphi' \times \flux   \, , 
\end{equation}
that is, for an unbalanced voltage, the Coriolis and Euler components are opposite and when summed up cancel out each other. 
Hence, the ``force'' that the capacitor is subject to is the current vector:
\begin{equation*}
\begin{aligned}
\bfg \ii 
&= 
C \beta_\varphi \bfg v_{\|}  +
C\bfg\omega_\varphi \times (\bfg\omega_\varphi \times \flux) \, , 
\end{aligned}
\end{equation*}
that is, the sum of the relative and centrifugal component.  The phase
currents applied to the capacitor are shown in
Fig.~\ref{fig:E1:force}.

\begin{figure}[h!]
\centering
  \begin{subfigure}{.49\linewidth}
  \centering
  \resizebox{\linewidth}{!}{\includegraphics{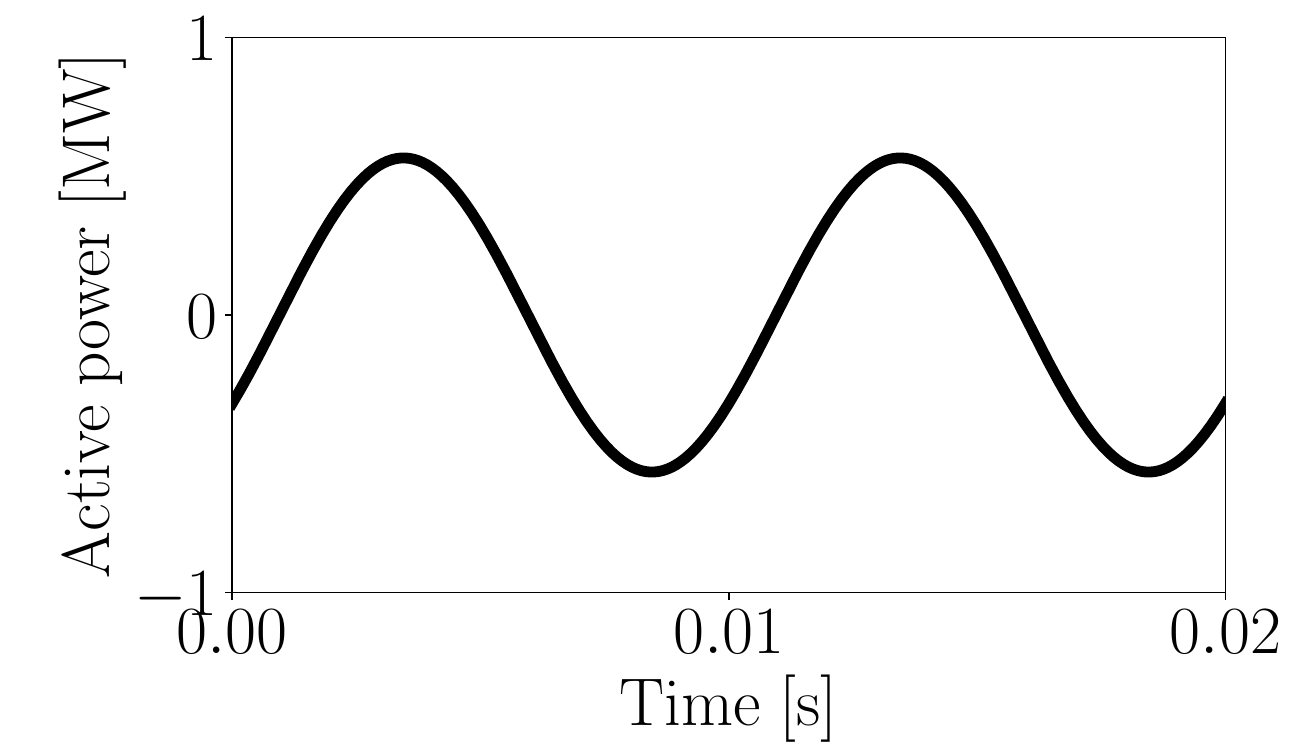}}
  \caption{Active power $p$}
  \label{fig:E1:p}
  \end{subfigure}
  \begin{subfigure}{.49\linewidth}
  \centering
  \resizebox{\linewidth}{!}{\includegraphics{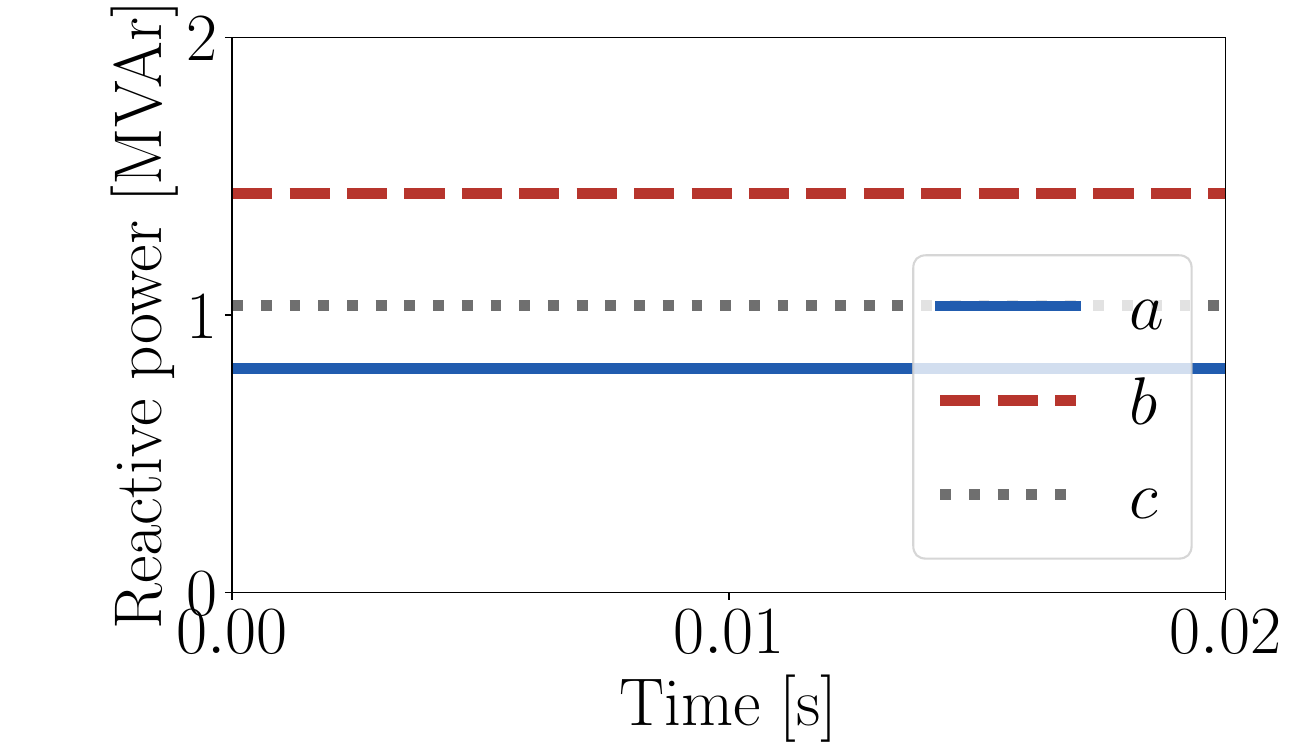}}
  \caption{Components of reactive power $\bfg Q$}
  \label{fig:E1:Q}
  \end{subfigure}
  \caption{3-phase capacitor
  in stationary unbalanced case: 
  instantaneous powers.}
  \label{fig:E1:pQ}
  \vspace{-3mm}
\end{figure}

The instantaneous active power and the components of the instantaneous
reactive power pseudovector are illustrated in Fig.~\ref{fig:E1:pQ}.
As expected, the active power has a period that is half of the period
of the voltage/current.

\begin{figure}[ht!]
\centering
  \begin{subfigure}{.49\linewidth}
  \centering
  \resizebox{\linewidth}{!}{\includegraphics{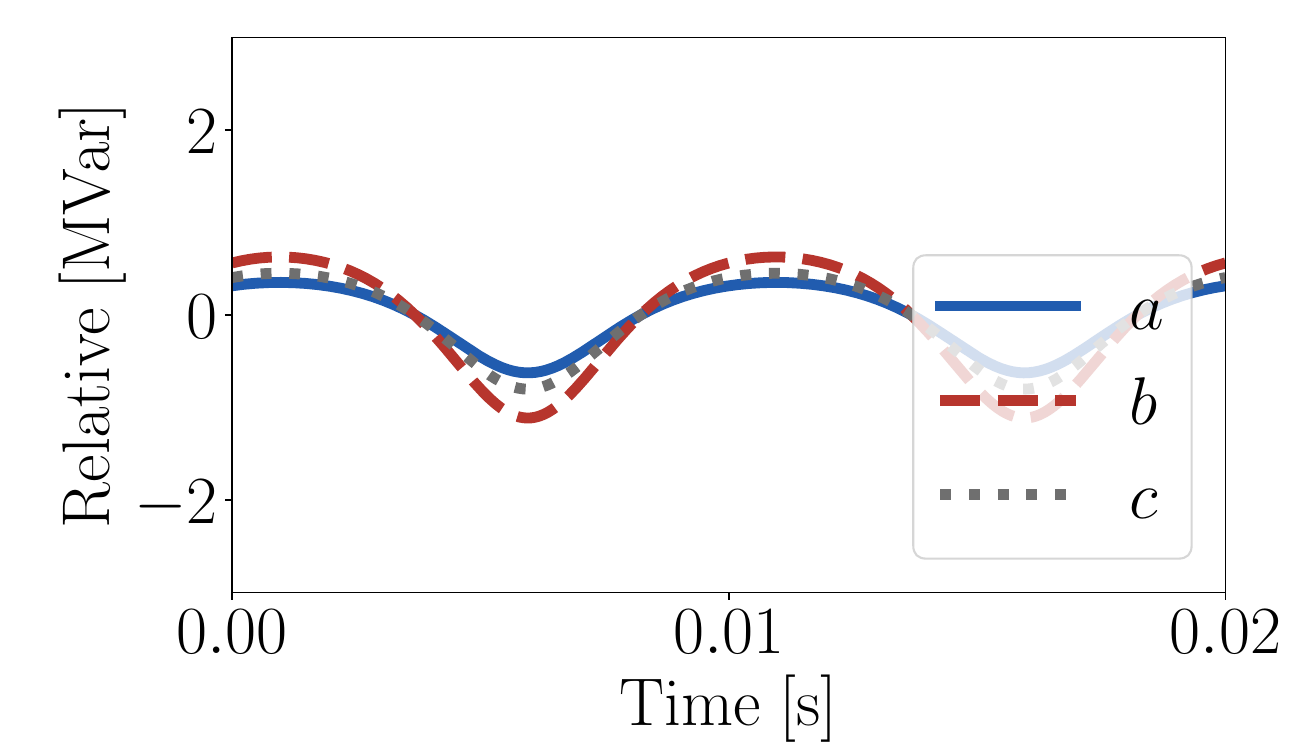}}
  \caption{Relative $\bfg v \times C \beta_\varphi \bfg v_{\|}$}
  \label{fig:E1:Qrel}
  \end{subfigure}
  \begin{subfigure}{.49\linewidth}
  \centering
  \resizebox{\linewidth}{!}{\includegraphics{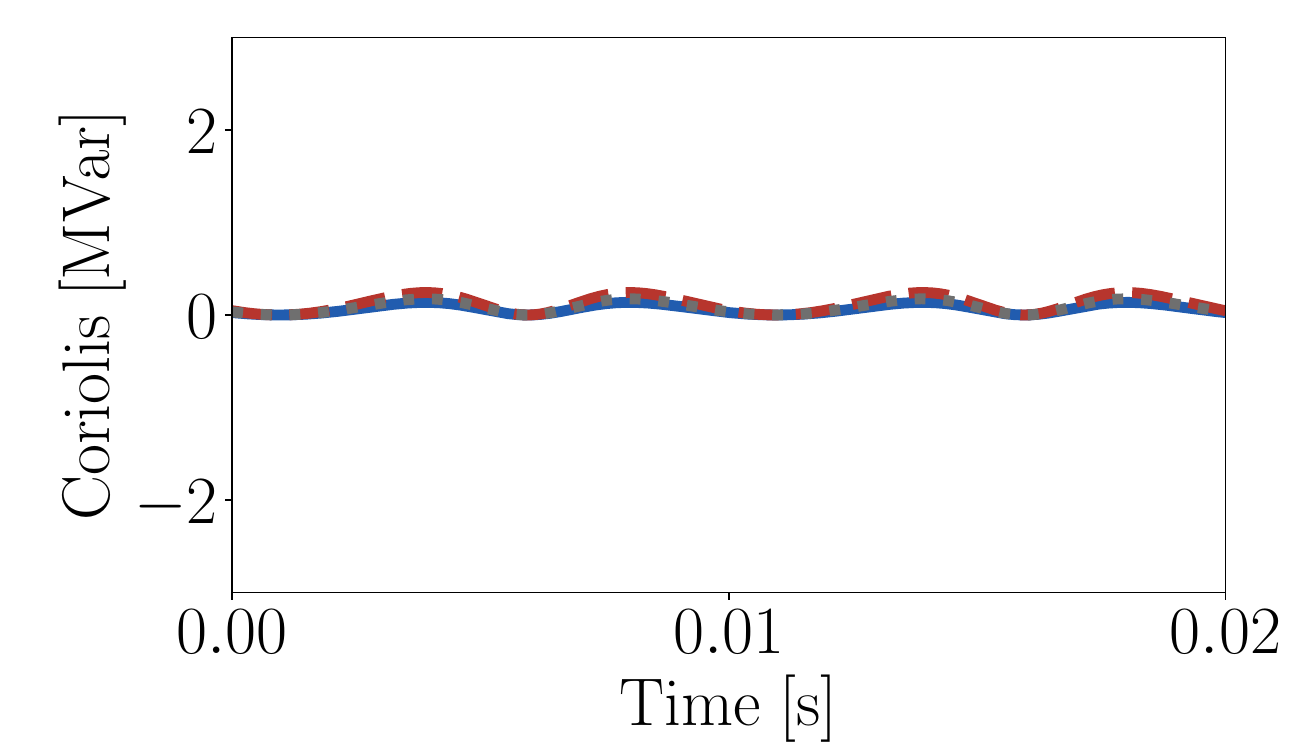}}
  \caption{Coriolis
    $ \bfg v \times (2C \bfg\omega_\varphi \times \bfg v_{\|}) $}
  \label{fig:E1:Qcor}
  \end{subfigure}
   \begin{subfigure}{.49\linewidth}
  \centering
  \resizebox{\linewidth}{!}{\includegraphics{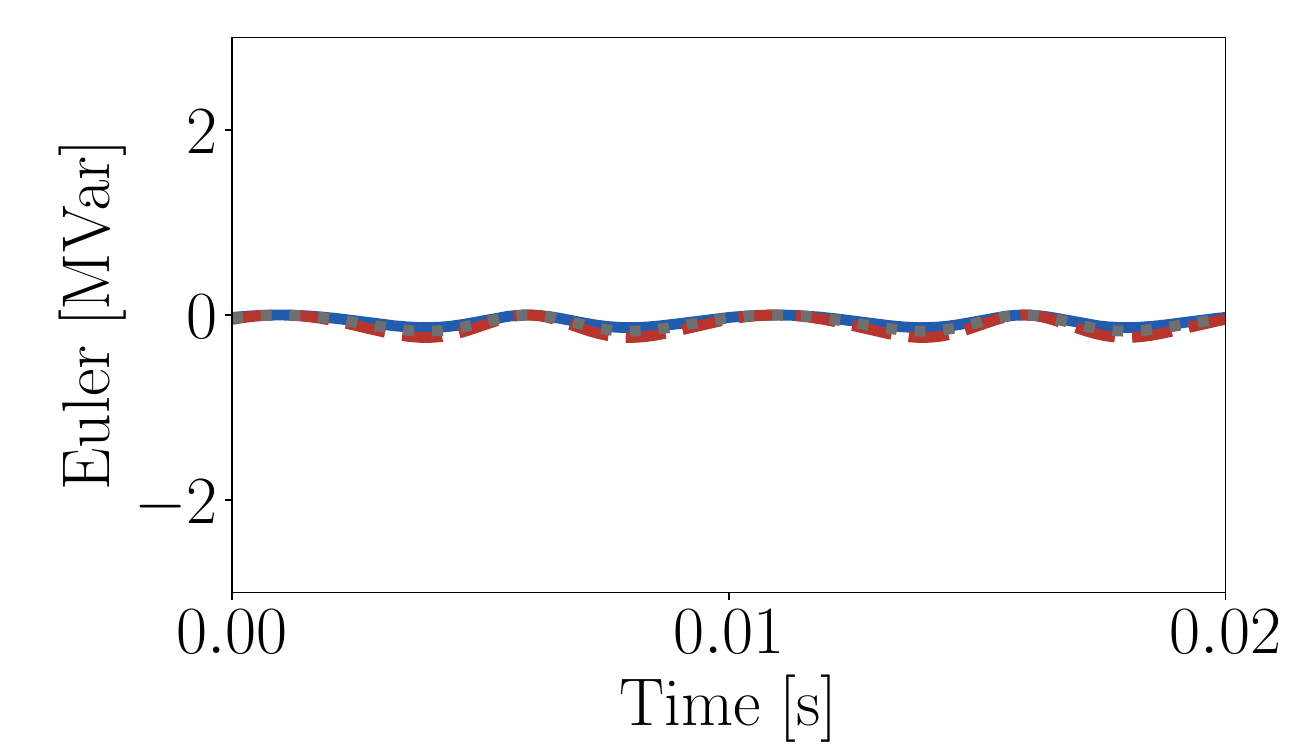}}
  \caption{Euler $\bfg v \times (C\bfg\omega_\varphi' \times \flux )$}
  \label{fig:E1:Qeul}
  \end{subfigure}
  \begin{subfigure}{.49\linewidth}
  \centering
  \resizebox{\linewidth}{!}{\includegraphics{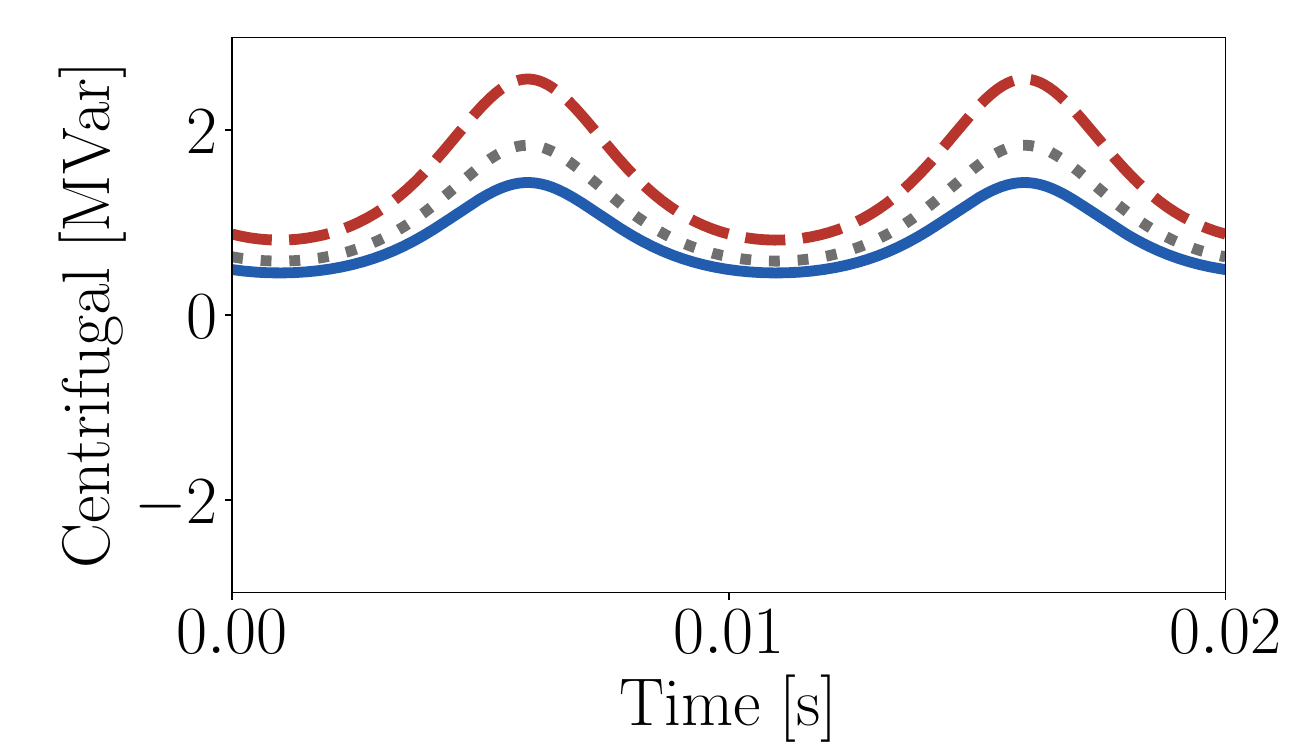}}
  \caption{Centrifugal
    $ \bfg v \times [C\bfg\omega_\varphi \times (\bfg\omega_\varphi
    \times \flux)]$}
  \label{fig:E1:Qcen}
  \end{subfigure}
  \caption{Balanced 3-phase capacitor with stationary unbalanced
    sinusoidal AC voltage: Relative, Coriolis, Euler and centrifugal
    components of reactive power pseudovector. The Coriolis and Euler
    components are opposite to each other.  The relative and
    centrifugal components sum up to a constant pseudovector.}
  \label{fig:E1:coriolis_theorem:Q}
  \vspace{-3mm}
\end{figure}

For the reactive power pseudovector, the relative, Coriolis, Euler and
centrifugal components of the reactive power pseudovector are shown in
Fig.~\ref{fig:E1:coriolis_theorem:Q}.  Following from
\eqref{eq:coreulopp}, the Coriolis and Euler components sum up to
zero. Moreover, we observe that the sum of the relative and
centrifugal sum up to a constant pseudovector, i.e.:
\begin{equation*}
\begin{aligned} 
\bfg v \times C \beta_\varphi \bfg v_{\|}  +
\bfg v \times [ C\bfg\omega_\varphi \times (\bfg\omega_\varphi \times \flux) ] 
=
c_1 \e{1} + c_2 \e{2} + c_3 \e{3} \, .
\end{aligned}
\end{equation*}
Therefore, the total instantaneous reactive power is also a constant pseudovector, 
in this case:
\begin{equation}
  \begin{aligned}
    \bfg Q = 0.807 \, \e{1} +
    1.437 \, \e{2} +
    1.034 \, \e{3} \ \ {\rm MVAr} \, ,
  \end{aligned}
  \nonumber
\end{equation}
which is also illustrated in 
Fig.~\ref{fig:E1:Q}.

\color{black}
 
\subsection{Stationary Unbalanced Load Case}

We have so far considered only point particles for which the moment of
inertia is a scalar, namely $I = m |\bfg r|^2$.  We can easily
generalize this notation to the general case of a set of point masses
rigidly connected or to masses with non-spherical shape.  All
definitions and developments presented remain the same, except for the
fact that the moment of inertia becomes a symmetrical second-order
tensor, called \textit{moment of inertia tensor}.  Thus, in 3D, one
has to define the six elements of the tensor.  The moment of inertia
tensor can be also conveniently thought as an operator that returns a
pseudovector.  for a system of discrete, rigidly-connected point
masses:
\begin{equation}
  \label{eq:Iw}
  \inertia{\orbit} = \sum_h m_h \, \bfg r_h \times (\orbit \times \bfg r_h) \, ,
\end{equation}
whereas, in a continuum, one has:
\begin{equation}
  \label{eq:Iwcont}
  \inertia{\orbit} = \int_{\tau} \rho(\bfg r) \, \bfg r \times (\orbit \times \bfg r) \, d \tau \, .
\end{equation}
The formulation in terms of the moment of inertia tensor $\inertia{\orbit}$ is useful in the study of unbalanced three-phase AC systems.  In particular, the case of a particle with three rigidly connected point particles with different masses or, equivalently, a mass with ellipsoidal shape, is equivalent to an unbalanced load.  If the particle is an ellipsoid, in fact, $\inertia{\orbit}$ is diagonal but the three diagonal elements have different values.  We can extend this analogy to a more complex case, i.e., the case for which $\inertia{\orbit}$ includes non-null diagonal elements.  This corresponds to the case, for example, of mutual-coupling among the inductances of a load.  
 
Using \eqref{eq:Iw}, the angular momentum and kinetic energy can be written as:
\begin{equation}
   \label{eq:LT}
   \bfg L = \inertia{\orbit} \, , \qquad T  = \frac{1}{2} \orbit \cdot \inertia{\orbit} \, .
\end{equation}
Similarly, one can define all other quantities previously described in the paper.  In particular, we observe that 
\eqref{eq:hatS2}, that is, the proposed expression of the instantaneous power, holds and, if the kinetic energy is expressed in terms of $\inertia{\orbit}$, it can be rewritten as:
\begin{equation}
   \boxed{\hat{S} = (\orbit \cdot \inertia{\orbit} ) \, \Wgeom{u} \, }
\end{equation}

The inertia tensor is useful when discussing three-phase AC systems.  For example, assuming that the coordinates of the system are the three phases, namely $(\e{a}, \e{b}, \e{c})$, for a balanced bank of capacitors with capacitance $C$ per phase, one obtains:
\begin{equation}
   \inertia{} = 
   C \, |\flux|^2 \begin{bmatrix}
   1 & 0 & 0 \\
   0 & 1 & 0 \\
   0 & 0 & 1 \\   
   \end{bmatrix} .
\end{equation}
Similarly, unbalanced banks of capacitors can be written as:
\begin{equation}
   \inertia{} = 
   |\flux|^2 \begin{bmatrix}
   C_{\rm a} & 0 & 0 \\
   0 & C_{\rm b} & 0 \\
   0 & 0 & C_{\rm c} \\   
   \end{bmatrix} .
\end{equation}
%

\begin{figure}[ht!]
\centering
  \begin{subfigure}{.49\linewidth}
  \centering
  \resizebox{\linewidth}{!}{\includegraphics{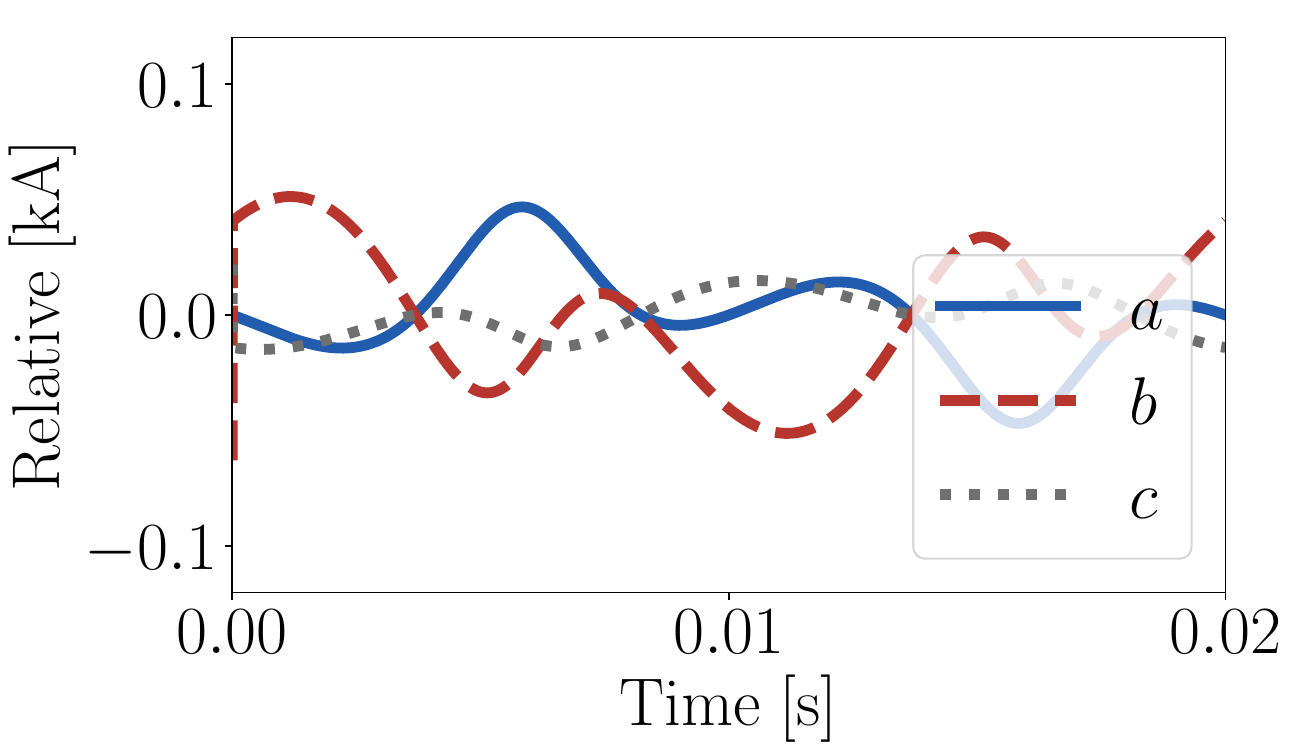}}
  \caption{Relative}
  \label{fig:E1b:frel}
  \end{subfigure}
  \begin{subfigure}{.49\linewidth}
  \centering
  \resizebox{\linewidth}{!}{\includegraphics{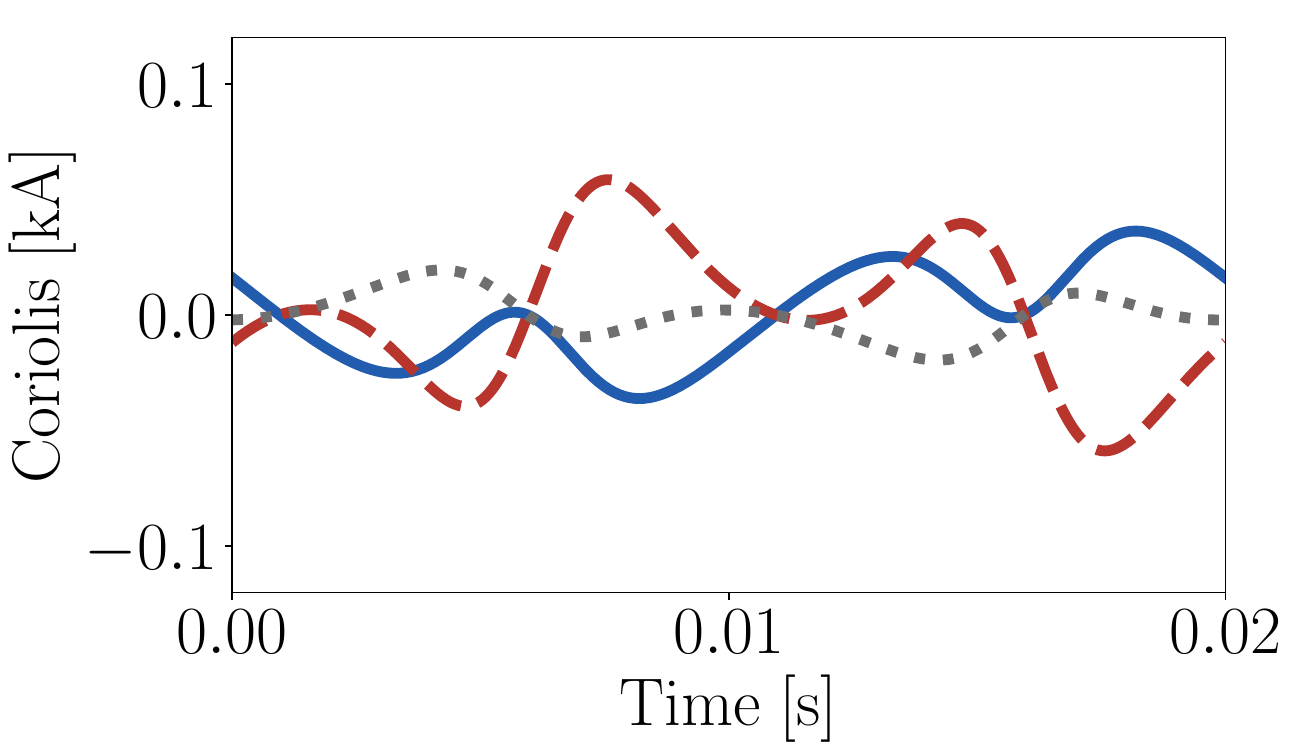}}
  \caption{Coriolis}
  \label{fig:E1b:fcor}
  \end{subfigure}
   \begin{subfigure}{.49\linewidth}
  \centering
  \resizebox{\linewidth}{!}{\includegraphics{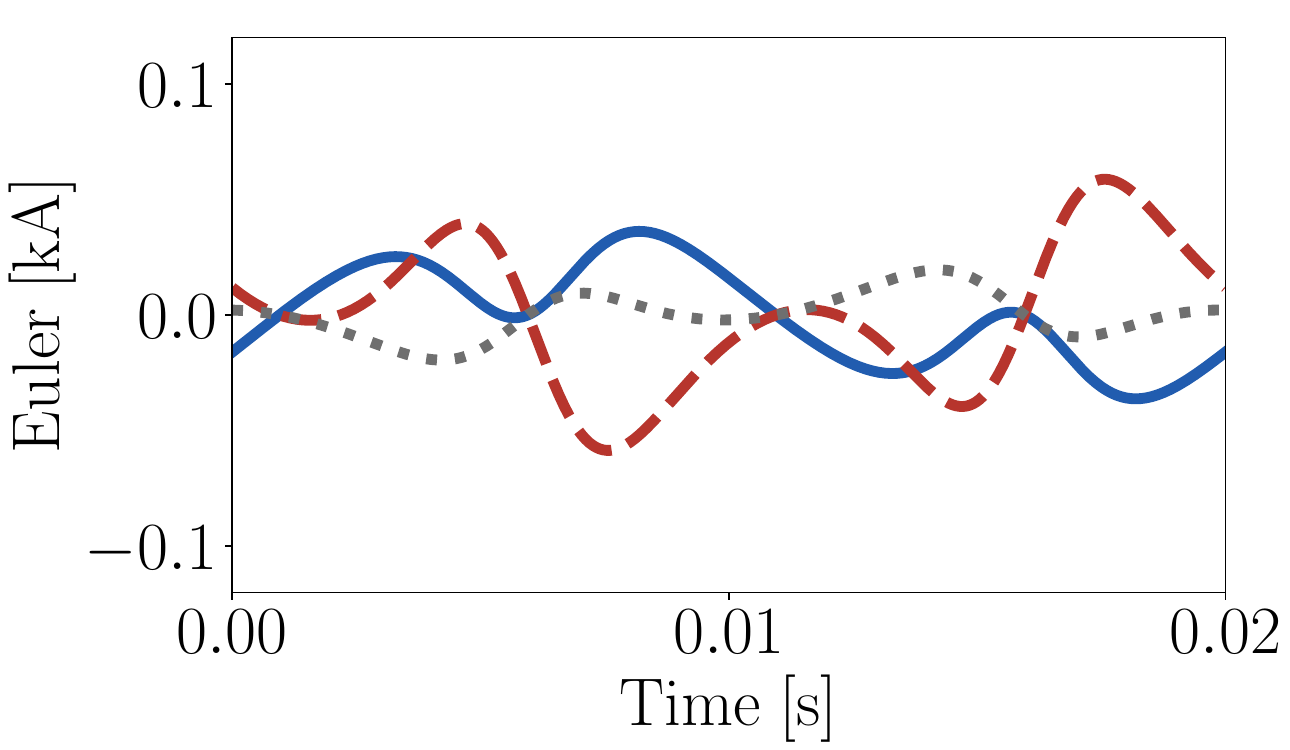}}
  \caption{Euler}
  \label{fig:E1b:feul}
  \end{subfigure}
  \begin{subfigure}{.49\linewidth}
  \centering
  \resizebox{\linewidth}{!}{\includegraphics{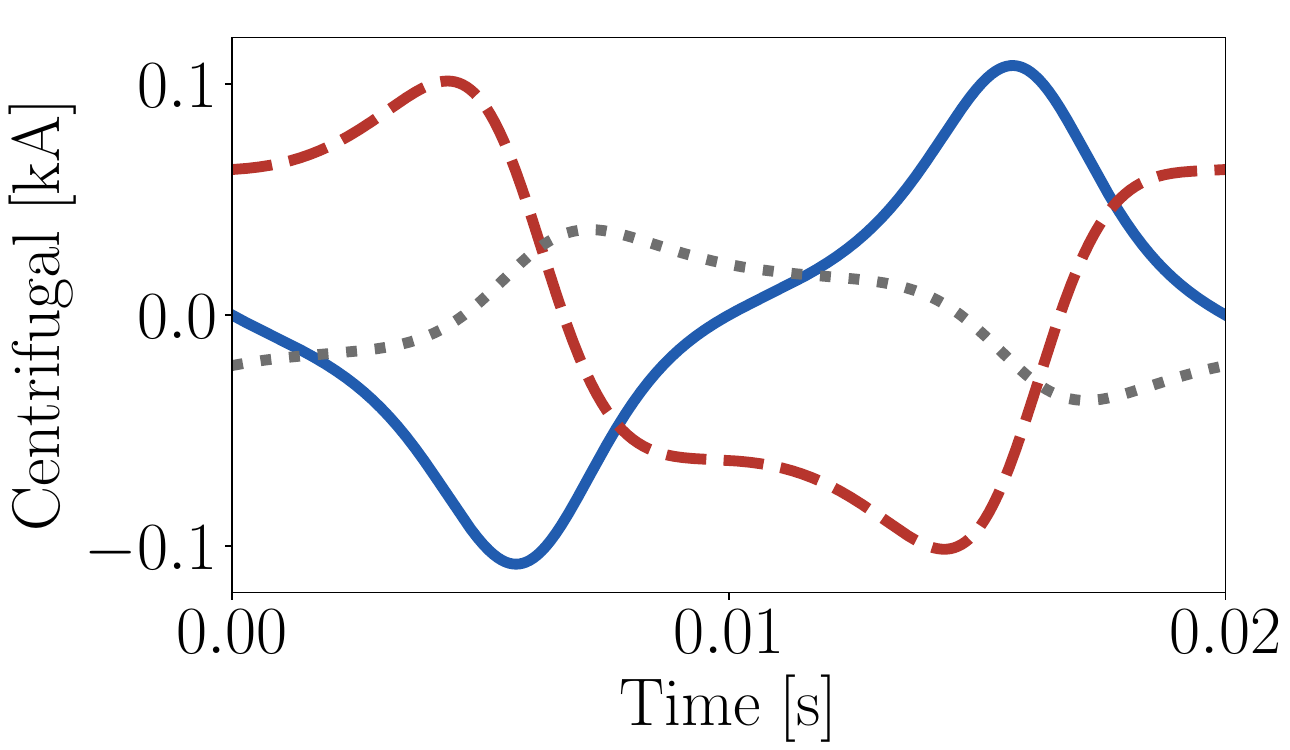}}
  \caption{Centrifugal}
  \label{fig:E1b:fcen}
  \end{subfigure}
  \caption{Unbalanced 3-phase capacitor with stationary unbalanced
    sinusoidal AC voltage: Relative, Coriolis, Euler and centrifugal
    components of current. The Coriolis and Euler components are
    opposite to each other.}
  \label{fig:E1b:coriolis_theorem}
  \vspace{-3mm}
\end{figure}

To further illustrate the application of our theory to unbalanced load systems, 
we assume a numerical example with the unbalanced voltage considered in Section~\ref{sec:unbalsin}, where in this case the three-phase capacitor is taken as unbalanced.  In particular, we use $C_a = 10$~$\mu \rm F$, $C_b = 20$~$\mu \rm F$,  $C_c = 5$~$\mu \rm F$.
The relative, Coriolis, Euler, and centrifugal components of the capacitor current $\bfg\ii_{\scriptscriptstyle C}$ 
in this case are illustrated in Fig.~\ref{fig:E1b:coriolis_theorem}.
Once again, the Coriolis and Euler components are opposite, i.e.~their sum is equal to zero.

\color{black}

\subsection{Stationary Balanced Non-Sinusoidal Case}

Let us assume again the balanced capacitor considered in
Sections~\ref{sec:balsin} and \ref{sec:unbalsin}. The capacitor
voltage is in this case stationary but now includes a fifth harmonic
component, as follows:
\begin{equation*}
  \begin{aligned}
    v_a &= V [\cos (\wo t) + 0.05 \cos (5\wo t)] \, , \\
    v_b &= V [\cos (\wo t - {2\pi}/{3})
          + 0.05 \cos (5\wo t - {2\pi}/{3})]\, , \\
    v_c &= V [\cos (\wo t + {2\pi}/{3})
          + 0.05\cos (5\wo t + {2\pi}/{3})]\, ,
  \end{aligned}
\end{equation*} 
where $V=20$~kV.  Moreover, to see the effect of losses, we also
consider that a conductance (with value $G$ per phase) is connected in
parallel to the capacitor.  The current $\bfg\ii$ applied to the
parallel combination of capacitor and conductance is then:
\begin{equation*}
\bfg\ii = \bfg \ii_{\scriptscriptstyle C} + G \, \bfg v  \, .
\end{equation*}

\begin{figure}[ht!]
\centering
  \begin{subfigure}{.49\linewidth}
  \centering
  \resizebox{\linewidth}{!}{\includegraphics{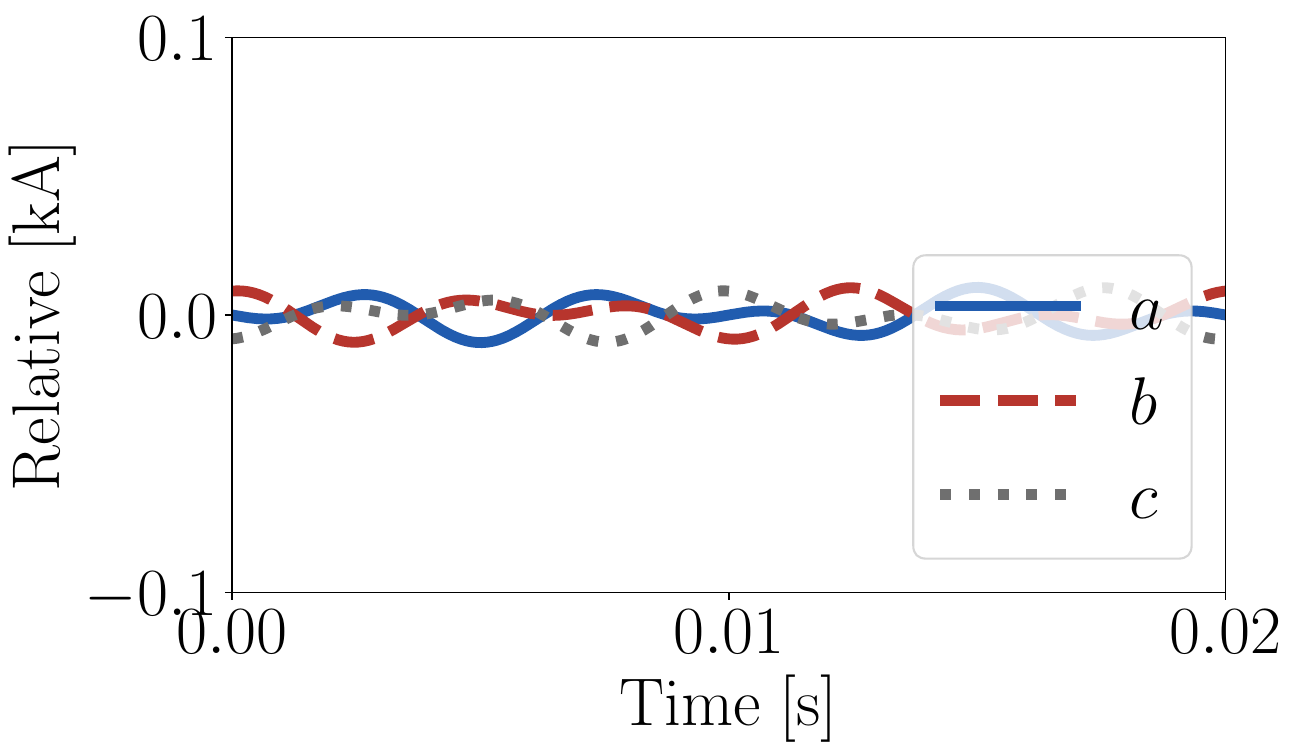}}
  \caption{Relative 
  $C \beta_\varphi \bfg v_{\|} $}
  \label{fig:E2:frel}
  \end{subfigure}
  \begin{subfigure}{.49\linewidth}
  \centering
  \resizebox{\linewidth}{!}{\includegraphics{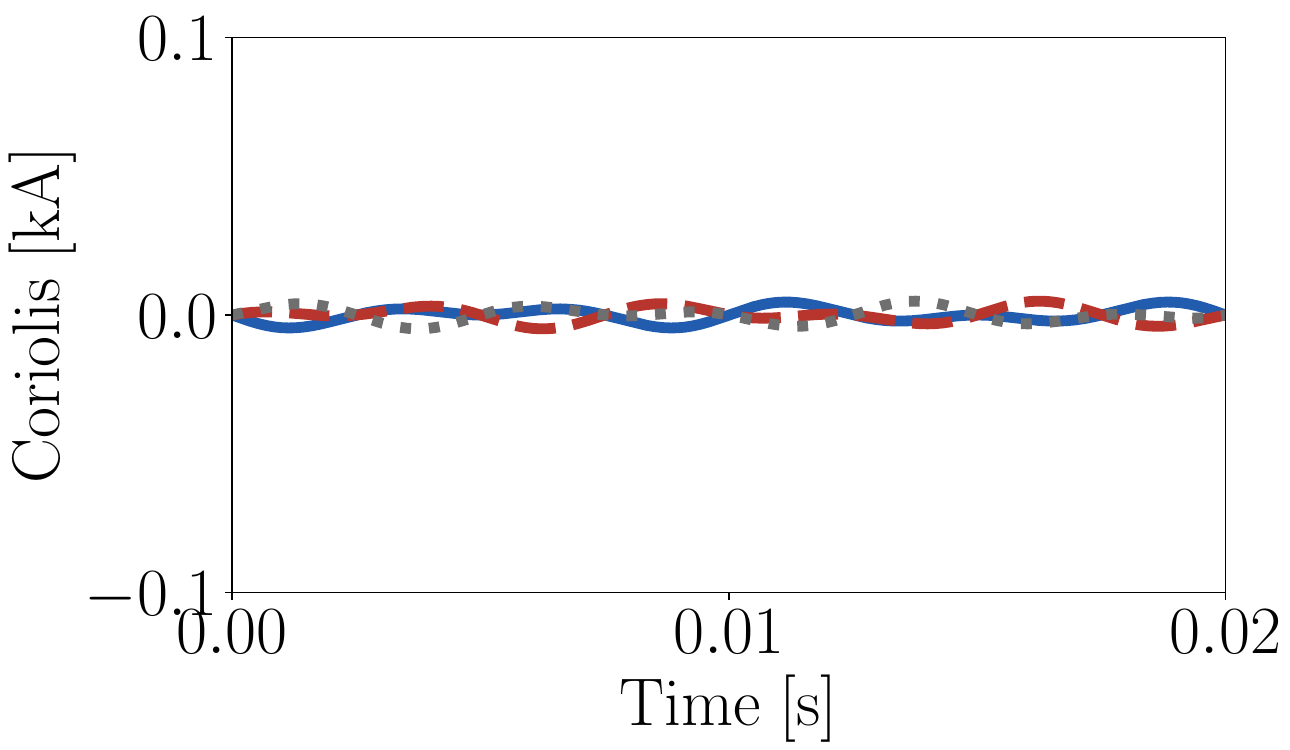}}
  \caption{Coriolis $
  2 C\bfg\omega_\varphi \times \bfg v_{\|} $}
  \label{fig:E2:fcor}
  \end{subfigure}
   \begin{subfigure}{.49\linewidth}
  \centering
  \resizebox{\linewidth}{!}{\includegraphics{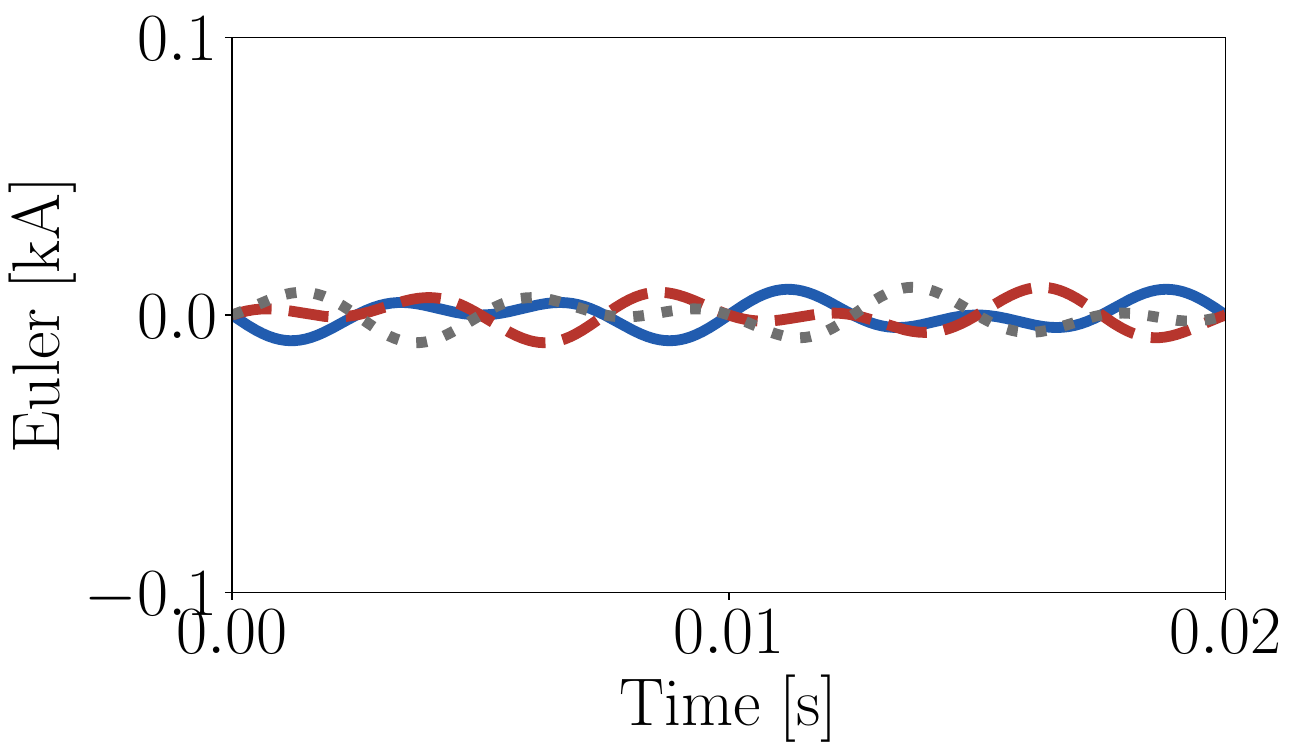}}
  \caption{Euler 
  $C\bfg\omega_\varphi' \times \flux $}
  \label{fig:E2:feul}
  \end{subfigure}
  \begin{subfigure}{.49\linewidth}
  \centering
  \resizebox{\linewidth}{!}{\includegraphics{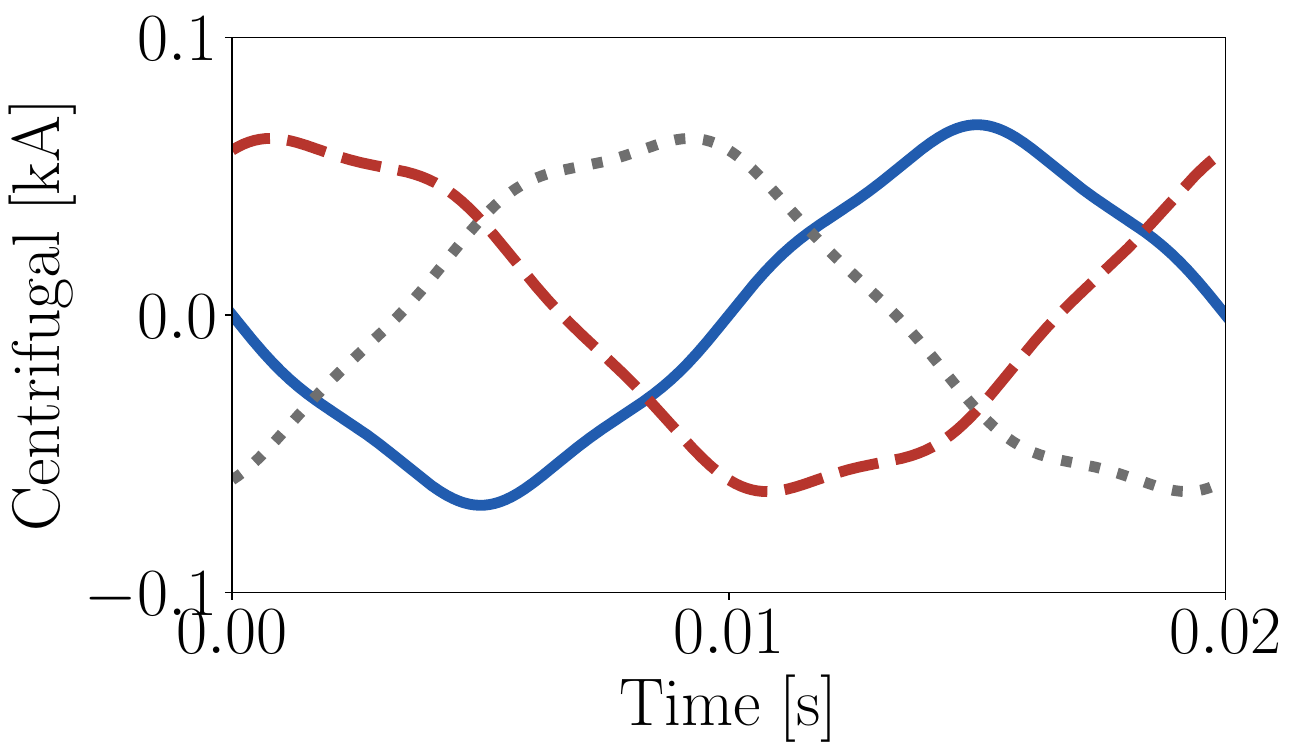}}
  \caption{Centrifugal $
  C\bfg\omega_\varphi \times (\bfg\omega_\varphi \times \flux)$}
  \label{fig:E2:fcen}
  \end{subfigure}
  \caption{ 3-phase capacitor with stationary balanced non-sinusoidal
    AC voltage: Relative, Coriolis, Euler and centrifugal components
    of applied current $\bfg\ii_{\scriptscriptstyle C}$.  }
\label{fig:E2:coriolis_theorem}
  \vspace{-3mm}
\end{figure}

\begin{figure}[ht!]
  \centering
  \resizebox{0.7\linewidth}{!}{\includegraphics{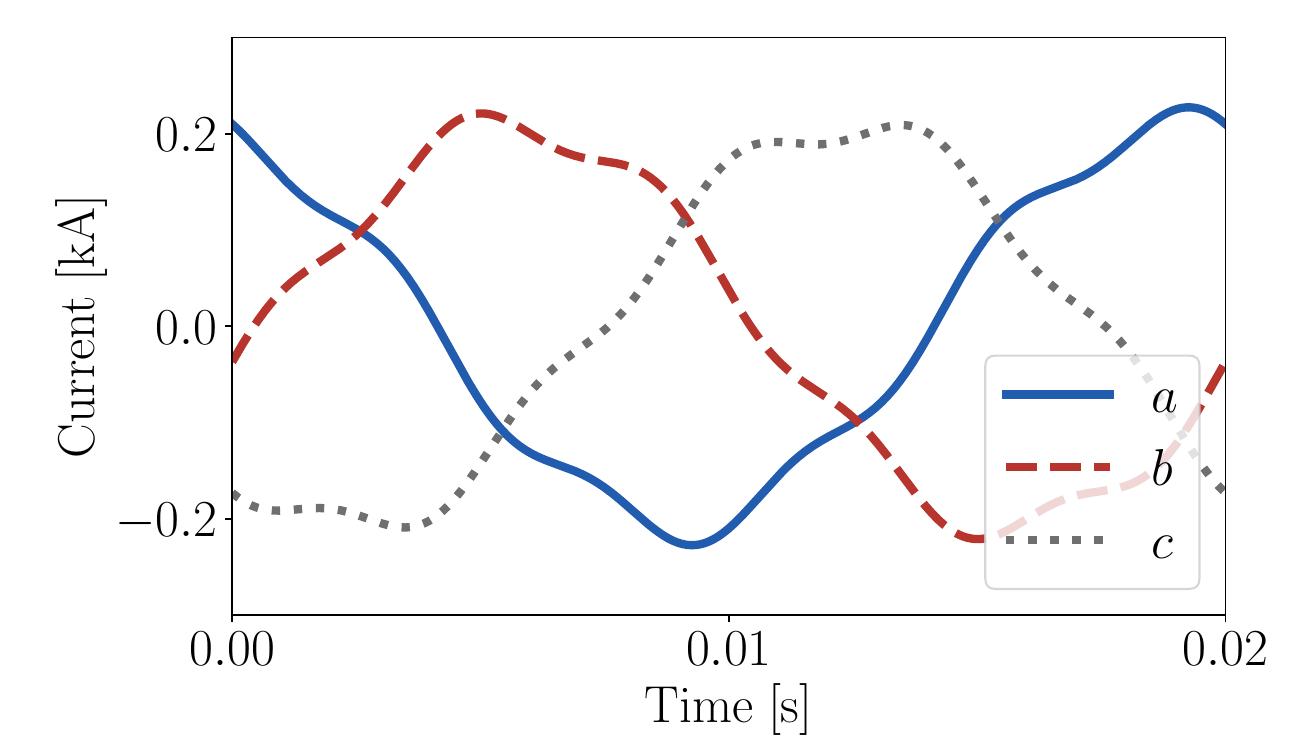}}
  \caption{Stationary balanced non-sinusoidal case: Current $\bfg \ii$
    applied to the combination of capacitor and conductance.}
  \label{fig:E2:force}
\end{figure}

The relative, Coriolis, Euler, and centrifugal components of the
current $\bfg\ii_{\scriptscriptstyle C}$ applied to the capacitor are
illustrated in Fig.~\ref{fig:E2:coriolis_theorem}.  All components are
observed to be non-zero and vary in time.  Assuming that $G=0.01$~S,
the current $\bfg \ii$ is shown in Fig.~\ref{fig:E2:force}.

\begin{figure}[ht!]
  \centering
  \resizebox{0.6\linewidth}{!}{\includegraphics{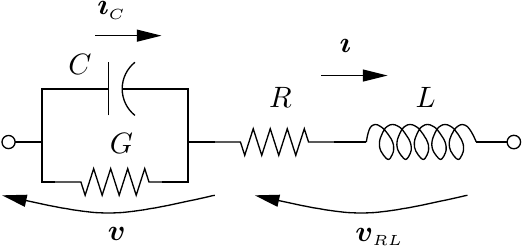}}
  \caption{Resistance ($R$) and inductor ($L$) connected in series to
    the parallel of a capacitor ($C$) and a conductance ($G$).}
  \label{fig:E2:rlc}
\end{figure}

Let us assume now that a three-phase inductor with $L=0.02$~H per
phase and a three-phase resistance with $R=8$~$\Omega$ per phase, are
connected in series with the parallel combination of the capacitor and
conductance, as shown in Fig.~\ref{fig:E2:rlc}.

The voltage vector $\bfg v_{\scriptscriptstyle RL}$ applied to the series of the inductor and resistance is:
\begin{equation*}
\bfg v_{\scriptscriptstyle RL} = \bfg v_{\scriptscriptstyle L} + R \, \bfg \ii = \bfg v_{\scriptscriptstyle L} + R \,
(\bfg \ii_{\scriptscriptstyle C} + G \bfg v )
\, .
\end{equation*}

\begin{figure}[ht!]
\centering
  \begin{subfigure}{.49\linewidth}
  \centering
  \resizebox{\linewidth}{!}{\includegraphics{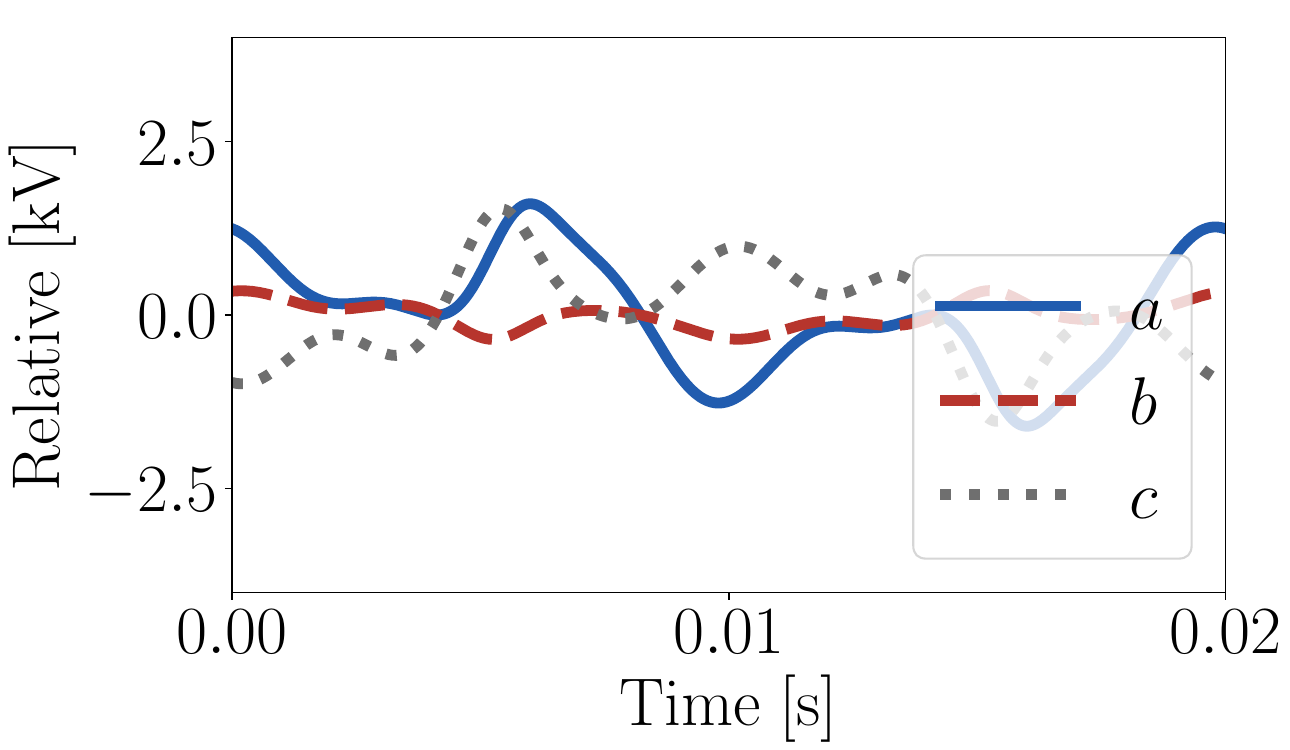}}
  \caption{Relative 
  $L\beta_q \bfg \ii_{\|} $}
  \label{fig:E2:frel:L}
  \end{subfigure}
  \begin{subfigure}{.49\linewidth}
  \centering
  \resizebox{\linewidth}{!}{\includegraphics{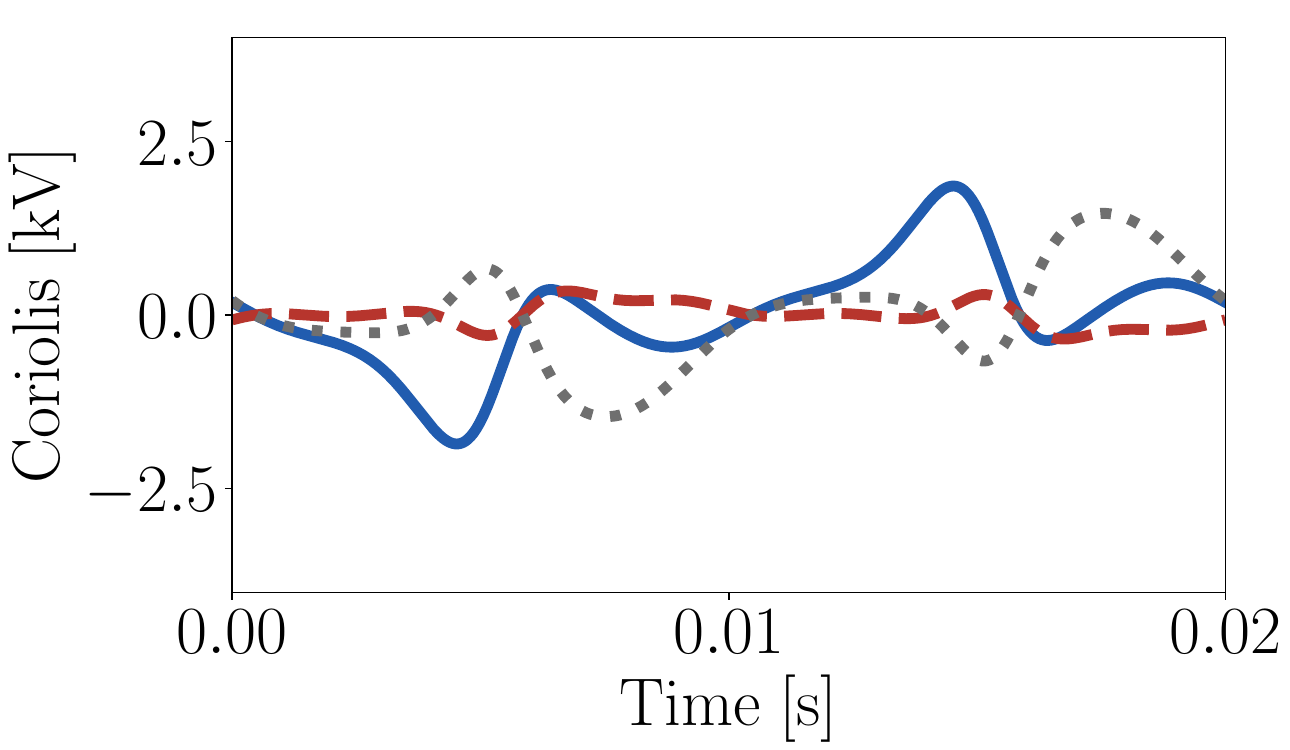}}
  \caption{Coriolis $
  2 L\bfg\omega_q \times \bfg \ii_{\|} $}
  \label{fig:E2:fcor:L}
  \end{subfigure}
   \begin{subfigure}{.49\linewidth}
  \centering
  \resizebox{\linewidth}{!}{\includegraphics{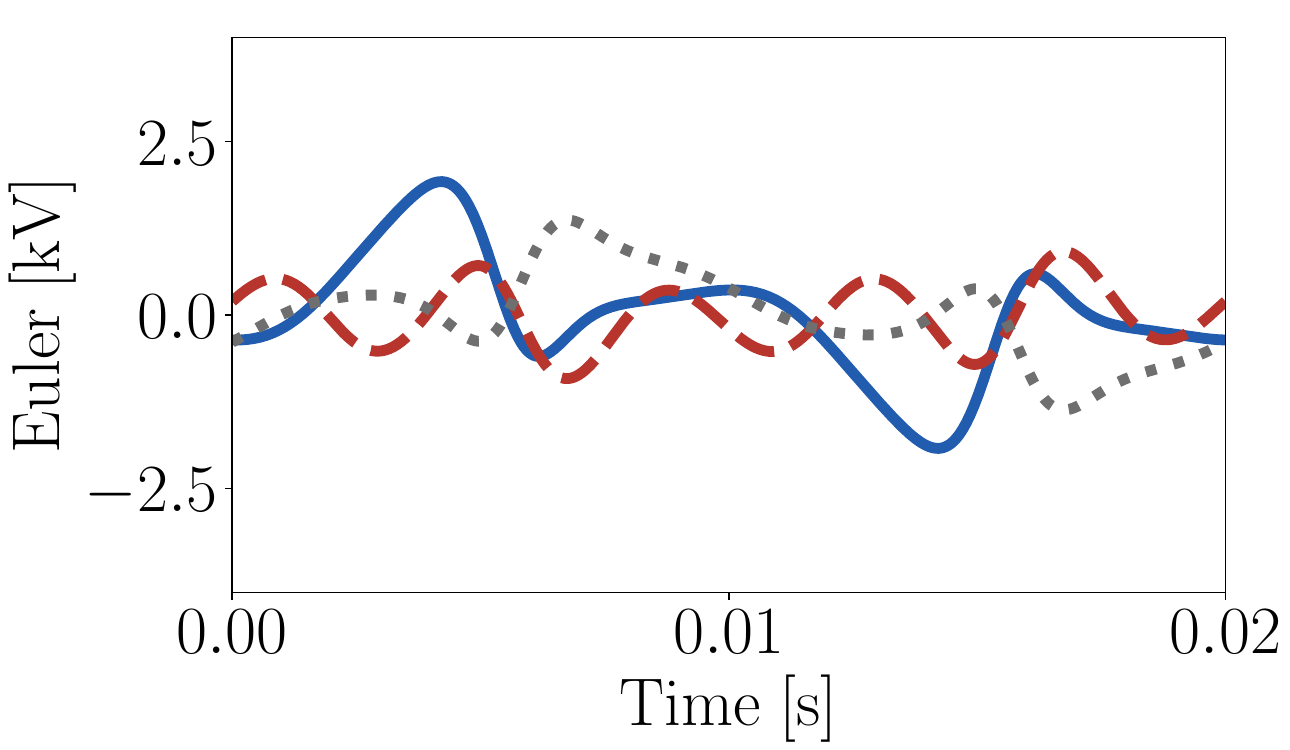}}
  \caption{Euler 
  $L\bfg\omega_q' \times \bfg q $}
  \label{fig:E2:feul:L}
  \end{subfigure}
  \begin{subfigure}{.49\linewidth}
  \centering
  \resizebox{\linewidth}{!}{\includegraphics{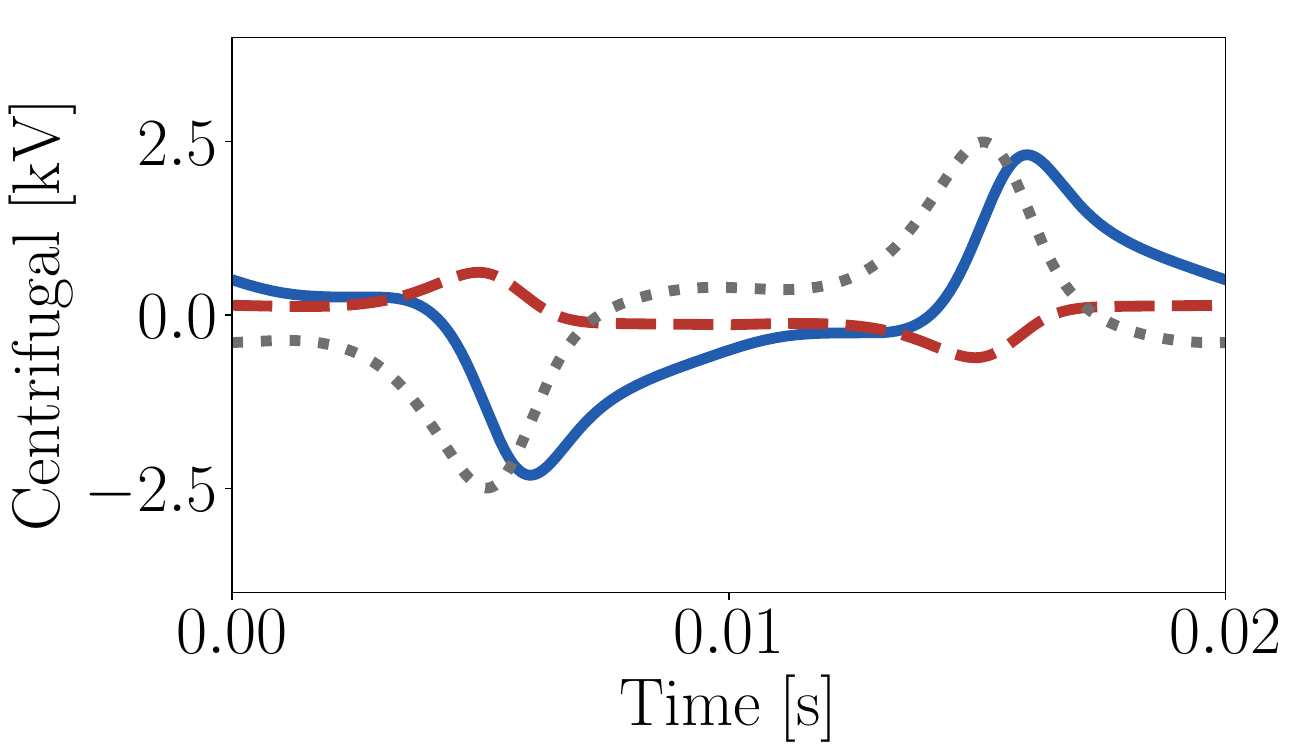}}
  \caption{Centrifugal $
  L\bfg\omega_q \times (\bfg\omega_q \times \bfg q)$}
  \label{fig:E2:fcen:L}
  \end{subfigure}
  \caption{ 3-phase inductor with stationary balanced non-sinusoidal
    AC current $\bfg \ii$: Relative, Coriolis, Euler and centrifugal
    components of applied voltage $\bfg v_{\scriptscriptstyle L}$.  }
\label{fig:E2:coriolis_theorem:L}
  \vspace{-3mm}
\end{figure}

\begin{figure}[ht!]
  \centering
  \resizebox{0.7\linewidth}{!}{\includegraphics{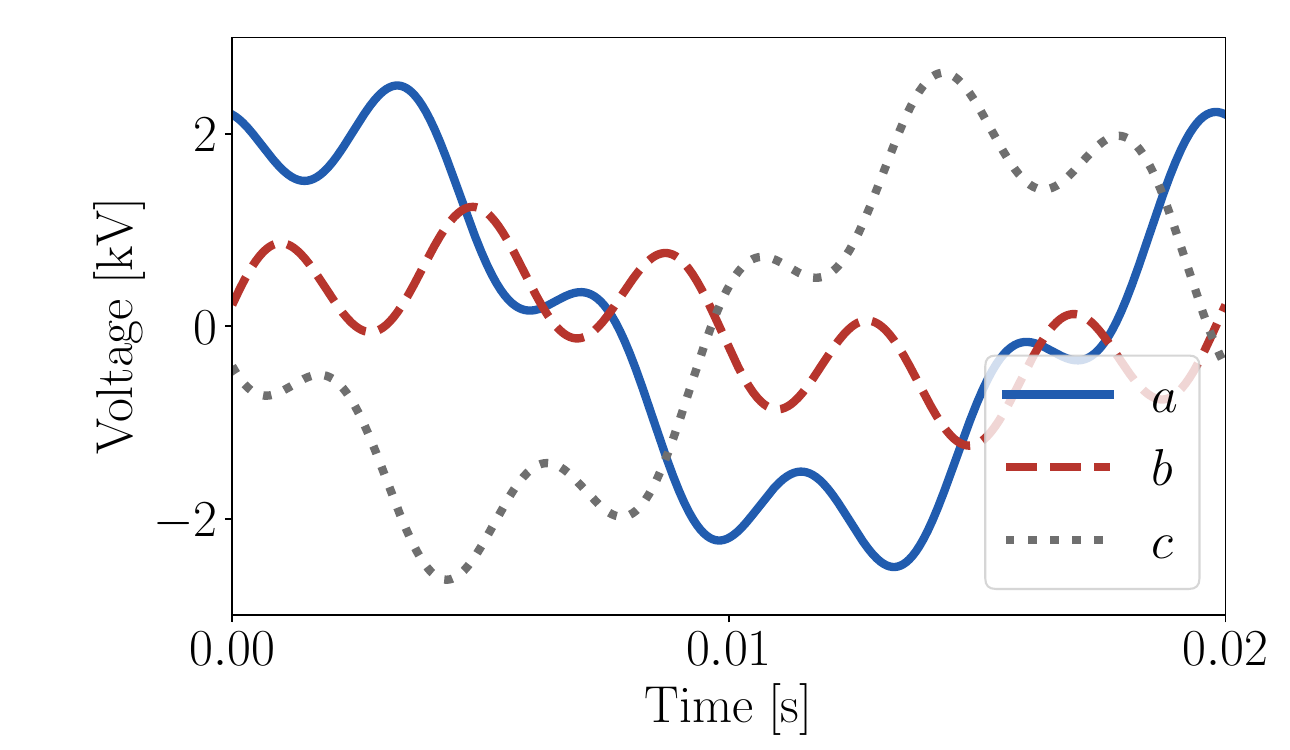}}
  \caption{Stationary balanced non-sinusoidal case: Voltage $\bfg v_{\scriptscriptstyle RL}$ applied to the series of inductor and resistance.}
  \label{fig:E2b:force}
\end{figure}

The relative, Coriolis, Euler and centrifugal components of the
voltage $\bfg v_{\scriptscriptstyle L}$ applied to the inductor are
plotted in Fig.~\ref{fig:E2:coriolis_theorem:L}. The figure shows that
the high levels of harmonic distortion considered in this example
result in all four force components having a more or less equal
contribution to the voltage $\bfg v_{\scriptscriptstyle L}$.  The
profile of the voltage $\bfg v_{\scriptscriptstyle RL}$ is shown in
Fig.~\ref{fig:E2b:force}.

Finally, the total instantaneous active and reactive powers of the
circuit considered in Fig.~\ref{fig:E2:rlc} are:
\begin{align*}
  p &= G |\bfg v|^2 + R |\bfg \ii|^2 \, , \\
  \bfg Q &=
           \bfg v \times \bfg \ii_{\scriptscriptstyle C}  +  \bfg \ii \times \bfg v_{\scriptscriptstyle RL} \, .  
\end{align*}

\begin{figure}[ht!]
\centering
  \begin{subfigure}{.49\linewidth}
  \centering
  \resizebox{\linewidth}{!}{\includegraphics{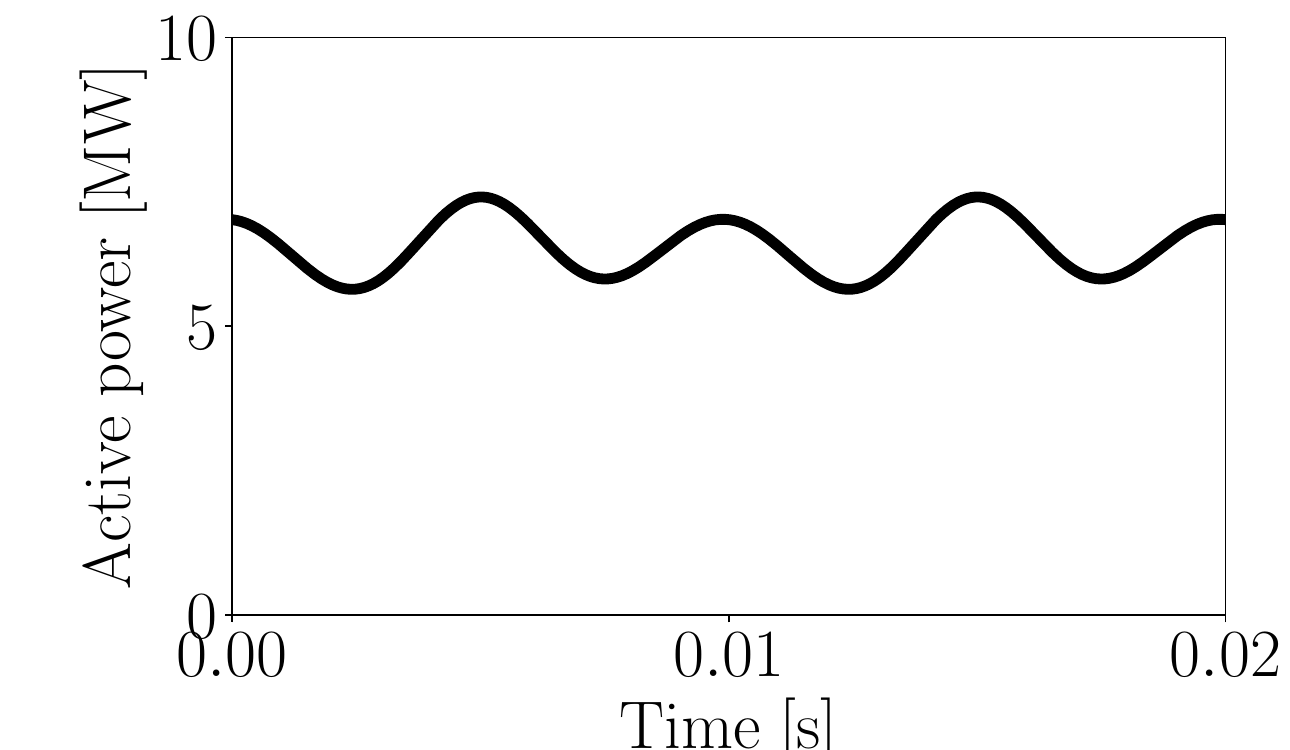}}
  \caption{Active power $p$}
  \label{fig:E2:p}
  \end{subfigure}
  \begin{subfigure}{.49\linewidth}
  \centering
  \resizebox{\linewidth}{!}{\includegraphics{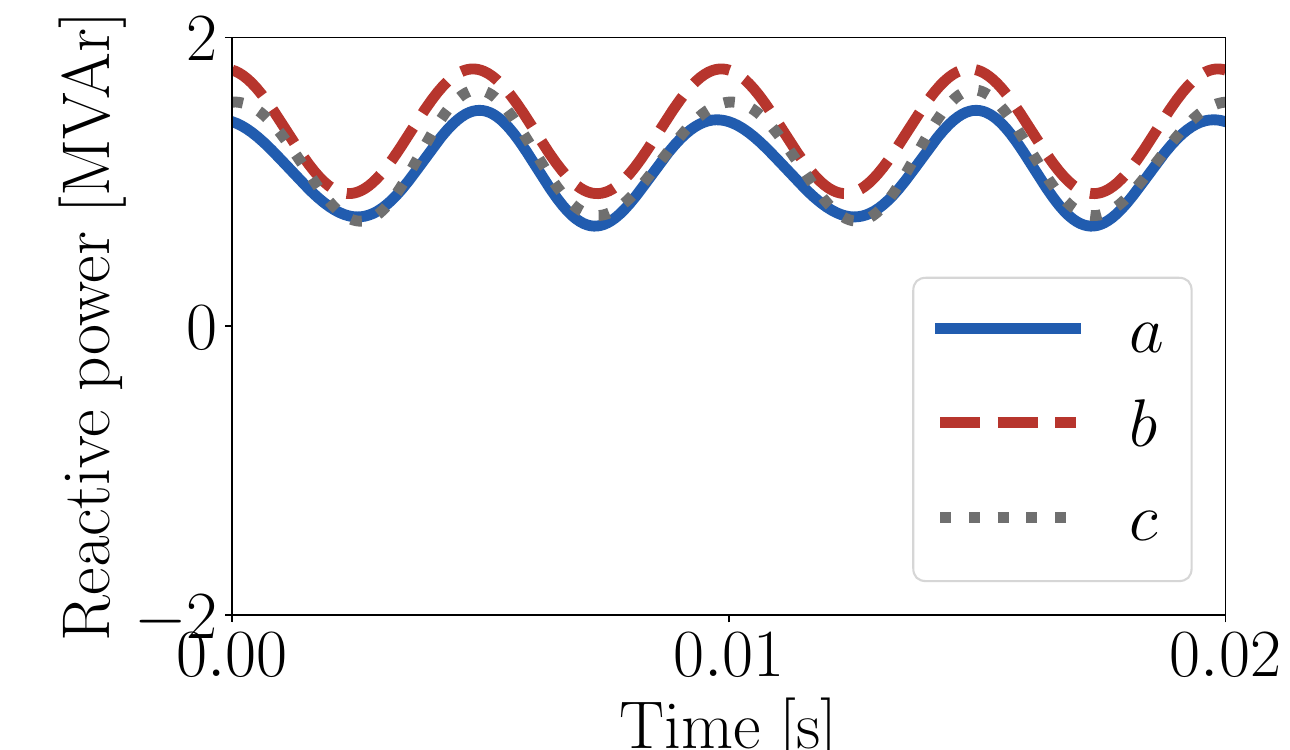}}
  \caption{Components of reactive power $\bfg Q$}
  \label{fig:E2:Q}
  \end{subfigure}
  \caption{Stationary non-sinusoidal case: instantaneous powers for
    circuit shown in Fig~\ref{fig:E2:rlc}.}
  \label{fig:E2:pQ}
  \vspace{-3mm}
\end{figure}

The profiles of $p$ and of the components of $\bfg Q$ for the
numerical example examined are presented in Fig.~\ref{fig:E2:pQ}.

\subsection{Non-Stationary Case}
\label{ex:nonbalanced}

\begin{figure}[ht!]
  \centering
  \resizebox{0.7\linewidth}{!}{\includegraphics{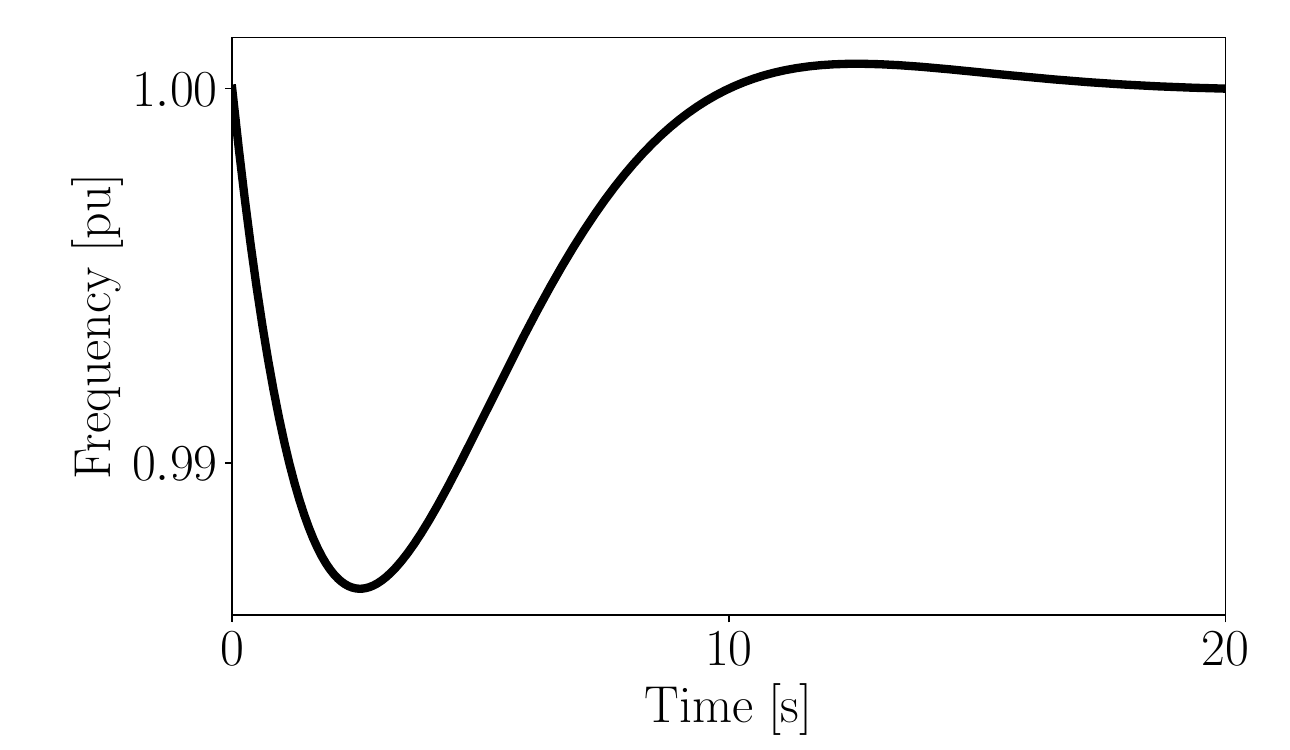}}
  \caption{Non-stationary case: frequency $\theta'(t)$ (in pu).}
  \label{fig:E3:frequency}
  \end{figure}

In this example, we consider again the same three-phase capacitor
studied above, which however is now assumed to be subject to a
non-stationary voltage vector, with:
\begin{equation}
  \begin{aligned}
v_a &= V \cos \theta(t) \, , \\
v_b &= V \cos (\theta(t)- {2\pi}/{3}) \, , \\
v_c &= V \cos (\theta(t) + {2\pi}/{3}) \, ,
  \end{aligned}
  \nonumber
\end{equation} 
where 
$\theta(t)= \wo t -0.04\wo e^{-0.3t}( 1.66 \cos\dfrac{\pi}{10} t + 1.59 \sin\dfrac{\pi}{10}t  )$.
The derivative of $\theta(t)$ is:
\begin{equation}
\begin{aligned}
\theta'(t) = \wo -\wo0.04e^{-0.3t} \sin(0.1\pi t)  \, , 
\end{aligned}
\nonumber
\end{equation} 
which emulates the profile of a typical underfrequency response following a negative active power imbalance in a power system (see also Fig.~\ref{fig:E3:frequency}).

\begin{figure}[ht!]
\centering
  \begin{subfigure}{.49\linewidth}
  \centering
  \resizebox{\linewidth}{!}{\includegraphics{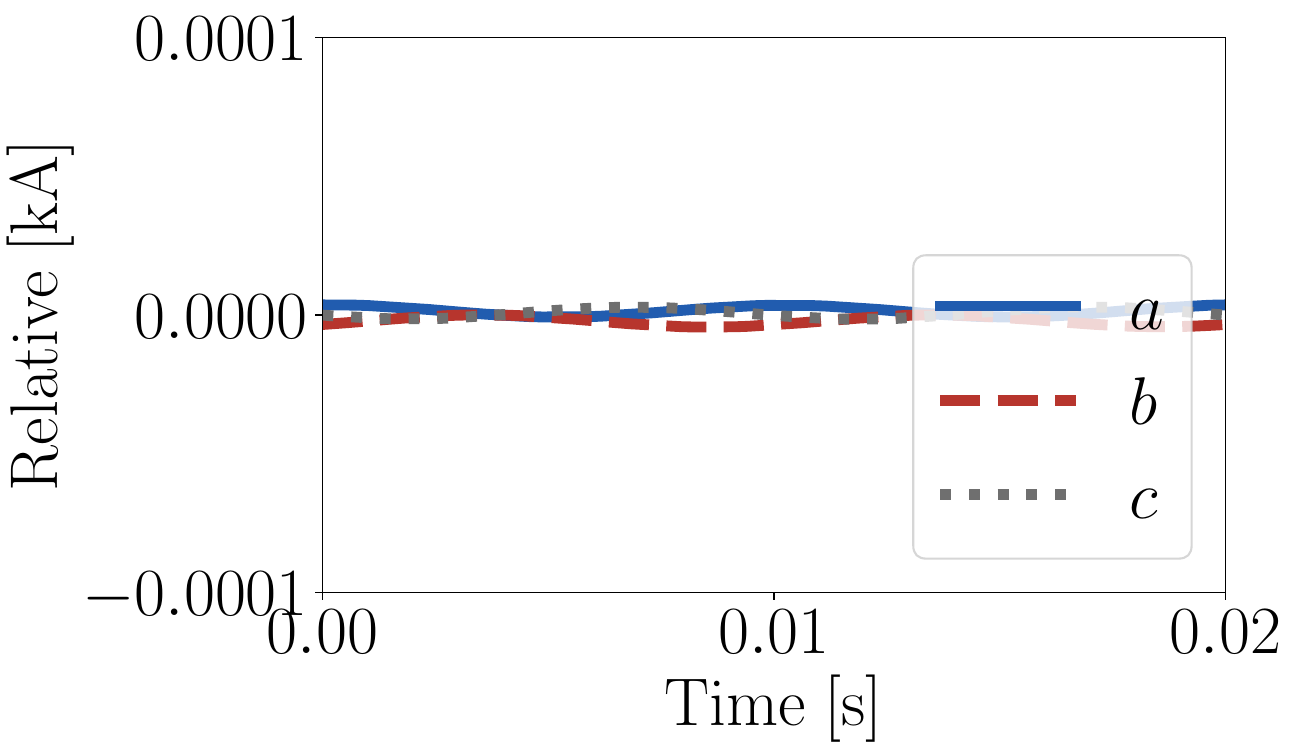}}
  \caption{Relative $C \beta_\varphi \bfg v_{\|}$}
  \label{fig:E3:frel}
  \end{subfigure}
  \begin{subfigure}{.49\linewidth}
  \centering
  \resizebox{\linewidth}{!}{\includegraphics{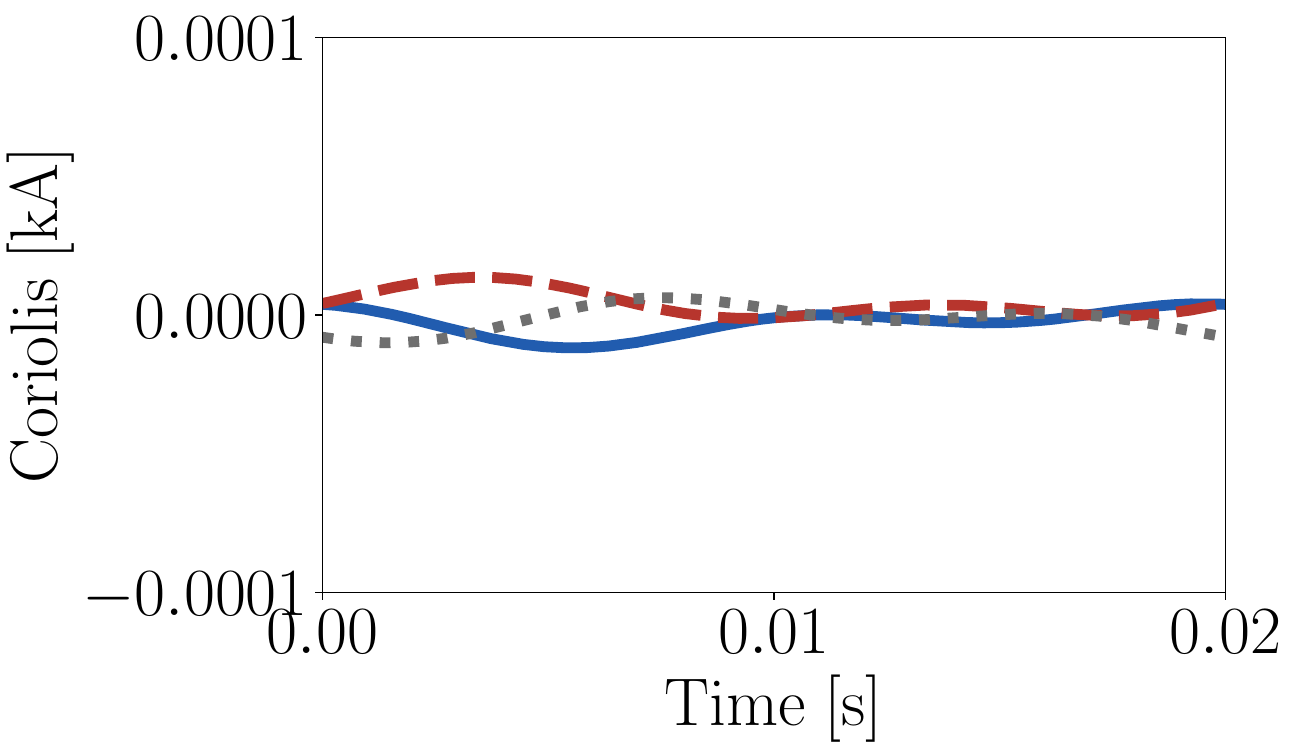}}
  \caption{Coriolis $
  2C \bfg\omega_\varphi \times \bfg v_{\|} $}
  \label{fig:E3:fcor}
  \end{subfigure}
   \begin{subfigure}{.49\linewidth}
  \centering
  \resizebox{\linewidth}{!}{\includegraphics{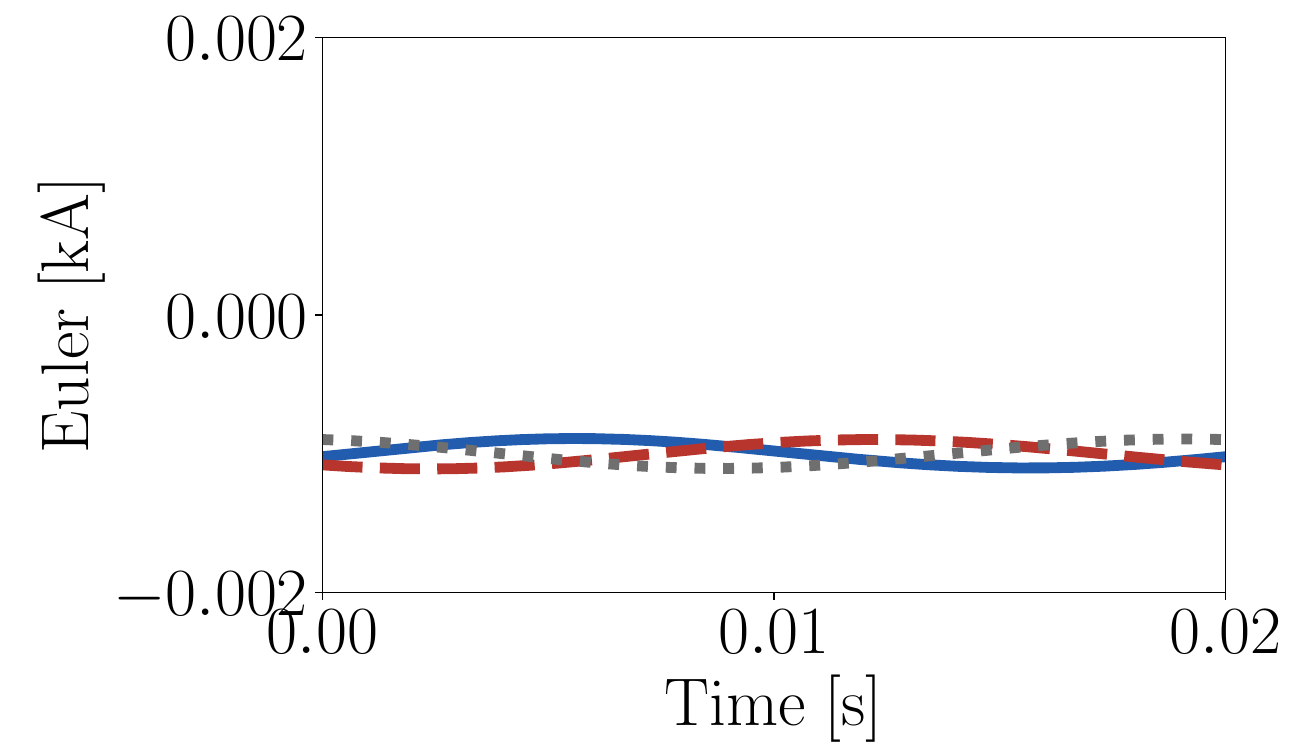}}
  \caption{Euler 
  $C\bfg\omega_\varphi' \times \flux $}
  \label{fig:E3:feul}
  \end{subfigure}
  \begin{subfigure}{.49\linewidth}
  \centering
  \resizebox{\linewidth}{!}{\includegraphics{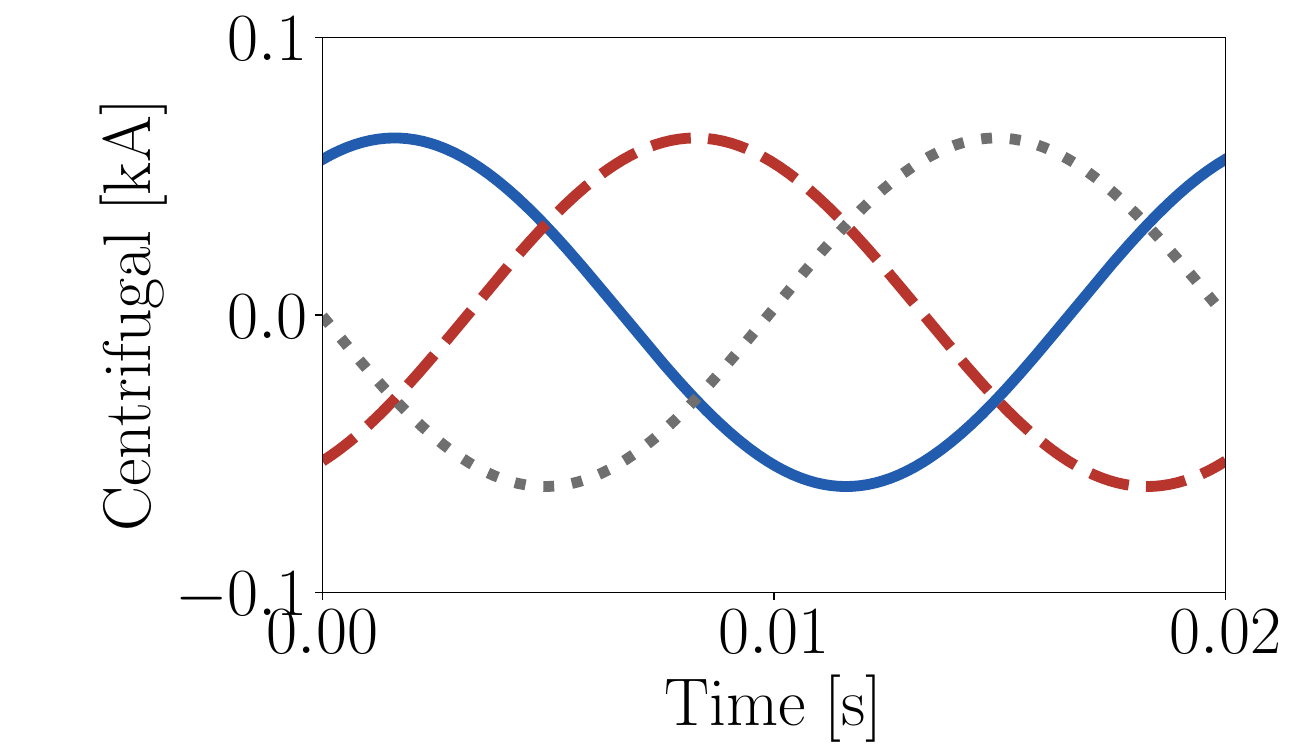}}
  \caption{Centrifugal $
  C\bfg\omega_\varphi \times (\bfg\omega_\varphi \times \flux)$}
  \label{fig:E3:fcen}
  \end{subfigure}
  \caption{3-phase capacitor with non-stationary balanced AC voltage:
    Relative, Coriolis, Euler and centrifugal components of current. }
  \label{fig:E3:coriolis_theorem}
  \vspace{-3mm}
\end{figure}

Figure~\ref{fig:E3:coriolis_theorem} shows the first period of the relative, Coriolis, Euler, and centrifugal components of the current applied to the capacitor. In this case, the centrifugal component largely determines the profile of the capacitor's current, with the contribution of the relative, Coriolis components being non-zero but practically negligible.  On the other hand, the Euler component contributes a small yet noticeable negative current.

\begin{figure}[ht!]
\centering
  \begin{subfigure}{.49\linewidth}
  \centering
  \resizebox{\linewidth}{!}{\includegraphics{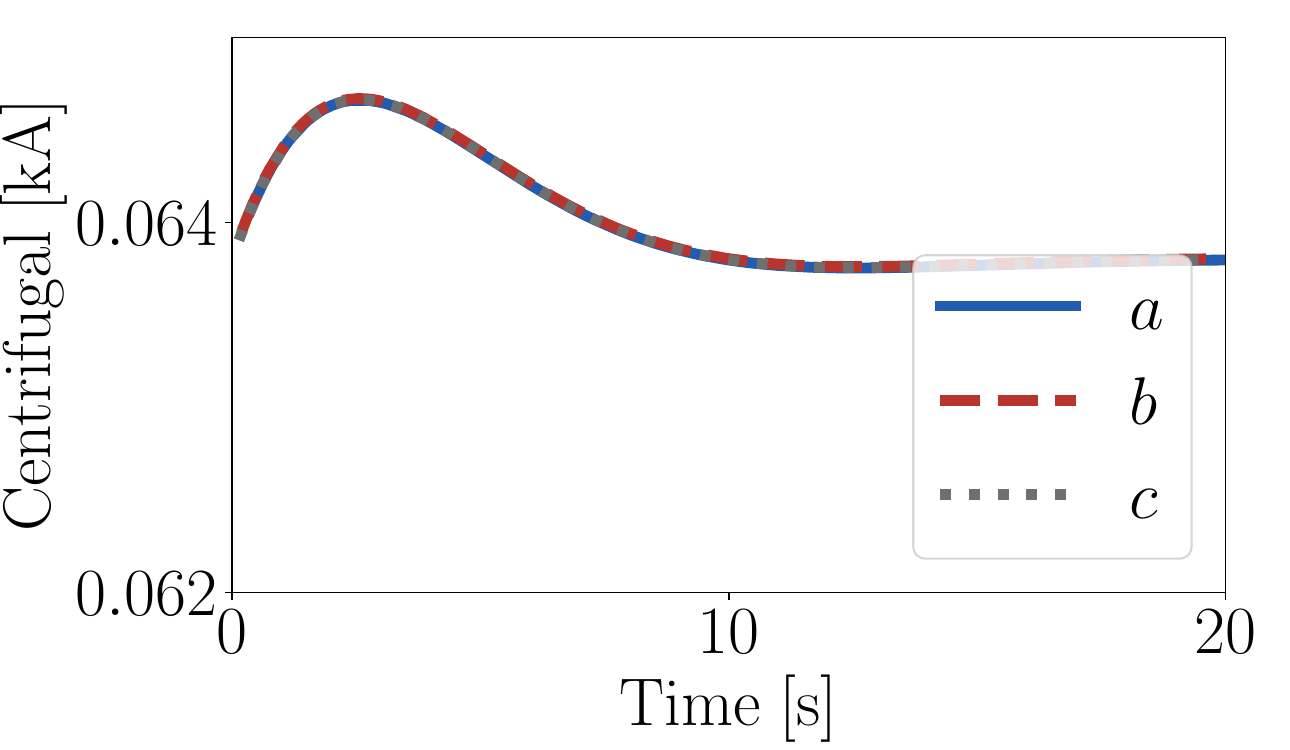}}
  \caption{Centrifugal}
  \label{fig:E3:fcen_env}
  \end{subfigure}
  \begin{subfigure}{.49\linewidth}
  \centering
  \resizebox{\linewidth}{!}{\includegraphics{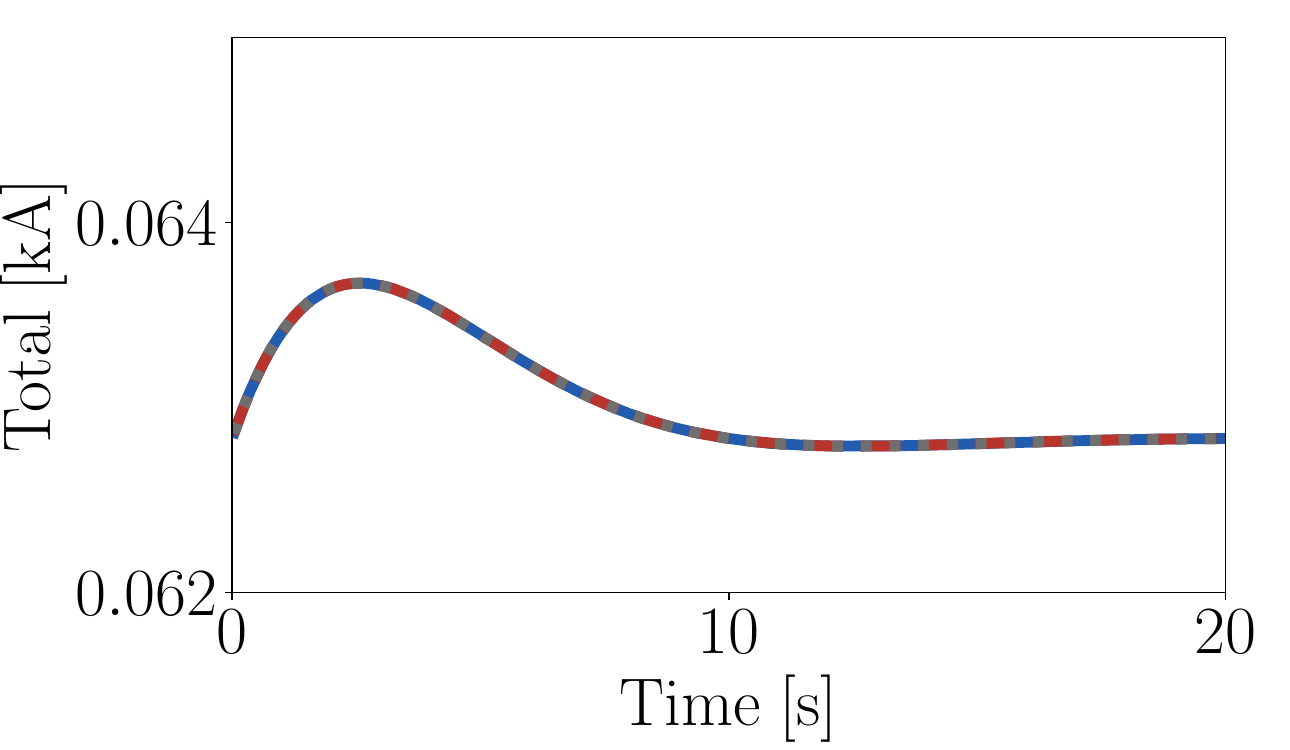}}
  \caption{Total}
  \label{fig:E3:ftot_env}
  \end{subfigure}
  \caption{3-phase capacitor with non-stationary AC voltage: upper
    envelopes of centrifugal and total current. The offset is due to
    the effect of a small negative Euler current.}
  \label{fig:E3:f_envelope}
  \vspace{-3mm}
\end{figure}

The upper envelopes of the centrifugal and total capacitor current for
the duration of the frequency transient are illustrated in
Fig.~\ref{fig:E3:f_envelope}.  It is seen that the total current has a
small offset of about $-1$~A with respect to the centrifugal, which is
again due to the effect of the Euler current.

\begin{figure}[ht!]
\centering
  \begin{subfigure}{.49\linewidth}
  \centering
  \resizebox{\linewidth}{!}{\includegraphics{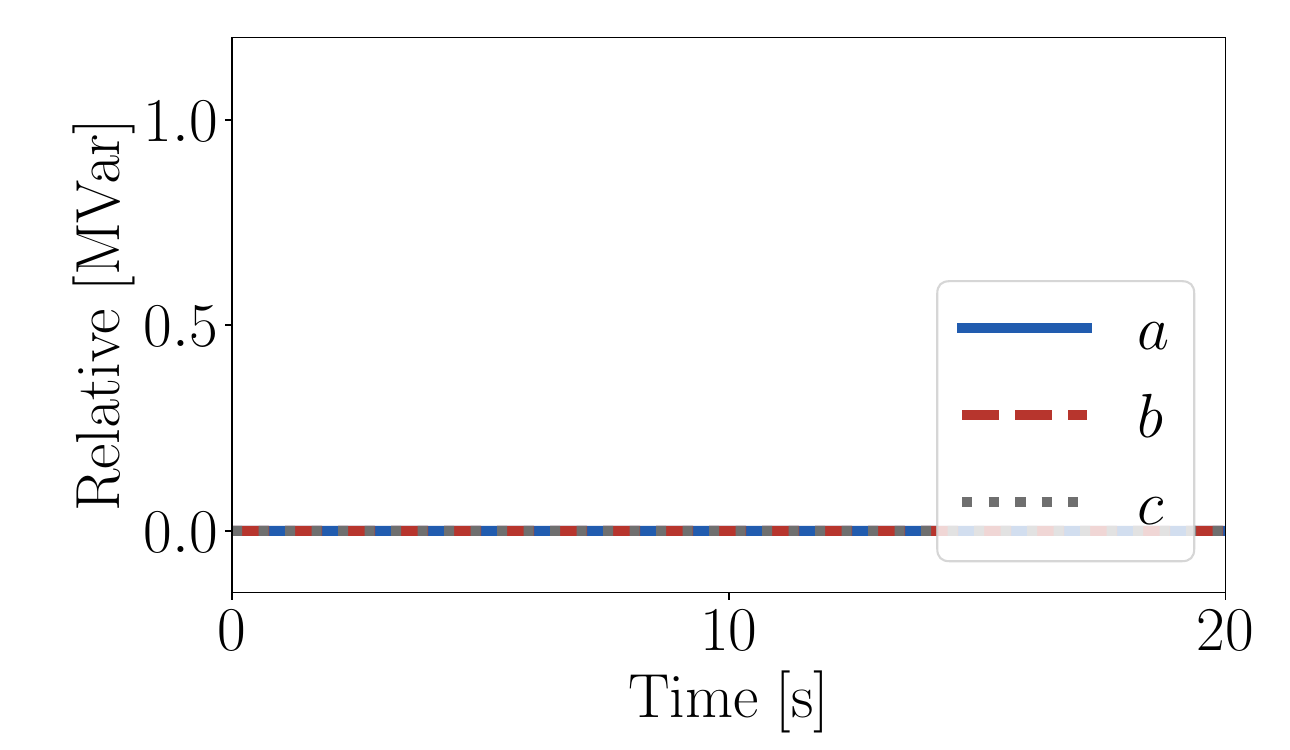}}
  \caption{Relative}
  \label{fig:E3:Qrel}
  \end{subfigure}
  \begin{subfigure}{.49\linewidth}
  \centering
  \resizebox{\linewidth}{!}{\includegraphics{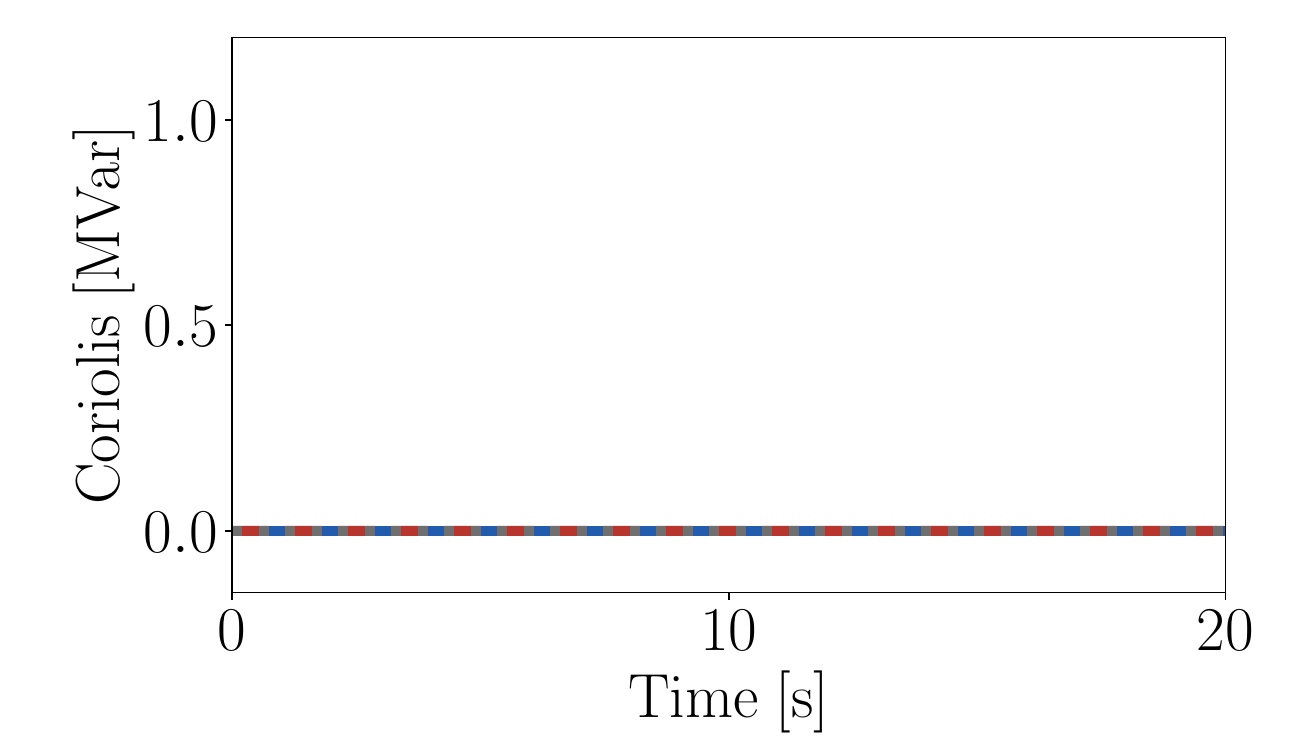}}
  \caption{Coriolis}
  \label{fig:E3:Qcor}
  \end{subfigure}
  \begin{subfigure}{.49\linewidth}
  \centering
  \resizebox{\linewidth}{!}{\includegraphics{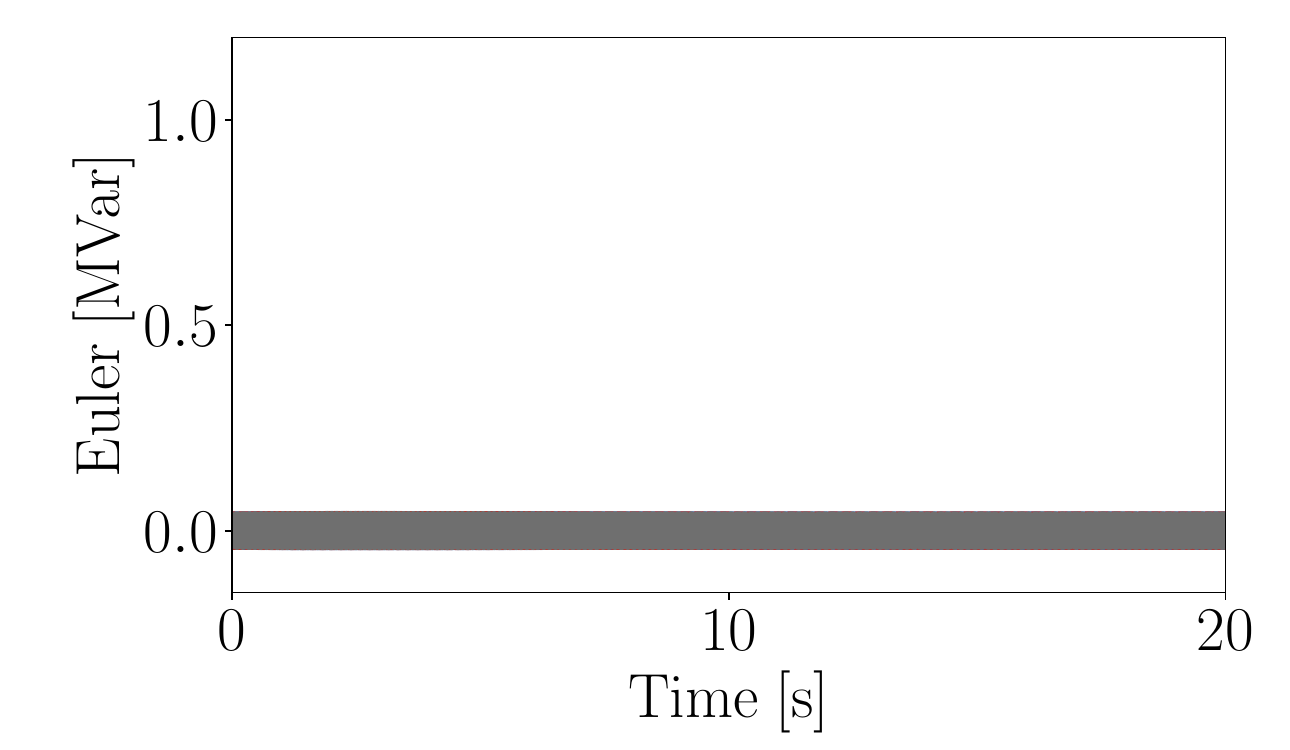}}
  \caption{Euler}
  \label{fig:E3:Qeul}
  \end{subfigure}
  \begin{subfigure}{.49\linewidth}
  \centering
  \resizebox{\linewidth}{!}{\includegraphics{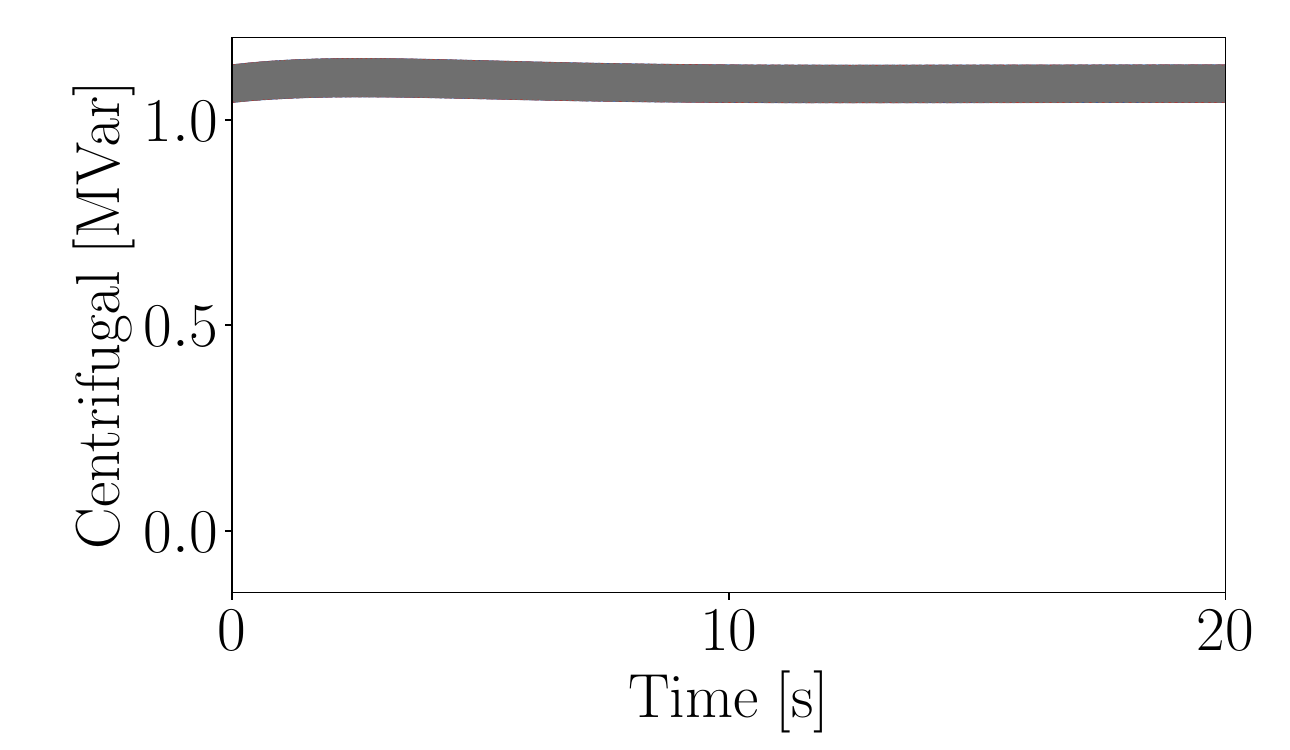}}
  \caption{Centrifugal}
  \label{fig:E3:Qcen}
  \end{subfigure}
  \caption{3-phase capacitor with non-stationary balanced AC voltage:
    Relative, Coriolis, Euler and centrifugal components of reactive
    power pseudovector.}
  \label{fig:E3:coriolis_theorem:Q}
  \vspace{-3mm}
\end{figure}

\begin{figure}[h!]
  \centering
  \resizebox{0.7\linewidth}{!}{\includegraphics{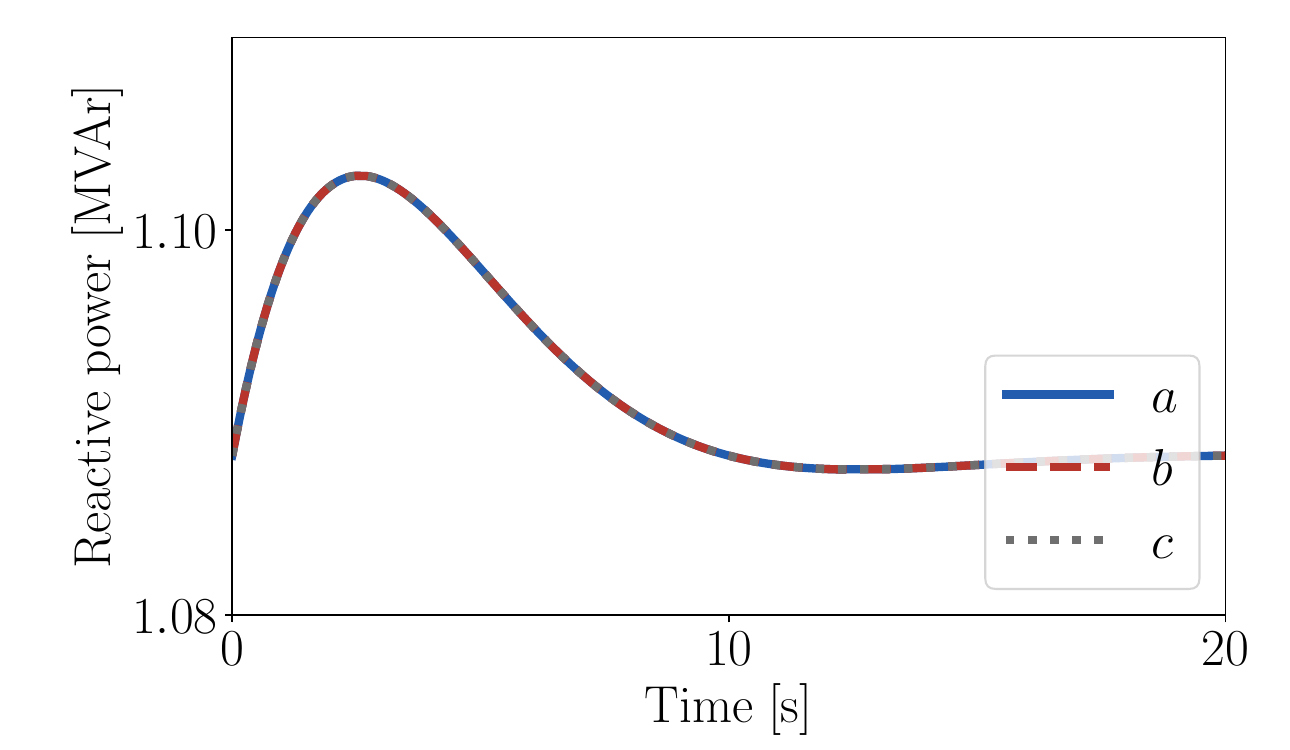}}
  \caption{Non-stationary case:
  components of capacitor reactive power $\bfg Q$.}
  \label{fig:E3:Q}
\end{figure}

Finally, for the instantaneous active power we have that $p=0$, as
expected.  On the other hand, as shown in
Fig.~\ref{fig:E3:coriolis_theorem:Q}, the instantaneous reactive power
pseudovector is non-zero and its response is dominated by the
centrifugal component. Following from the time-varying frequency
examined, the phase components of the total instantaneous reactive
power pseudovector $\bfg Q$ go through a transient, which is clearly
illustrated in Fig.~\ref{fig:E3:Q}.

\color{black}

\subsection{Comparison with Conventional Approaches to Define the Reactive Power}

The proposed framework to provides an unconventional, yet, physically
informed way to decompose the expression of the reactive power.
Compared to the conventional reactive power definitions provided in
the standard \cite{IEEEStd1459}, the following appear as the main
differences and advantages.
\begin{itemize}
\item The standard provides various definitions, depending on the
  operating conditions (balanced/unbalanced,
  sinusoidal/non-sinusoidal), and for each operating condition,
  defines various types of reactive and apparent powers.  In this
  work, we show that the reactive power has an underlying physical
  meaning and, as such, its definition must be unique.
\item The standard assumes stationary conditions.  More crucially, as
  also highlighted in \cite{Kirkham:2022}, all definitions require
  knowledge of the frequency.  On the other hand, the proposed
  approach provides the value of the frequency as part of the
  calculations.  As a consequence, the frequency does not need to be
  known \textit{a priori}, nor needs to be constant.  This is a major
  advantage of the proposed framework with respect to existing active
  and reactive power definitions.
\item The proposed framework allows a decomposition of the reactive
  power in terms that have a physical meaning.  Also the standard
  proposes a decomposition of the power into various components
  (fundamental, harmonics, nonactive, etc.).  However, the terms
  defined in the standard are a consequence of the Fourier transform
  of voltages and currents.  If a different expansion were utilised
  (e.g., Hilbert-Huang transform), different terms would be obtained.
  The proposed framework provides a decomposition that is linked to
  geometric invariants which are not affected by the chosen reference
  frame.
\item We also observe that for the balanced sinusoidal case, the
  proposed framework is more precise than the definition of the
  standard because the frequency does not need to be assumed equal to
  the synchronous reference.  This issue of conventional definitions
  is even more apparent for the balanced, slowly time-varying
  conditions described in the example discussed in
  Section~\ref{ex:nonbalanced}.  In this case, the standard definition
  of reactive power would not only be imprecise due to the fact that
  the frequency is not constant, but it would also not able to capture
  the terms due to relative, Coriolis and Euler effects, which are
  small but not null during the transient.
\end{itemize}

\section{Conclusions}
\label{sec:conc}

In this work, we propose a novel definition of instantaneous power.
This definition is based on generalized Lagrangian coordinates and
well-known concepts borrowed from classical mechanics and basic
differential geometry.  The proposed expression \eqref{eq:hatS2} for
the instantaneous power reveals new insights on its physical meaning
as it is shown to be the product of kinetic energy and a geometric
frequency operator.  We also provide a rigorous mathematical framework
to this interpretation, and show the instantaneous power's dependence
on the geometric invariants provided by the Frenet frame apparatus
such as curvature and torsion.  The proposed approach also allows
decomposing active and reactive powers into various components, each
with a precise physical meaning based on apparent forces.  For the
stationary sinusoidal case, for example, we show that the reactive
power is exclusively due to a centrifugal acceleration.  Other terms,
such as Coriolis and Euler accelerations, appear in unbalanced and
non-sinusoidal conditions.  The proposed approach has two main
advantages compared with currently available definitions of the
instantaneous power: (i) it allows a better and, as based on a
mechanical analogy, intuitive understanding of what reactive power is;
and (ii) it provides a mathematical framework that can help improve
the performance and design better controllers for unbalanced and
non-sinusoidal systems.  This appears particularly important in modern
power systems that are dominated by power electronic converters.  In
future work, we aim at extending the proposed approach to nonlinear
circuit components as well as exploiting its features to improve power
system dynamic performance.

\appendices

\section{Algebra of Multivectors}
\label{app:clifford}

Multivectors extend the concept of complex numbers, quaternions and
vectors to a collection of quantities that include scalars, vectors,
bivectors (or pseudo-vectors), trivectors, etc. (see, for example, the
first chapter in \cite{Jancewicz:1989}).  Since in this work we do not
use more than three dimensions, it suffices to consider multivectors
composed exclusively of scalars and vectors.  Moreover, pseudo-vectors
can be operated as vectors through Hodge duality.  In turn, the
multivectors considered in this paper are equivalent to Hamiltonian
quaternions:
\begin{equation}
  \hat{X} = \lambda + \bfg X \, ,
\end{equation}
where $\lambda$ is a scalar and $\bfg X$ is a vector or pseudo-vector.  
Then, the \textit{geometric product} is equivalent to the Hamiltonian product of quaternions and  can be written using only the inner and cross products, as follows:
\begin{equation}
  \label{eq:geomprod}
  \begin{aligned}
    \hat{X} \otimes \hat{Y}
    &= (\lambda + \bfg X) \otimes (\mu + \bfg Y) \\
    &= ( \lambda \, \mu - \bfg X \cdot \bfg Y) + (\lambda \, \bfg Y + \mu \, \bfg X  + \bfg X \times \bfg Y  )\, .
  \end{aligned}
\end{equation}
Same rules apply if one of any of the two multivectors includes a vector instead of a pseudo-vector, e.g., $\hat{X} = \lambda + \bfg x$.
The conjugate of a multivector is defined as:
\begin{equation}
  \hat{X}^* = \lambda - \bfg X \, .
\end{equation}
Then, one has:
\begin{equation}
  (\hat{X} \otimes \hat{Y})^* = \hat{Y}^* \otimes \hat{X}^* \, ,
\end{equation}
and
\begin{equation}
  \hat{X} \otimes \hat{X}^* = |\hat{X}|^2 = \lambda^2 + |\bfg X|^2 \, .
\end{equation}
%
%

\section{Frenet Frame of Space Curves}
\label{app:frenet}

Let us consider a space curve
$\bfg x:[0,+\infty)\rightarrow\mathbb{R}^3$ with
$\bfg x = (x_1, x_2, x_3)$. Where $x_1=x_1(t)$, $x_2=x_2(t)$,
$x_3=x_3(t)$, is the set of parametric equations for the
curve.  Equivalently:
\begin{equation}
  \bfg x = x_1 \, \e{1} + x_2 \, \e{2} + x_3 \, \e{3} \, ,
\end{equation}
where $(\e{1}, \e{2}, \e{3})$ is an orthonormal basis.  The arc length $s$
of the curve is defined as:
\begin{equation}
  s = \int_0^t \sqrt{\bfg u(r) \cdot \bfg u(r)} \, dr + s_0 \, ,
\end{equation}
from which one obtains the expression:
\begin{equation}
  \label{eq:s}
  s' = \frac{ds}{dt} = \sqrt{\bfg u \cdot \bfg u} = |\bfg u| \, ,
\end{equation}
where 
\begin{equation}
  \bfg u =
  \frac{d}{dt} (x_1 \, \e{1}) +
  \frac{d}{dt} (x_2 \, \e{2}) +
  \frac{d}{dt} (x_3 \, \e{3}) \, ,
\end{equation}
and $\cdot$ denotes the inner product of two vectors, which in
three dimensions, for $\bfg a = (a_1, a_2, a_3)$, $\bfg b = (b_1, b_2, b_3)$, becomes:
\begin{equation}
  \label{eq:inner}
  \bfg a \cdot \bfg b = a_1b_1 + a_2b_2 + a_3b_3 \, .
\end{equation}
The arc length $s$ is an invariant of the curve.  It is relevant to
observe that, according to the chain rule, the derivative of $\bfg x$
with respect to $s$ can be written as:
\begin{equation}
  \label{eq:xdot}
  \bfd x =
  \frac{d \bfg x}{d s} =
  \frac{d\bfg x}{dt} \frac{dt}{ds} = \frac{\bfg u}{s'} =
  \frac{\bfg u}{|\bfg u|} \, .
\end{equation}
The vector $\bfd x$ has magnitude 1 and is tangent to the curve
$\bfg x$.

The Frenet frame is defined by the tangent vector $\T$, the normal
vector $\N$ and the binormal vector $\B$, as follows:
\begin{equation}
\label{eq:TNB}
  \begin{aligned}
    \T &= \bfd x \, , \qquad
    \N &= \frac{\bfdd x}{|\bfdd x|} \, , \qquad
    \B &= \T \times \N \, , 
  \end{aligned}
\end{equation}
where $\times$ represents the cross product, which in three dimensions
can be written as the determinant of a matrix, as follows:
\begin{equation}
    \bfg a \times \bfg b = 
    \left | 
    \begin{matrix}
    \e{1} & \e{2} & \e{3} \\
    a_1 & a_2 & a_3 \\
    b_1 & b_2 & b_3 \\
    \end{matrix} 
    \right | \, .
\end{equation}
%
%
%
%
%
%

In \cite{freqfrenet}, the azimuthal ($\wk$) and torsional ($\wt$)
frequencies are defined as follows:
\begin{align}
  \label{eq:kappa}
  \wk &= \frac{|\bfg u \times \bfg u'|}{|\bfg u|^2} \, , \\
  \label{eq:tau}
  \wt & = \frac{\bfg u \cdot (\bfg u' \times \bfg u'')}{ \wk^2 \, |\bfg u|^3} \, .
\end{align}
These frequencies link the Frenet frame with its time derivatives, as
follows:
\begin{equation}
  \begin{aligned}
    \label{eq:serret}
    \T' &= \wk \N \, , \\
    \N' &= -\wk \T + \wt \B \, , \\
    \B' &= -\wt \N \, ,
  \end{aligned}
\end{equation}
which are the well-known Frenet-Serret equations.

Finally, \eqref{eq:TNB} can be equivalently expressed as:
\begin{equation}
  \label{eq:frenet2}
  \begin{aligned}
    \bfg u &= |\bfg u| \, \T  \, , \\
    \bfg n_u &= \bfg u' - \wr \, \bfg u =
             \sqrt{|\bfg u'|^2 - (|\bfg u|')^2} \, \N  \, , \\
    \sw &= \bfg u \times \bfg n_u =
          \frac{\bfg u \times \bfg u'}{|\bfg u|^2} =
          \wk \, \B \, ,
  \end{aligned}
\end{equation}
where $\wr = |\bfg u|'/|\bfg u|$ is the \textit{radial frequency}
defined in \cite{freqfrenet}.
Noting that $\bfg n_u = \sw \times \bfg u$, one can obtain the
following expression for the first time derivative of 
$\bfg u$
\cite{freqfrenet}:
\begin{align}
  \label{eq:du}
  \begin{aligned}
    \bfg u' &=
    \wr \, \bfg u + \sw \times \bfg u \, .
  \end{aligned}
\end{align}
%


\section{Vector Product Identities}
\label{app:identities}

The following identities for triple and quadruple vector products are relevant for the derivations presented in the work.
\begin{itemize}

\item The dot (inner) product is commutative:
\begin{equation}
  \label{eq:vector1}
  \bfg a \cdot \bfg b = 
  \bfg b \cdot \bfg a \, .
\end{equation}
\item The cross product is anticommutative:
\begin{equation}
  \label{eq:vector2}
  \bfg a \times \bfg b = -
  \bfg b \times \bfg a \, .
\end{equation}
\item If $\bfb R$ is a matrix that satisfies ${\rm det}(\bfb R) = 1$, then:
\begin{equation}
  (\bfb R \, \bfg a) \times (\bfb R \, \bfg b) = \bfb R \, ( \bfg a \times \bfg b) \, .
\end{equation}
\item Scalar triple vector product circular shift:
\begin{equation}
  \label{eq:triple}
  \bfg a \cdot (\bfg b \times \bfg c) = 
  \bfg c \cdot (\bfg a \times \bfg b) = 
  \bfg b \cdot (\bfg c \times \bfg a) \, .
\end{equation}
\item Lagrange identity:
\begin{equation}
  \label{eq:lagrange}
  \bfg {a} \times (\bfg {b} \times \bfg {c} ) =
  (\bfg {a} \cdot \bfg {c} ) \, \bfg {b} -
  (\bfg {a} \cdot \bfg {b} ) \, \bfg {c} \, .
\end{equation}
\item Jacobi identity:
\begin{equation}
  \label{eq:jacobi}
  \triple{a}{b}{c} + \triple{b}{c}{a} + \triple{c}{a}{b} = \bfg 0 \, .
\end{equation}
\item Scalar quadruple product:
\begin{equation}
  \label{eq:quad1}
  (\bfg a \times \bfg b) \cdot (\bfg c \times \bfg d) =
  (\bfg a \cdot \bfg c)(\bfg b \cdot \bfg d) -
  (\bfg a \cdot \bfg d)(\bfg b \cdot \bfg c) \, . 
\end{equation}
%
%
\end{itemize}

\section{Link to Geometric Invariants}
\label{sec:relative}

In this section, we derive the analytical expressions of $\orbit$ and $T$ in terms of geometric invariants, such as curvature and torsion provided by the Frenet apparatus.  With this aim, we express the instantaneous power multivector 
using the relative coordinates defined by the moving Frenet frame.

\begin{figure}[ht!]
  \centering
  \resizebox{0.5\linewidth}{!}{\includegraphics{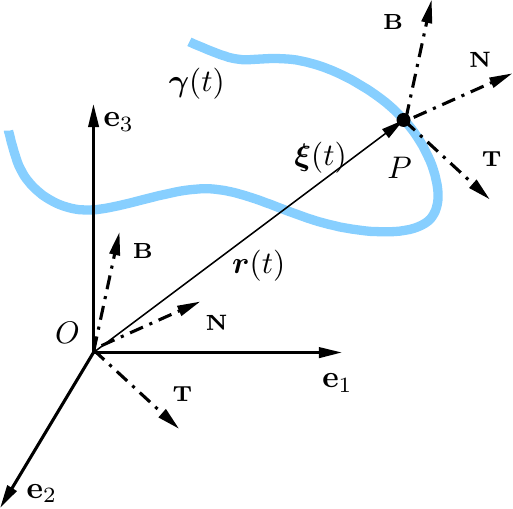}}
  \caption{Coordinates of a space curve $\bfg \gamma(t)$ in absolute and Frenet
    coordinates.}
  \label{fig:frenet}
\end{figure}

Let us consider the space curve $\bfg \gamma$ represented in Fig.~\ref{fig:frenet}.  This curve has coordinates $\bfg r$ with respect to the origin $O$ of fixed Cartesian axes $(\e{1}, \e{2}, \e{3})$.
The change of coordinates on the moving Frenet frame is given by:
\begin{equation}
  \bfg \xi = \bfb F \, \bfg r \, ,
\end{equation}
where $\bfg \xi$ is the position in Frenet-frame coordinates and: 
\begin{equation}
  \bfb F = [\T, \N, \B]\Tr \, ,
\end{equation}
i.e., the rows of $\bfb F$ are the tangent, normal and binormal vectors of the Frenet frame (see Appendix~\ref{app:frenet}).  
Note that $\T$, $\N$, $\B$ are defined in
Cartesian coordinates.
Then, $\bfg r$ is rewritten as:
\begin{equation}
  \bfg r = \bfb F^{-1} \, \bfb \xi = \bfb F\Tr \, \bfg \xi \, .
\end{equation}

Figure \ref{fig:frenet} illustrates the coordinate transformation.
The position of point $P$ along the curve $\bfg \gamma(t)$ is
represented by the vector $\bfg r$, which describes the distance
$\overline{OP}$ in the $(\e{1}, \e{2}, \e{3})$ coordinates.  The basis
$(\T, \N, \B)$ of the Frenet frame which is determined based on the
curve $\bfg \gamma(t)$ can be also utilized as coordinates to define
the segment $\overline{OP}$.  


%

We are interested in expressing the velocity $\bfg u$ and acceleration
$\bfg u'$ as functions of $\bfg \xi$, the velocity $\bfg \xi'$ and
acceleration $\bfg \xi''$.  Since $\bfg \xi$ is defined on a
non-inertial frame, namely the moving Frenet frame, its motion is
relative.  One has thus to take into account in the motion also the
coordinates $(\T, \N, \B)$.  First note that, by construction,
$\bfb F$ is orthonormal, that is, its transpose is equal to its
inverse ($\bfb F\Tr = \bfb F^{-1}$) and its determinant is
${\rm det}(\bfb F) = 1$.  Then, from Cartan's theory on moving frames
\cite{Needham:2021}, the following result holds:
\begin{equation}
  \label{eq:Wdef}
  \bfg \Omega_{\rm d} = \bfb F \, (\bfb F')\Tr = - \bfb F' \bfb F\Tr = 
  \begin{bmatrix}
    0 & -\wk & 0 \\
    \wk & 0 & -\wt \\
    0 & \wt & 0 \\ 
  \end{bmatrix} ,
\end{equation}
and $\wt$ and $\wk$ are the torsional and azimuthal frequencies
\cite{freqfrenet, paradoxes}.  Equation \eqref{eq:Wdef} can be written
as:
\begin{equation}
  \label{eq:serret:F'}
  \bfb F' = -\bfg \Omega_{\rm d} \, \bfb F \, ,
\end{equation}
which is the matrix form of the well-known Frenet-Serret equations
\eqref{eq:serret} given in Appendix~\ref{app:frenet}.
The skew-symmetric matrix $\bfg \Omega_{\rm d}$ can be written as:
\begin{equation}
  \label{eq:W}
  \bfg \Omega_{\rm d} = \darboux \times \, ,
\end{equation}
where $\darboux$ is the \textit{Darboux vector}:
\begin{equation}
  \label{eq:darboux}
  \darboux = \wt \, \T + \wk \, \B \, .
\end{equation}

We can now express the velocity and acceleration using the coordinates
$\tnb$.  The velocity can be written as:
\begin{equation}
  \label{eq:dx}
  \bfg u = \bfb F\Tr \bfg \xi' + (\bfb F')\Tr \bfg \xi \, ,
\end{equation}
and, multiplying by $\bfb F$ both sides:
\begin{equation}
  \label{eq:Pdx}
  \bfg \nu = \bfb F \, \bfg u =
  \bfb F \bfb F\Tr \bfg \xi' + \bfb F (\bfb F')\Tr \bfg \xi \, ,
\end{equation}
or, equivalently,
\begin{equation}
  \label{eq:nu}
  \begin{aligned}
    \bfg \nu
    &= \bfg \xi' + \darboux \times \bfg \xi \, .
  \end{aligned}
\end{equation}

The acceleration is given by:
\begin{equation}
  \label{eq:alpha0}
  \begin{aligned}
  \bfg \alpha &= \bfb F \bfb x'' \\ &= \bfg \xi'' +
  2 \darboux \times \bfg \xi' +
  \darboux' \times \bfg \xi +
  \darboux \times (\darboux \times \bfg \xi) \, .
  \end{aligned}
\end{equation}
In fact, from \eqref{eq:Pdx}, one has:
\begin{equation}
   \begin{aligned}
   \bfg \nu' &= \bfb F' \bfg u + \bfb F \bfg u' \\
   &= \bfb F' \bfb F\Tr (\bfg \xi ' + \darboux \times \bfg \xi) + \bfg \alpha \\
   &= -\darboux \times (\bfg \xi ' + \darboux \times \bfg \xi) + \bfg \alpha \\ 
   &= -\darboux \times \bfg \xi ' -\darboux \times \darboux \times \bfg \xi + \bfg \alpha \, ,  
   \end{aligned}
\end{equation}
where we have defined $\bfg \alpha = \bfg F \bfg u'$ and, from \eqref{eq:nu}:
\begin{equation}
   \begin{aligned}
   \bfg \nu' &= \bfg \xi'' + \darboux' \times \bfg \xi + \darboux \times \bfg \xi \, .
   \end{aligned}
\end{equation}
Moreover, using the Lagrange formula for the triple cross product \eqref{eq:lagrange}, 
equation \eqref{eq:alpha0} can be equivalently written as:
\begin{equation}
  \label{eq:alpha}
  \bfg \alpha = \bfg \xi'' +
  2 \darboux \times \bfg \xi' +
  \darboux' \times \bfg \xi + (\bfg \xi \cdot \darboux) \, \darboux -
  \omega_{\rm d}^2 \, \bfg \xi \, ,
\end{equation}
where $\omega_{\rm d}^2 = \wk^2 + \wt^2$ is the magnitude of the
Darboux vector and is referred to as \textit{Lancret curvature} for
unit-speed curves \cite{menninger}.  
Similarly to \eqref{eq:coriolis}, equation \eqref{eq:alpha} is yet
another version of the Coriolis theorem.  This version features a
rotation vector $\darboux$ that defines the intrinsic rotation of the
curve $\bfg r$ and is defined by the curvature and torsion of the
curve itself.

\subsection{Momentum Density}

In this section, we provide the expressions of the momentum density
and angular momentum using the relative coordinates on the moving
Frenet frame.  This operation allows decomposing the energy and
instantaneous power into terms that have specific meaning, similar to
the terms into which the acceleration is decomposed in
\eqref{eq:alpha}.  The goal is to be able to identify unequivocally
and based on a precise physical interpretation the active and reactive
powers in various operating conditions.

We begin with the momentum density $\ell$, the magnitude of which, as
any scalar, is unaltered in relative coordinates.  In fact, defining
the momentum in $(\T, \N, \B)$ coordinates:
\begin{equation}
  \label{eq:pi}
  \bfg \pi = \bfb F \mom = m \, \bfg \nu \, , 
\end{equation}
the momentum density becomes:
\begin{equation}
  \begin{aligned}
    \ell &= \bfg r \cdot \mom  = m \, \bfg r \cdot \bfg u \\ &=
    (\bfb F\Tr \bfg \xi) \cdot (\bfb F\Tr \bfg \pi) =
    (\bfb F\Tr \bfg \xi)\Tr (\bfb F\Tr \bfg \pi) \\ &= 
    \bfg \xi\Tr \bfb F \bfb F\Tr \bfg \pi =
    \bfg \xi\Tr \bfg \pi = \bfg \xi \cdot \bfg \pi \\
    &= m \, \bfg \xi \cdot \bfg \nu = m \, \bfg \xi \cdot \bfg \xi' \, ,
  \end{aligned}
\end{equation}
where the last expression has been obtained using \eqref{eq:nu} and
$\bfg \xi \perp (\darboux \times \bfg \xi)$.
Then, note that:
\begin{equation}
   \varrho_r = \frac{\bfg r \cdot \bfg u}{\bfg r \cdot \bfg r} = \frac{\bfg \xi \cdot \bfg \xi'}{\bfg \xi \cdot \bfg \xi} \, ,
\end{equation}
and that:
\begin{equation*}
  \begin{aligned}
    |\bfg r|^2
    = \bfg r\Tr \bfg r 
    = (\bfb F\Tr \bfg \xi)\Tr (\bfb F\Tr \bfg \xi) 
    = \bfg \xi\Tr \bfb F \bfb F\Tr \bfg \xi 
    = \bfg \xi\Tr \bfg \xi 
    = |\bfg \xi|^2 \, ,
  \end{aligned}
\end{equation*}
where we have used that $\bfb F$ is orthonormal and, hence,
$\bfb F \bfb F\Tr$ is the identity matrix.\footnote{This result is
  consistent as the coordinate transformation from
  $(\e{1}, \e{2}, \e{3})$ to $(\T, \N, \B)$ consists only in a
  rotation and, thus, preserves lengths.}  The moment of inertia is:
\begin{equation}
  I = m \, \bfg r \cdot \bfg r = m \, |\bfg r|^2 = m \, |\bfg \xi |^2 \, .
\end{equation}
From the latter three equations descends that the momentum density is
$\ell = I \, \ur$, which is the same expression as \eqref{eq:ell2}.
In conclusion, since the Frenet frame imposes only a rotation of the
coordinates, all scalar quantities have the same value in absolute and
relative coordinates.

\subsection{Angular Momentum}

The angular momentum in the Frenet frame becomes:
\begin{equation}
  \label{eq:Lambda}
  \begin{aligned}
    \boldsymbol{\mathit{\Lambda}} 
    &= \bfb F \bfg L = \bfb F (\bfg r \times \mom) \\
    &= (\bfb F \bfg r ) \times (\bfb F \mom) = m \, (\bfb F \bfg r ) \times (\bfb F \bfg u) \\
    &= m\, \bfg \xi \times \bfg \nu = m \, \bfg \xi \times (\bfg \xi' + \darboux \times \bfg \xi) \\
    &= m \, \bfg \xi \times \bfg \xi' + m \, |\bfg \xi|^2 \darboux - m \, (\darboux \cdot \bfg \xi) \, \bfg \xi \, .
  \end{aligned}
\end{equation}
Note that the identity between first and second line of
\eqref{eq:Lambda} holds because $\bfb F$ is orthonormal and because of
the property of the cross product \eqref{eq:vector2} given in
Appendix~\ref{app:identities}.
Dividing and multiplying the first and third terms of the last
expression in \eqref{eq:Lambda} by $|\bfg \xi|^2$ leads to:
\begin{equation}
  \label{eq:Lambda2}
    \boldsymbol{\mathit{\Lambda}} = I \, [ \bfg \omega_{\xi} + \darboux - \bfg \nu_{\rm d} ] \, ,
\end{equation}
where we have defined the relative angular velocity $\bfg \omega_{\xi}$ as:
\begin{equation}
  \label{eq:wxi}
  \bfg \omega_{\xi} = \frac{\bfg \xi \times \bfg \xi'}{\bfg \xi \cdot \bfg \xi} \, ,
\end{equation}
and the component of the relative position along the axis of rotation given by the Darboux vector $\bfg \nu_{\rm d}$ as:
\begin{equation}
   \bfg \nu_{\rm d} = \frac{\darboux \cdot \bfg \xi}{\bfg \xi \cdot \bfg \xi} \, \bfg \xi \, .
\end{equation}

From \eqref{eq:Lambda2}, we obtain the expression of the orbital angular velocity in the Frenet frame:
%
%
%
\begin{equation}
\label{eq:orbit}
\boxed{
  \bfb F \, \orbit = \bfg \omega_{\xi} + \darboux - \bfg \nu_{\rm d}
  }
\end{equation}
that shows that the orbital angular velocity is, in general, different
from the Darboux vector and coincides with it --- except for the
coordinate change due to the Frenet frame --- if and only if
$\bfg \xi \, \| \, \bfg \xi'$ and $\darboux \perp \bfg \xi$.
Moreover, in the first example discussed in Section
\ref{sec:examples}, we show that
$\bfg \omega_{\xi} = \bfg \nu_{\rm d} = \bfg 0$, and hence
$\orbit = \darboux$, hold in balanced stationary conditions.

\subsection{Kinetic Energy}

The kinetic energy can be written as follows:
\begin{equation}
  \label{eq:Trel}
  \begin{aligned}
  T 
  &= \frac{1}{2} I |\orbit|^2 = \frac{1}{2} m |\bfg u|^2 = \frac{1}{2} m |\bfg \nu|^2 \\
  &= \frac{1}{2} m  \{ |\bfg \xi'|^2 + |\bfg \xi|^2 \lancret^2 - (\darboux \cdot \bfg \xi)^2 + 2 \bfg \xi' \cdot (\darboux \times \bfg \xi) \} \\
  &= \frac{1}{2} I \, (\lancret^2 - |\bfg \nu_{\rm d}|^2 ) + \frac{1}{2} m \, \{ |\bfg \xi'|^2 + 2 \darboux \cdot (\bfg \xi \times \bfg \xi') \} \\
  &= \frac{1}{2} I \, (\lancret^2 - |\bfg \nu_{\rm d}|^2 + 2 \darboux \cdot \bfg \omega_{\xi}) + \frac{1}{2} m \, |\bfg \xi'|^2  \, ,
  \end{aligned}
\end{equation}
where we have utilized \eqref{eq:triple} and \eqref{eq:quad1} --- see Appendix \ref{app:identities}. 
Using \eqref{eq:wxi} and \eqref{eq:quad1} again, one has:
\begin{equation}
  \bfg \omega_{\xi} \cdot \bfg \omega_{\xi} = \frac{|\bfg \xi|^2 |\bfg \xi'|^2 - (\bfg \xi \cdot \bfg \xi')^2}{|\bfg \xi|^4} \, ,
\end{equation}
hence, \eqref{eq:Trel} can be rewritten as:
\begin{equation}
  \label{eq:Trel2}
  \begin{aligned}
  T 
  &=  \frac{1}{2} I \, (\lancret^2 - |\bfg \nu_{\rm d}|^2 + 2 \darboux \cdot \bfg \omega_{\xi} + |\bfg \omega_{\xi}|^2 + \ur^2) \, ,
  \end{aligned}
\end{equation}
or, equivalently:
\begin{equation}
  \label{eq:Tfinal}
  \boxed{T = \frac{1}{2} I ( |\bfg \omega_{\xi} + \darboux - \bfg \nu_{\rm d}|^2 + \ur^2)}
\end{equation}
The latter equation is, as expected, equivalent to the expression of the kinetic energy given in \eqref{eq:T0} and allows expressing the instantaneous power in \eqref{eq:hatS2} as a function of geometric invariants, i.e., $\ur$, $\darboux$ and $\Wgeom{u}$, as well as of the relative angular and linear velocities $\bfg \omega_{\xi}$ and $\bfg \nu_{\rm d}$, respectively.



\vspace{-3mm}


\begin{biography}[{\includegraphics[width=1in, height=1.25in, clip, keepaspectratio]{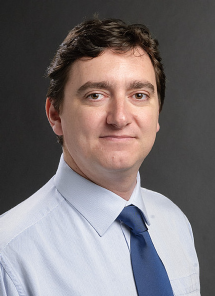}}]
  {Federico Milano} (F'16) received from the University of Genoa,
  Italy, the ME and Ph.D.~in Electrical Engineering in 1999 and 2003,
  respectively.  In 2013, he joined the University College Dublin,
  Ireland, where he is currently a full professor.  He is an IEEE PES
  Distinguished Lecturer, a senior editor of the IEEE Transactions on
  Power Systems, an IET Fellow and editor in chief of the IET
  Generation, Transmission \& Distribution.  He is the chair of the
  IEEE Power System Stability Controls Subcommittee and of the
  Technical Programme Committee of the 23th Power System Computation
  Conference.  His research interests include power system modelling,
  control and stability analysis.
\end{biography}

\vspace{-3mm}

\begin{biography}[{\includegraphics[width=1in, height=1.25in, clip, keepaspectratio]{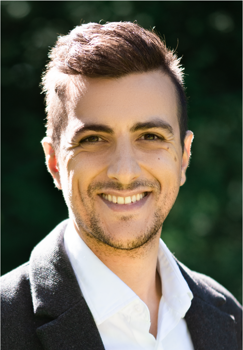}}]
  {Georgios Tzounas} (M’21) received the Diploma (M.E.) in Electrical
  and Computer Engineering from the National Technical Univ.~of
  Athens, Greece, in 2017, and the Ph.D.~from University College
  Dublin (UCD), Ireland, in 2021.  In Jan.-Apr.~2020, he was a
  visiting researcher at Northeastern Univ., Boston, MA.  From 2021 to
  2023, he was a postdoctoral researcher, first with UCD (2021-2022)
  and the with ETH Z\"urich (2022-2023).  Since Apr.~2023, he has been
  an Assistant Professor with the School of Electrical and Electronic
  Engineering at UCD.  His primary research area is power system
  dynamics.
\end{biography}

\vspace{-3mm}

\begin{biography}[{\includegraphics[width=1in, height=1.25in, clip, keepaspectratio]{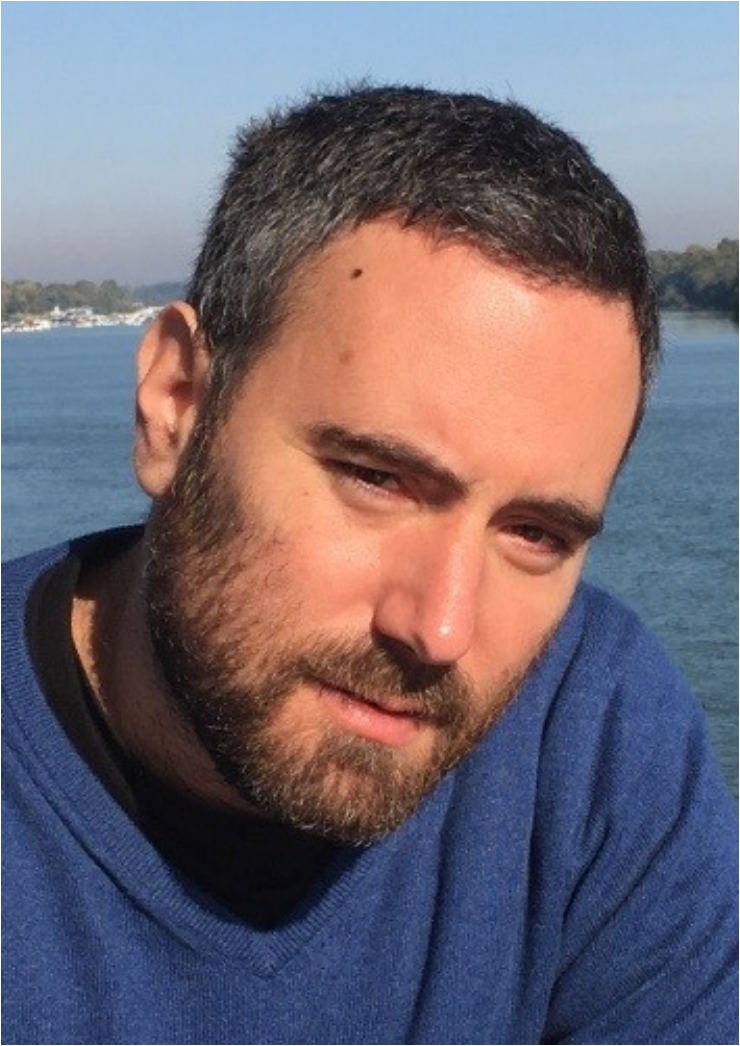}}]
  {Ioannis Dassios} received his Ph.D. in Applied Mathematics from the
  Dpt of Mathematics, Univ.~of Athens, Greece, in 2013. He worked as a
  Postdoctoral Research and Teaching Fellow in Optimization at the
  School of Mathematics, Univ.~of Edinburgh, UK.  He also worked as a
  Research Associate at the Modelling and Simulation Centre,
  University of Manchester, UK, and as a Research Fellow at MACSI,
  Univ.~of Limerick, Ireland.  He is currently a UCD Research Fellow
  at UCD, Ireland.
\end{biography}

\vfill

\end{document}